\documentclass[printer]{aa}
\usepackage{graphicx}
\usepackage{txfonts}

\def\ha{\relax \ifmmode {\rm H}\alpha\else H$\alpha$\fi}
\def\arcsec{\hbox{$^{\prime\prime}$}}

\def\nii{\relax \ifmmode {\rm N\,{\sc ii}}\else N\,{\sc ii}\fi}
\def\hii{\relax \ifmmode {\rm H\,{\sc ii}}\else H\,{\sc ii}\fi}
\def\hi{\relax \ifmmode {\rm H\,{\sc i}}\else H\,{\sc i}\fi}
\def\deg{\hbox{$^{\circ}$}}

\begin{document}

\title{On the morphology of sigma-drop galaxies}
\author{S.~Comer\'on\inst{1}
\and J.~H.~Knapen\inst{1}
\and J.~E.~Beckman\inst{1,2}}

\institute{Instituto de Astrof\'isica de Canarias, E-38200 La Laguna,Spain
\and Consejo Superior de Investigaciones Cient\'ificas, Spain}

\abstract{

{\it Context} Local reductions of the stellar velocity
dispersion in the central regions of galaxies are known as sigma-drops ($\sigma$-drops). Knowing the origin of
these features can lead to better understanding of inner galactic
dynamics.

{\it Aims} We present a sample of 20 $\sigma$-drop galaxies matched with a
control sample of galaxies without $\sigma$-drop in order to search for correlations between $\sigma$-drops
and the properties, primarily morphological, of the nuclear zones and discs of their host galaxies.

{\it Methods} We study the dust and \ha{} distribution at 0.1\,arcsec scale, using {\it Hubble Space Telescope} imaging, in
the circumnuclear zones of the two samples of galaxies, searching for differences
and trying to establish a link between the nuclear kinematics and the host morphology. We have also considered the CO and \hi{}
emission of the galaxies and their luminosity profiles.

{\it Results} We classify the two samples following both
morphological parameters and the luminosity profiles. We find a larger fraction of nuclear dust spirals and \ha{} rings in the $\sigma$-drop sample. We also find that the fraction of Seyfert galaxies in the $\sigma$-drop sample is bigger than that of LINERs and that the reverse is true for the control sample.

{\it Conclusions} Our findings are evidence that a $\sigma$-drop is very probably due to inflow-induced star formation in a dynamically cool disc, or in a gas ring, shock focused by an inner Lindblad resonance above a certain critical density level. The same mechanism that feeds the nuclear ring or the nuclear disc is probably reponsible for the higher rate of Seyfert galaxies among the $\sigma$-drop hosts.


}

\maketitle

\section{Introduction}

It is normally considered physically reasonable that the stars closest to centre of a galaxy, close to the bottom of the potential well, should have the highest velocity dispersions. In other words, a radial plot of velocity dispersion across a galaxy should peak on the nucleus. This prediction, however, would be predicated on the isotropy of the stellar orbits in a spheroid in equilibrium under gravity with its mass centrally concentrated. As the angular resolution of spectroscopic data has improved, detailed measurements on nearby galaxies have shown in some galaxies an unexpected drop in the stellar velocity dispersion as the nucleus is approached, a phenomenon referred to as a dispersion drop or $\sigma$-drop (Emsellem et al.~2001; \cite{EM04}). One of the first known examples of
this phenomenon can be found in \cite{JA88}, later results include
those reported by \cite{JA91}, Bottema (1993), Van der Marel (1994),
Bertola et al.~(1996), Bottema \& Gerritsen (1997), Fisher (1997), Simien \& Prugniel (1997a, b, and c), H\'eraudeau \&
Simien (1998), Simien \& Prugniel (1998), H\'eraudeau et al.~(1999), Simien \& Prugniel (2000), Emsellem et al.~(2001),
Simien \& Prugniel (2002), de Zeeuw et al.~(2002), Aguerri et
al.~(2003), M\'arquez et al.(2003), Shapiro et al.~(2003), Chung \&
Bureau (2004), Falc\'on-Barroso et al.~(2006), Ganda et al.~(2006), Emsellem et al.~(2006), and Dumas et al.~(2007).

The frequency of $\sigma$-drops in galaxy populations has still not been determined with precision, but estimates currently go as high as 50\% of disc galaxies (from the previously cited articles). They have been found in a
wide variety of galaxy types in both active and
non-active galaxies.
A few cases have been observed in lenticular and
elliptical galaxies (Emsellem 2006).

Explanations have been proposed for this phenomenon with theoretical arguments and also using numerical simulations. For disc galaxies a scenario has been proposed in which the stellar population in a $\sigma$-drop galaxy was formed from a circumnuclear rapidly rotating dynamically cool gaseous component. The stars inherit the velocity pattern of the gas from which they form, so their velocity dispersion is lower than that of the older, `underlying' stars. A young stellar population dominates an older population in luminosity, so the resulting spectrum is dominated by the lower velocity dispersion of the younger stars (see, e.g., \cite{WO03}, also
Allard et al.~2005, 2006). This effect can be enhanced by a falling velocity dispersion of the gas towards the centre of a galaxy, due to a strong accumulation of gas in a dissipative disc that would cool the gas in a cold component (i.e., a cold disc as shown in Falc\'on-Barroso et al.~2006). Hydro/$N$-body simulations (\cite{WO06}) have shown that a $\sigma$-drop can form in less than 500\,Myr, but that its lifetime can exceed 1\,Gyr if the nuclear zone is continually fed with gas to sustain star formation. This continued formation of stars is required to maintain a long term $\sigma$-drop, which in turn is required for the observed large $\sigma$-drop fraction among the galaxy population. When star formation ceases, dynamical relaxation will cause the $\sigma$-drop to dissipate in a time-scale that goes from few hundred Myr for short-lived $\sigma$-drops (circumnuclear volume feeding during less than 500\,Myr) to the order of 1\,Gyr or even more for long-lived $\sigma$-drops
(see Fig.~2 in \cite{WO06}).

Although this model is considered not only plausible but probable, there are a number of alternatives. A massive, concentrated dark matter halo could remove kinetic energy from the stellar component. Numerical simulations of this effect can be found in Athanassoula \& Misiriotis (2002), their Fig.~13. In this model the massive halo gives rise to a strong bar, and the $\sigma$-peak is suppressed compared with galaxies having smaller halos. However, these numerical simulations are only done with particles (stars) and do not take into account the effects of gas concentration in the galactic centre. It has also been argued (\cite{DR90}) that a $\sigma$-drop might be a symptom of the absence of a central supermassive black hole, but \cite{MA94} argues against this that a cause might even be the widening of spectral line wings in the presence of a black hole, giving rise to an apparent reduction in $\sigma$.

In this article we present a statistical study of 20 of these $\sigma$-drop galaxies, and match the sample with a control sample to see if we can identify characteristics which distinguish the overall morphologies of the $\sigma$-drop hosts.

In Section 2 we explain the sample selection criteria, and give a brief description of the resulting samples. In Section 3 we show how we processed the data, in Section 4 we study the $\sigma$-drop properties such as size, in Section 5 we describe the morphological classification adopted, in Section 6 we study the presence of activity in nuclei, and in Section 7 we show a test based on the use of luminosity profiles of the sample galaxies. In Section 8 we add \hi +CO radio observations to the previous optical and near IR data sets, applying these to the task of differentiating $\sigma$-drop galaxies from the control sample, and in Section 9 we discuss the material presented in this article, and present our conclusions.

\section{Sample selection}

\begin{table*}
\begin{center}
\caption{\label{source} Source of the velocity dispersion profiles or 2D maps that we used to construct the samples.}
\begin{tabular}{l c | l c}
\hline
\multicolumn{2}{c|}{$\sigma$-drop sample}&\multicolumn{2}{c}{control sample}\\
Galaxy & Source & Galaxy & Source\\
\hline
NGC~1068 & Dumas et al.~(2007) & NGC~488 & Ganda et al.~(2006)\\
NGC~1097 & Emsellem et al.~(2001) & NGC~864 & Ganda et al.~(2006)\\
NGC~1808 & Emsellem et al.~(2001) & NGC~1566 & Bottema (1993)\\
NGC~2460 & Shapiro et al.~(2003) & NGC~2964 & Ganda et al.~(2006)\\
NGC~2639 & M\'arquez et al.~(2003) & NGC~2985 & Dumas et al.~(2007)\\
NGC~2775 & Shapiro et al.~(2003) & NGC~3031 & H\'eraudeau \& Simien (1998)\\
NGC~3021 & H\'eraudeau et al.~(1999) & NGC~3169 & H\'eraudeau \& Simien (1998)\\
NGC~3412 & Aguerri et al.~(2003) & NGC~3198 & Bottema (1993)\\
NGC~3593 & Bertola et al.~(1996) & NGC~3227 & Dumas et al.~(2007)\\
NGC~3623 & Falc\'on-Barroso et al.~(2006) & NGC~3368 & H\'eraudeau et al.~(1999)\\
NGC~3627 & Dumas et al.~(2007)& NGC~3675 & H\'eraudeau \& Simien (1998)\\
NGC~4030 & Ganda et al.~(2006) & NGC~3718 & H\'eraudeau \& Simien (1998)\\
NGC~4303 & H\'eraudeau \& Simien (1998) & NGC~4102 & Ganda et al.~(2006)\\
NGC~4477 & Jarvis et al.~(1988) & NGC~4151 & Dumas et al.~(2007)\\
NGC~4579 & Dumas et al.~(2007) & NGC~4459 & Dumas et al.~(2007)\\
NGC~4725 & H\'eraudeau et al.~(1999) & NGC~5055 & Dumas et al.~(2007)\\
NGC~4826 & H\'eraudeau \& Simien (1998) & NGC~6340 & Bottema (1993) \\
NGC~6503 & Bottema \& Gerritsen (1997) & NGC~6501 & Falc\'on-Barroso et al.~(2006)\\
NGC~6814 & M\'arquez et al.~(2003) & NGC~7331 & H\'eraudeau et al.~(1999)\\
NGC~6951 & Dumas et al.~(2007) & NGC~7742 & Falc\'on-Barroso et al.~(2006)\\
\hline
\hline
\end{tabular}
\end{center}
\end{table*}

Our $\sigma$-drop galaxy sample is a sub-sample of the selection of
galaxies listed by \cite{EM06} as having spectroscopic evidence for
the presence of a drop in the central stellar velocity dispersion. We added one galaxy, NGC~4826 (\cite{HE98}) to bring the number up to 20 galaxies. We
selected those galaxies that had been imaged with the {\it Hubble Space Telescope}  ({\it  HST}) and
which were not classified as edge-on in the RC3 catalogue
(\cite{VA91}). For all of these galaxies we retrieved WFPC2 or ACS red
images from the {\it HST} archive, and whenever available also \ha\ images taken with the WFPC2 or ACS cameras.

\begin{figure*}
\begin{center}
\begin{tabular}{c}
\includegraphics[width=0.45\textwidth]{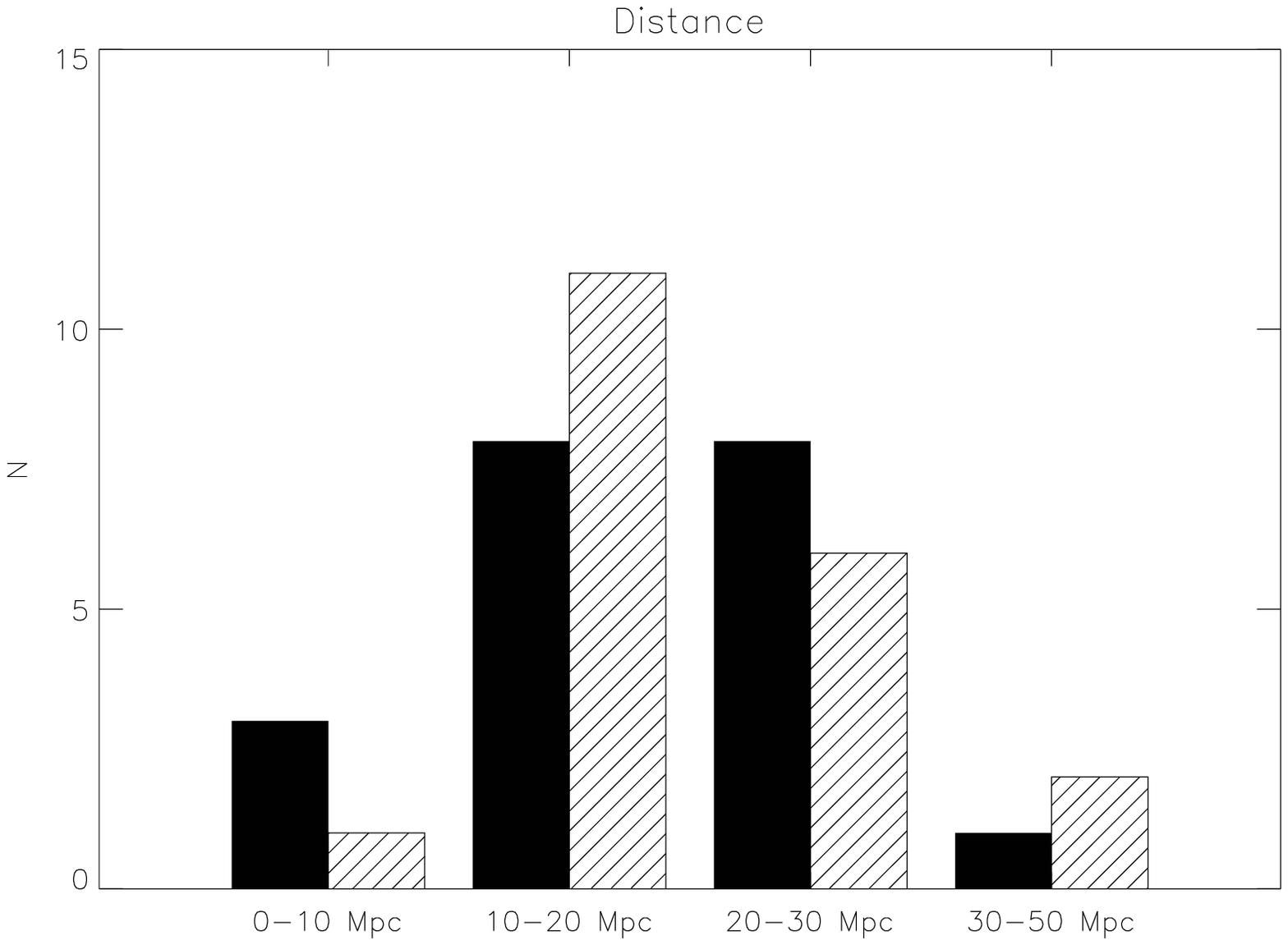}
\includegraphics[width=0.45\textwidth]{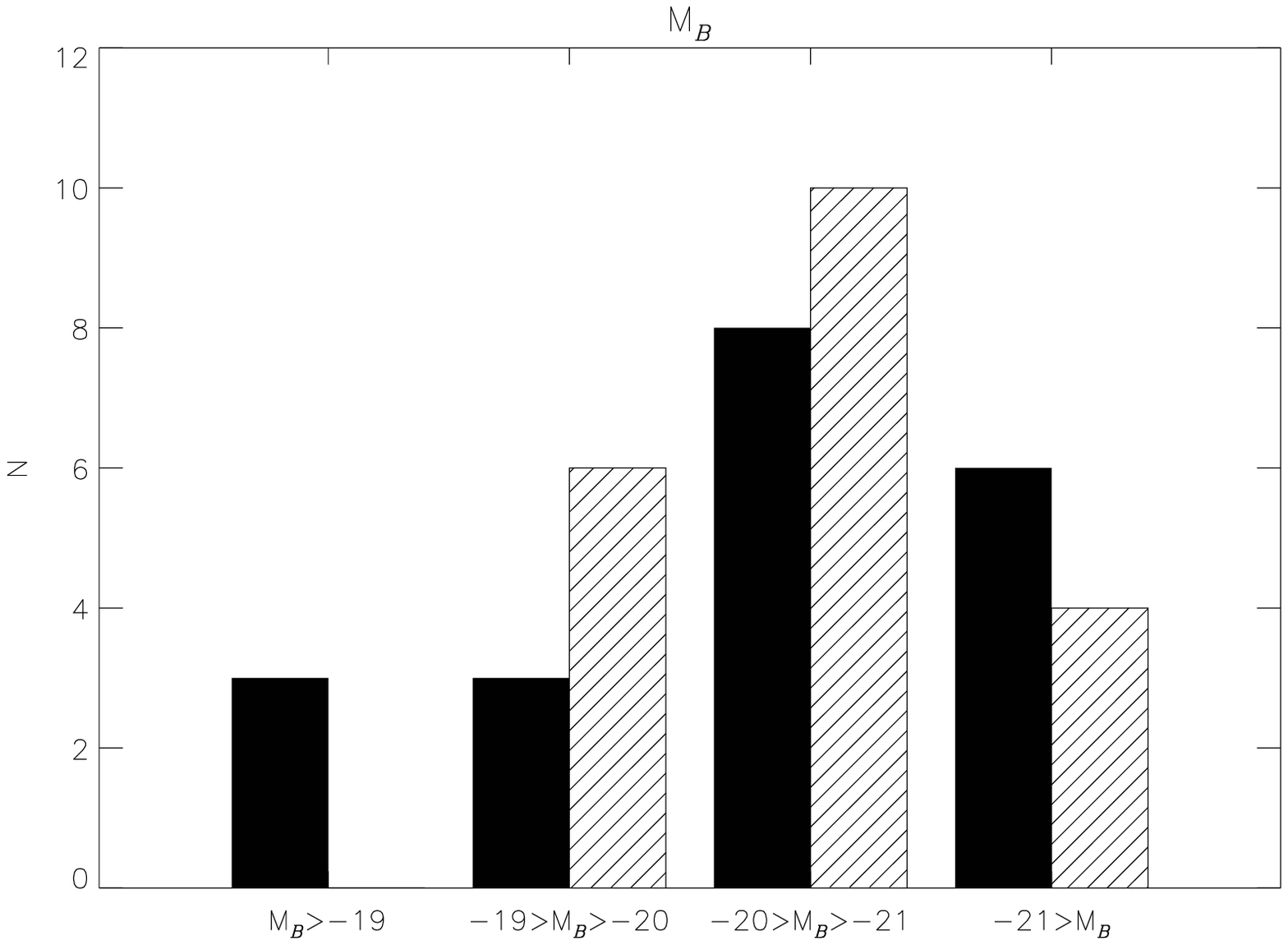}
\\
\includegraphics[width=0.45\textwidth]{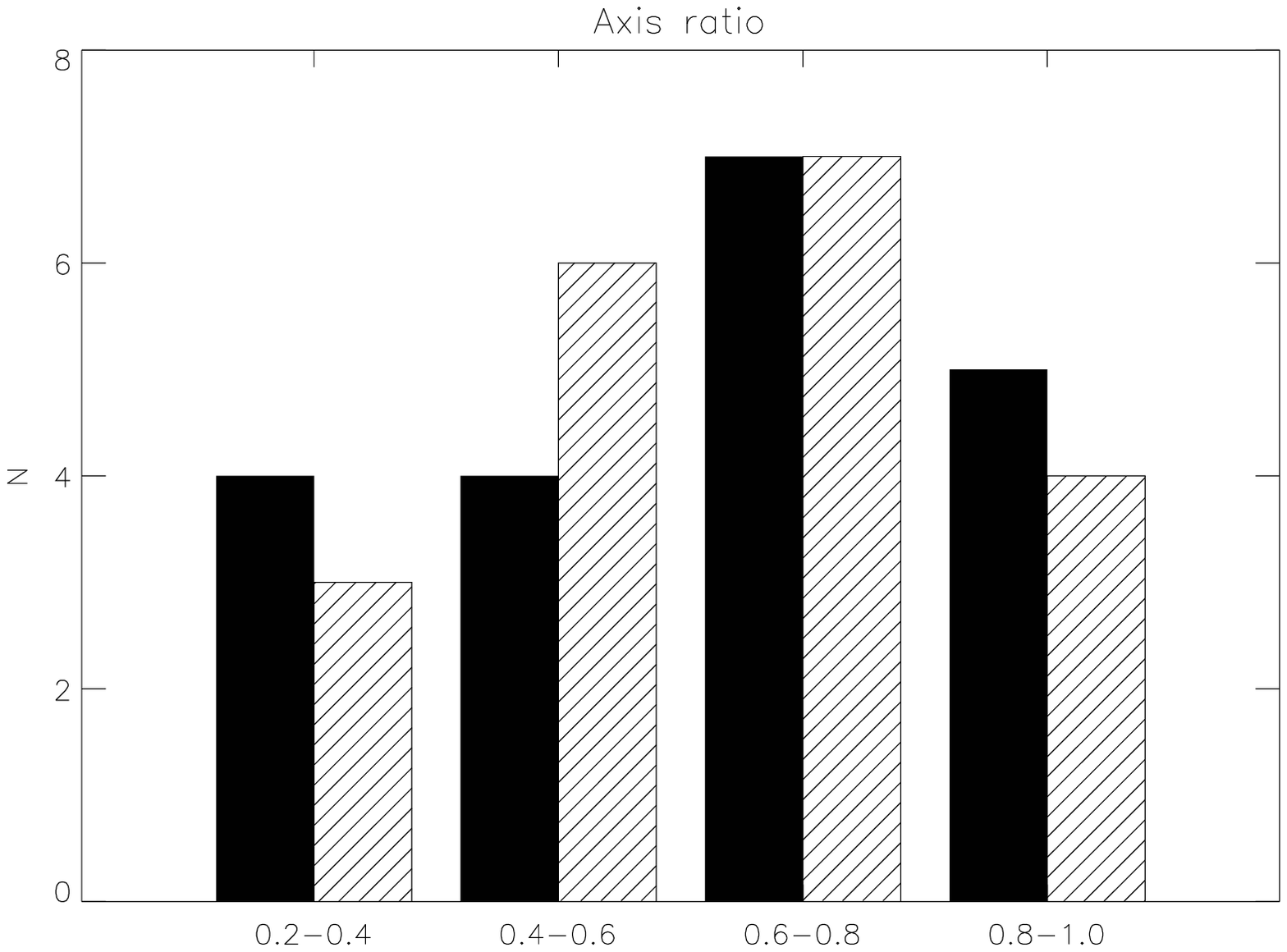}
\includegraphics[width=0.45\textwidth]{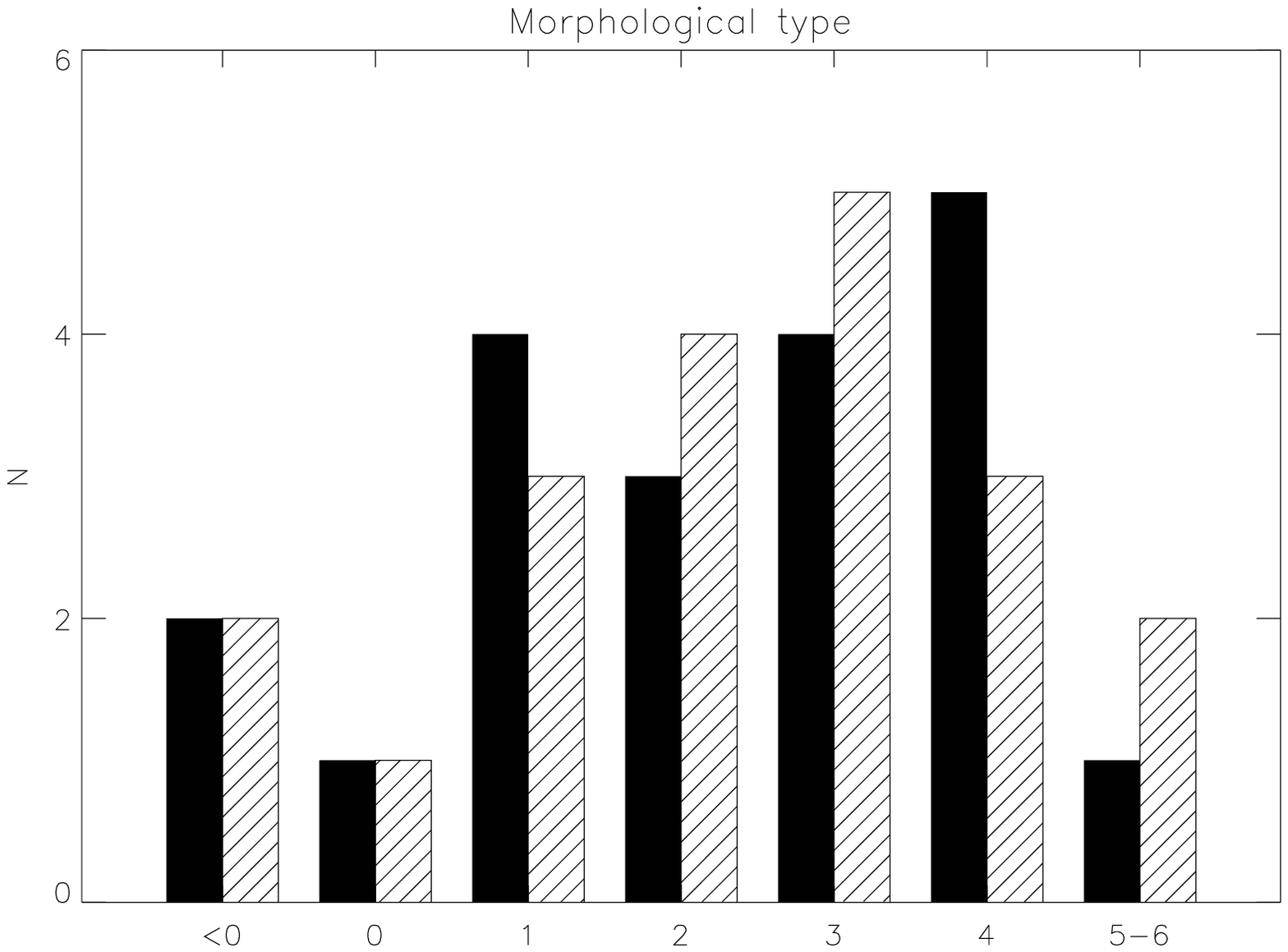}
\end{tabular}
\caption{\label{fig:1} Matching of the $\sigma$-drop and the control
sample with respect to four parameters of the host galaxy:
distance (top left), $B$ absolute magnitude (top right), axis ratio
$d/D$ (bottom left), and morphological type (bottom right). The filled
bars correspond to the $\sigma$-drop sample and the dashed bars denote
the control sample.}
\end{center}
\end{figure*}

We then constructed a control sample that matches almost exactly the
$\sigma$-drop sample in the statistical distribution across four key
parameters of the host galaxies. The four parameters are:

\begin{itemize}

\item{Distance (from radial velocity corrected for Virgo infall with a Hubble constant $H_{0}=70\,\rm{km\,s^{-1}\,Mpc^{-1}}$ for galaxies with radial velocities greater than $1400\,\rm{km\,s^{-1}}$, and from distance determinations from the literature, where possible, in other cases),}

\item{Absolute magnitude (from the LEDA database), corrected for galactic extinction and internal extinction of the target galaxies,}

\item{Ratio of minor and major axis diameter, $d/D$, of the galaxy,
which is directly related to the inclination (from LEDA, except
for NGC~1097 and NGC~6814 where the axis ratio measurement given there is
obviously biased by the strong bar. In the first case we determined
the axis ratio by measuring the kpc-scale ring of star formation and
assuming that it is intrinsically round. In the second case the inner morphology suggests that this is an almost face-on galaxy),}

\item{Morphological type of the galaxy in the Hubble sequence (from
\cite{VA91}).}
\end{itemize}

We started the sample matching process by classifying the
$\sigma$-drop host galaxies into several bins for each parameter. To
create the control sample, we then selected from the {\it HST} archive
all galaxies that have at least visible, red or near-IR imaging obtained with the WFPC2
and/or the ACS. Our control sample is made up of galaxies with evidence for the absence of a $\sigma$-drop in the literature. The stellar velocity dispersion profiles and 2D maps we used for this can be found in Jarvis et al.~(1998), Bottema (1993), Fisher (1997), H\'eraudeau \& Simien (1998), H\'eraudeau et al.~(1999), Falc\'on-Barroso et al.~(2006), Ganda et al.~(2006), and Dumas et al.~(2007), as detailed in Table~\ref{source}. A possible caveat in our control sample is that several dispersion profiles are measured only along the major axis of the galaxies (Bottema 1993; H\'eraudeau \& Simien 1998; H\'eraudeau et al. 1999; seven of our galaxies in total) while $\sigma$-drops sometimes only are detectable when measuring along the minor axis (Fisher 1997, in four of the seven galaxies with $\sigma$-drop), which could generate some contamination in our control sample. This contamination is probably below 20\%, especially considering that in the kinematical studies the proportion of $\sigma$-drop hosts is around 50\% and that only six galaxies in the control sample have only single-slit spectra.

It is easy to match a $\sigma$-drop sample with a control sample for early-type galaxies but it is much more difficult to do so for late-type galaxies. This is due to the lack of late-type galaxies without $\sigma$-drop in the literature; probably a selection effect because Fisher (1997), Simien \& Prugniel (2002), de Zeeuw et al.~(2002), Aguerri et al.~(2003), and Falc\'on-Barroso et al.~(2006) include only early-type galaxies, and the work of Ganda et al.~(2006) is the only relevant one devoted to late-type galaxies. 19 of 20 of our $\sigma$-drops come from the list of Emsellem (2006), in which the selection is independent of morphological type, contrary to the works we used to make the control sample.

From the list produced by this first stage in the
selection, we proceeded to take random galaxies and populate the bins
until achieving the best possible match between the two samples. In
some cases, several combinations of galaxies yielded similarly good
overall matching. In those cases we favoured the distribution in
morphological type over the other three parameters because we are
primarily interested in investigating possible links between
morphology and dynamics, and thus prefer to minimise any
biases in that area. The limits of the bins we selected are (see also
Fig.~\ref{fig:1}):

\begin{itemize}
\item{For distance, we split the samples at distances $R<$10\,Mpc,
$10-20$\,Mpc, $20-30$\,Mpc, and $30-50$\,Mpc.}
\item{For luminosity, we split the samples at magnitudes $B<-21.00$,
$-21.00<B<-20.00$, $-20.00<B<-19.00$, and $B>-19.00$.}
\item{For the  axis ratio $d/D$ we split the samples at
$d/D=0.2-0.4, 0.4-0.6, 0.6-0.8$, and $0.8-1.0$.}
\item{For morphological type, we created  a bin for the lenticular
galaxies and further bins for galaxy types $0, 1, 2, 3, 4$ and $5-6$.} 
\end{itemize}
\begin{table*}
\begin{center}
\caption{\label{sigmaproperties}Results of the morphological
classification of dust structure and \ha{} nuclear and circumnuclear
structure of the 20 galaxies from the $\sigma$-drop sample. ID of the
galaxy (Col.~1); morphological type (RC3, Col.~2); nuclear activity
(from NED, Col.~3); distance in Mpc, (Col.~4); $B$ magnitude
(NED, Col.~5); axis ratio $d/D$ (NED, except
in NGC~1097 where we measured $d/D$ from the \ha{} ring and NGC~6814 that looks face-on), (Col.~6); dust
structure within the radius $r_{\textrm{c}}$ (Col.~7); nuclear \ha{}
structure (Col.~8); circumnuclear \ha{} structure (Col.~9). Statements
such as \textquoteleft Ring+Patchy' describe in which order we find each
\ha{} structure, starting at the nucleus.}
\begin{tabular}{l c c c c c c c c}
\hline
ID & Type & Activity & Distance & $B$ & $d/D$ & Dust & Nucl \ha{} & Circ. nucl\\
&(RC3)&(NED)&(Mpc)&Magnitude& &structure & structure &\ha{} structure\\
\hline
NGC~1068&RSAT3& Sy1 Sy2   &15.3&$-$21.47&0.87&TW&   Amorphous&  Filaments   \\
NGC~1097&.SBS3&    Sy1    &15.2&$-$20.89&1.00&LW&    Peak    & Diffuse+Ring \\
NGC~1808&RSXS1&    Sy2    &10.9&$-$19.91&0.34&LW&    Peak    &  Ring+Patchy \\
NGC~2460&.SAS1&    --     &23.8&$-$19.75&0.74&C &     --     &      --      \\
NGC~2639&RSAR1&  Sy1.9    &48.7&$-$21.10&0.76&LW&     --     &      --      \\
NGC~2775&.SAR2&    --     &19.2&$-$20.60&0.79&N &     --     &      --      \\
NGC~3021&.SAT4&    --     &23.9&$-$19.79&0.59&TW&     --     &      --      \\
NGC~3412&.LBS0&    --     &11.3&$-$18.97&0.55&N &     --     &      --      \\
NGC~3593&.SAS0&\hii{} Sy2 & 9.9&$-$18.28&0.39&C &    Peak    &     Ring     \\
NGC~3623&.SXT1&   LINER   &12.6&$-$20.94&0.26&C &    Peak    &     None     \\
NGC~3627&.SXS3& LINER Sy2 &12.8&$-$21.46&0.45&C &   Amorphous&     None     \\
NGC~4030&.SAS4&    --     &21.1&$-$20.76&0.71&TW&     --     &      --      \\
NGC~4303&.SXT4&\hii{} Sy2 &23.1&$-$21.84&0.89&LW&     --     &      --      \\
NGC~4477&.LBS.&    Sy2    &21.0&$-$20.43&0.87&TW&     --     &      --      \\
NGC~4579&.SXT3&LINER Sy1.9&23.0&$-$21.69&0.78&CS&Peak+Bipolar Structure+UCNR&Filaments\\
NGC~4725&.SXR2&    Sy2    &13.2&$-$21.69&0.69&C &     --     &      --      \\
NGC~4826&RSAT2&    Sy2    & 7.5&$-$20.64&0.49&TW&    Peak    &    Patchy    \\
NGC~6503&.SAS6& \hii/LINER& 5.2&$-$18.70&0.33&C &    Peak    &  Ring+Patchy \\
NGC~6814&.SXT4&    Sy1.5  &22.6&$-$20.48&1.00&GD&     --     &      --      \\
NGC~6951&.SXT4& LINER Sy2 &24.4&$-$21.86&0.71&LW&    Peak    &     Ring     \\
\hline
\hline
\end{tabular}
\end{center}
\end{table*}

\begin{table*}
\begin{center}
\caption{\label{controlproperties} As Table~\ref{sigmaproperties}, now
for the 20 control sample galaxies.}
\begin{tabular}{l c c c c c c c c}
\hline
ID & Type & Activity & Distance & $B$ & $d/D$ & Dust & Nucl \ha{} & Circ. nucl\\
&(RC3)&(NED)&(Mpc)&Magnitude& &structure & structure &\ha{} structure\\
\hline
NGC~488 &.SAR3&    --     &32.1&$-$21.76&0.72&LW&     --     &     --       \\
NGC~864 &.SXT5&    --     &21.8&$-$20.57&0.69&CS&     --     &     --       \\
NGC~1566&.SXS4&    Sy1    &17.4&$-$21.43&0.51&CS&     --     &     --       \\
NGC~2964&.SXR4&    \hii   &20.6&$-$19.81&0.74&C &     --     &     --       \\
NGC~2985&PSAT2&   LINER   &22.6&$-$20.82&0.81&TW&  Peak+UCNR &   Diffuse    \\
NGC~3031&.SAS2&LINER Sy1.8& 3.6&$-$20.76&0.48&CS&Peak+Nuclear Spiral&    None      \\
NGC~3169&.SAS1&\hii{} LINER&17.7&$-$20.32&0.62&C&    Peak    &Diffuse+Patchy\\
NGC~3198&.SBT5&    --     &14.5&$-$19.54&0.28&C &     --     &     --       \\
NGC~3227&.SXS1&  Sy1.5    &23.6&$-$20.74&0.47&CS&    Peak    &  Filaments   \\
NGC~3368&.SXT2&  Sy LINER &10.4&$-$20.31&0.66&CS&    Peak    &  Diffuse     \\
NGC~3675&.SAS3&\hii{} LINER&13.8&$-$20.21&0.55&TW&   Peak    &Patchy+Diffuse\\
NGC~3718&.SBS1&  Sy1 LINER&18.6&$-$20.21&0.49&C &    Peak    &    None      \\
NGC~4102&.SXS3&\hii{} LINER&18.6&$-$19.71&0.56&C&     --     &     --       \\
NGC~4151&PSXT2&    Sy1.5  &17.1&$-$20.17&0.78&CS&    Peak    &  Filaments   \\
NGC~4459&.LAR+&\hii{} LINER&16.1&$-$19.79&0.76&TW&    --     &     --       \\
NGC~5055&.SAT4& \hii/LINER&16.5&$-$21.32&0.36&C &    Peak    &Diffuse+Patchy\\
NGC~6340&PSAS0&   LINER   &21.6&$-$19.92&0.95&N &    Peak    &    None      \\
NGC~6501&.LA.+&    --     &43.5&$-$20.51&0.85&LW&     --     &     --       \\
NGC~7331&.SAS3&   LINER   &15.1&$-$21.73&0.39&C &     --     &     --       \\
NGC~7742&.SAR3&LINER \hii &24.2&$-$19.92&0.95&N &     --     &     --       \\
\hline
\hline
\end{tabular}
\end{center}
\end{table*}

The final $\sigma$-drop and control samples are listed in Tables \ref{sigmaproperties} and \ref{controlproperties}.
Figure \ref{fig:1} and Table \ref{samples}, which shows the mean and the
median for each parameter and for each sub-sample, indicate that the two
samples are very similar. The only noticeable differences are that the
median luminosity and axis ratio for the control galaxies are smaller, and that the lower magnitude distribution end is sub-populated in the control sample. This is because there is only one galaxy without a $\sigma$-drop with $B>-19$ in the papers with velocity dispersion studies in the literature, and this galaxy has not been included because it is almost edge-on.

\begin{table}
\begin{center}
\caption{\label{samples} Statistical properties of the $\sigma$-drop and
control samples, showing mean and median values of the four parameters
used for matching the samples.}
\begin{tabular}{l | c c | c c}
\hline
 &\multicolumn{2}{c|}{$\sigma$-drop sample} &\multicolumn{2}{c}{control sample}
\\
Property & mean & median & mean & median \\
\hline
Distance (Mpc)         &  18.24 &  17.25 &  19.47 &  18.15\\
$B$ Magnitude       & $-$20.53 & $-$20.70 & $-$20.48 & $-$20.32\\
$d/D$                    &   0.66 &   0.71 &   0.63 &   0.64\\
Type &   2.2 &   2.5 &   2.3 &   2.5\\
\hline
\hline
\end{tabular}
\end{center}
\end{table}

\section{Data processing}

\subsection{Structure maps}

The structure map technique (Pogge \& Martini 2002) allows one to
observe the distribution of structure, and in particular that of dust, in a
galaxy on the scale of the PSF using only one image in one
filter. The structure map is mathematically defined as

\begin{equation}
S=\left(\frac{I}{I\otimes P}\right) \otimes P^{\rm t}
\end{equation}

where $S$ is the structure map, $I$ is the image, $P$ is the
point-spread function (PSF), $P^{\rm t}$ is the transform of the PSF, and
$\otimes$ is the convolution operator (\cite{PO02}).  We used
synthetic PSFs which were created with the Tiny Tim software
(\cite{KR99}).

We used images taken through green or red optical filters to apply the structure map
operator, namely $F547M, F606W, F791W$, and $F814W$ for WFPC2 images,
and $F625W$ and $F814W$ for ACS  images. The resulting images are shown
in Appendices~A (for the $\sigma$-drop sample) and B (for the control
sample, the appendices are online-only).

\subsection{\ha\ images}

For 10 of the 20 galaxies in our $\sigma$-drop sample and for ten more
in our control sample we could retrieve \ha\ narrow-band imaging from the {\it HST}
archive. We used \ha\ images taken through the narrow filters
$F656N$ or $F658N$, and a continuum image taken through a red
broad-band filter. In the few cases where images taken through both
\ha{} filters were available in the archive, we chose the $F658N$
image, because the \ha\ line of our sample galaxies was better
centred in its passband. The images used for the continuum
subtraction were those from which we derived the structure
maps.

After selecting the images from the archive, the continuum and
narrow-band images were aligned using standard IRAF software. We then
plotted for each pixel in the whole image the number of counts in the narrow-band filter
versus that in the continuum filter.  In the absence of \ha{} line
emission, the number of counts in the narrow-band filter will be
proportional to the number of counts in the continuum, and the
constant of proportionality is the factor by which the continuum
has to be scaled before subtracting from the narrow-band image. Since
even for a star-forming galaxy most pixels trace only continuum
emission, those pixels without \ha{} emission can easily be recognised
in the graph, and a fit to them yields the continuum scaling factor
(see Knapen et al.~2005, 2006 for details and a graphic
illustration). In most cases, there is a contribution from the [\nii]
line in the \ha{} images but this is not a problem because we do not
use the \ha\ images for photometry, and [\nii] emission is
unlikely to affect the morphology on the scales we are interested in
here. The resulting continuum-subtracted \ha\ images are shown in
Appendices A and B (for the $\sigma$-drop and control samples,
respectively; on-line only).

\section{Size of the $\sigma$-drop region}

\begin{table}
\begin{center}
\caption{\label{size}Sizes of the $\sigma$-drop and $r_{\rm{c}}$.}
\begin{tabular}{l c c c c c} 
\hline
Galaxy & \multicolumn{2}{c}{$\sigma$-drop radius} & \multicolumn{2}{c}{$r_{\rm{c}}$ size} & $\sigma$-drop radius/$r_{\rm{c}}$\\
& (\arcsec) & (pc) & (\arcsec) & (pc) &\\
\hline
NGC~1068 & 3 & 220 & 4.06 & 300 & 0.74\\
NGC~1097 & 4 & 290 & 6.43 & 470 & 0.62\\
NGC~1808 &1.5&  80 & 3.30 & 170 & 0.46\\
NGC~2460 &3.5& 400 & 1.09 & 130 & 3.21\\
NGC~2639 &1.5& 350 & 0.97 & 230 & 1.55\\
NGC~2775 & 5 & 470 & 2.56 & 240 & 1.95\\
NGC~3021 & 5 & 580 & 0.81 &  90 & 6.17\\
NGC~3412 & 3 & 160 & 2.50 & 140 & 1.20\\
NGC~3593 & 20& 960 & 2.94 & 140 & 6.80\\
NGC~3623 & 5 & 310 & 4.55 & 280 & 1.10\\
NGC~3627 & 4 & 250 & 5.35 & 330 & 0.75\\
NGC~4030 & 3 & 310 & 2.28 & 230 & 1.32\\
NGC~4303 & 5 & 560 & 3.79 & 420 & 1.32\\
NGC~4477 & 4 & 410 & 2.23 & 230 & 1.79\\
NGC~4579 & 4 & 450 & 3.37 & 380 & 1.19\\
NGC~4725 & 5 & 320 & 5.73 & 370 & 0.87\\
NGC~4826 &7.5& 270 & 6.00 & 220 & 1.25\\
NGC~6503 & 10& 250 & 3.70 &  90 & 2.70\\
NGC~6814 & 4 & 440 & 1.85 & 200 & 2.16\\
NGC~6951 &7.5& 890 & 1.99 & 240 & 3.77\\
\hline
\hline
\end{tabular}
\end{center}
\end{table}

\begin{figure*}
\begin{center}
\begin{tabular}{c}
\includegraphics[width=0.45\textwidth]{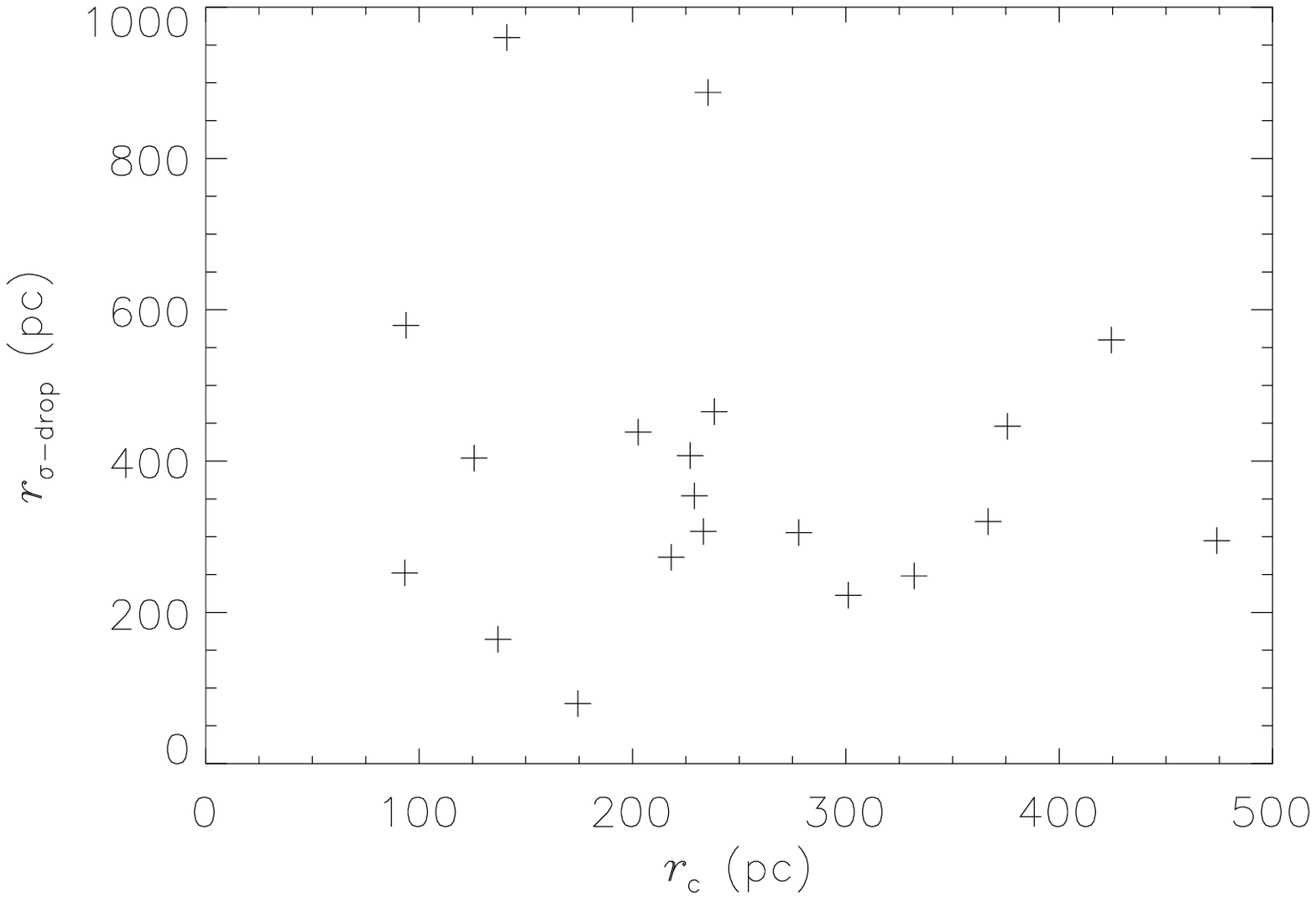}
\includegraphics[width=0.45\textwidth]{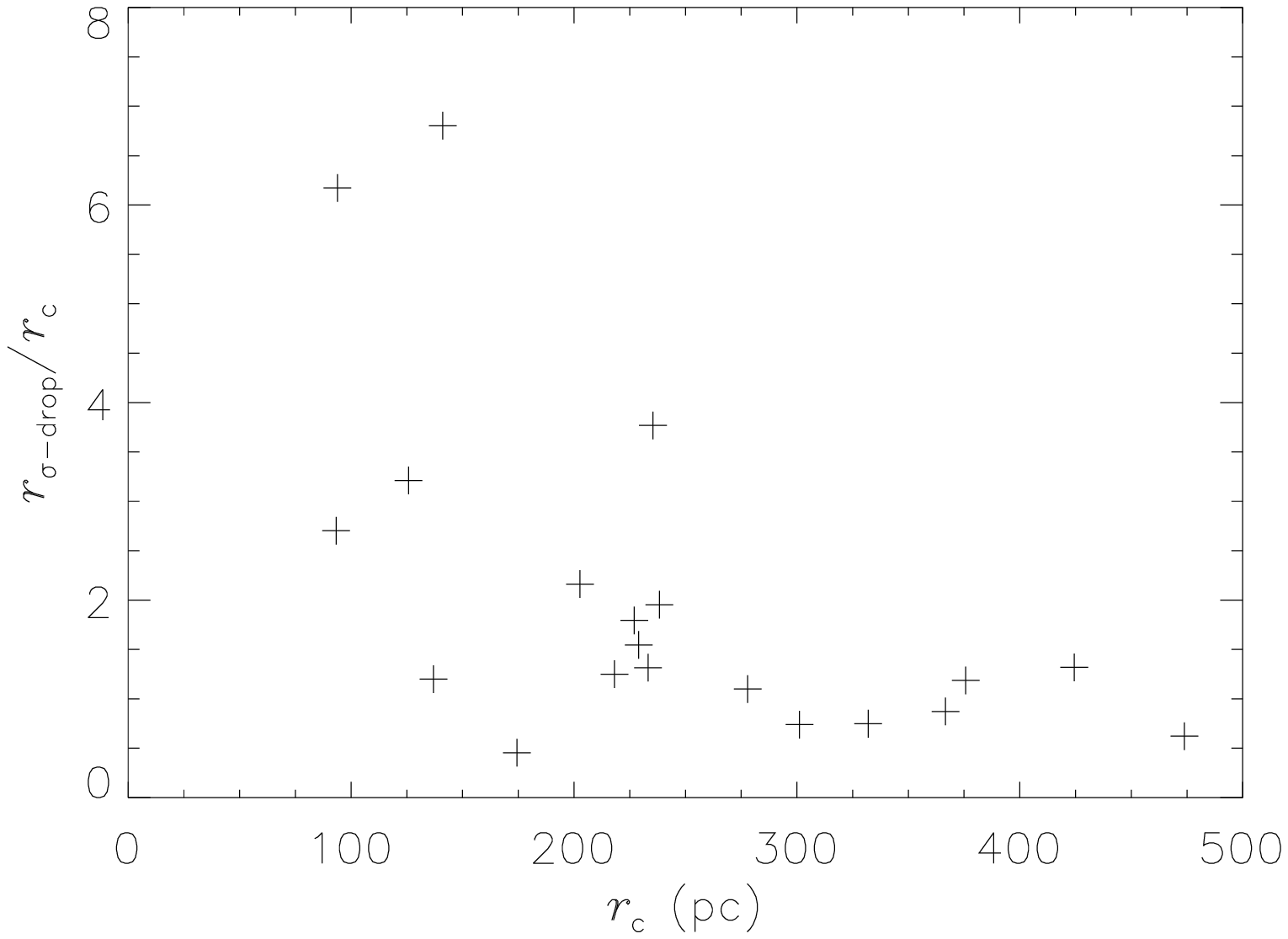}
\end{tabular}
\caption{\label{sigmasizes} In the left panel, radius of the $\sigma$-drop ($r_{\sigma\rm{-drop}}$) versus the photometric radius $r_{\rm{c}}$, and in the right panel, the ratio of $r_{\sigma\rm{-drop}}$ and $r_{\rm{c}}$ versus $r_{\rm{c}}$.}
\end{center}
\end{figure*}

We have searched for correlations between size of the $\sigma$-drop region and other properties of their host galaxies, such as differences in morphology or in radio emission from gas (see further sections  for details on how we study those properties). We have found no correlation between any of these properties and the size of the $\sigma$-drops.

We have also made a comparison between the radius of the $\sigma$-drop ($r_{\sigma\rm{-drop}}$) and 
$r_{\rm{c}}$, a measure of radius which is directly related to the visible size of the galaxy and 
which we define as 2\,\% of the radius of the galaxy (where the radius
is half of $D_{25}$ as taken from the RC3). We use the definition for
$r_{\rm{c}}$ because it is similar to the effective radius of the
bulge of a spiral galaxy (\cite{LA04})---it thus provides an objective
but simple and effective working definition of the central region.

The data used appear in Table~\ref{size} and the plots are in Fig.~\ref{sigmasizes}. The radii of the $\sigma$-drops were taken from the velocity dispersion plots in the papers cited in Section~2. These dispersions are not always very well defined because in several papers the spectra are taken only along one axis of the galaxy (usually the major axis) and this does not provide as good a measure of the $\sigma$-drop radius as $IFU$ spectroscopy. We note that the highest value of the ratio $r_{\sigma\rm{-drop}}/r_{\rm{c}}$, NGC~3593, is a `strange' galaxy with two counter-rotating discs (Bertola et al.~1996). We find some hint that higher $r_{\sigma\rm{-drop}}/r_{\rm{c}}$ ratios are found in smaller galaxies and vice versa. That would indicate that $r_{\sigma\rm{-drop}}$ {\it is not directly related to the size of the galaxy}, which is confirmed by the plot in the right panel of Fig.~\ref{sigmasizes}.

\section{Morphological classification}

\subsection{Bars}

Because we ignored bars in our sample selection procedure (Sect.~2),
we can as a first step investigate the presence and properties of the
bars in our $\sigma$-drop and control sample galaxies. The interest in
this lies primarily in the fact that bars can facilitate the inflow of
cool gas into the inner regions of galaxies by removing angular
momentum from the gas in the outer parts. If $\sigma$-drops are related
to star formation from dynamically cold gaseous material close to the
nucleus, it is reasonable to conjecture that bars may play an important
role in the process. In Table~\ref{sigmaproperties} and
Table~\ref{controlproperties} we therefore summarise the bar
classification of the galaxies. This is based primarily on the RC3
catalogue (de Vaucouleurs et al. 1991), where, as is conventional, SA
galaxies are those without a bar, SAB (also known as SX) are weakly or
moderately barred, and SB galaxies have a prominent bar in the
optical.

Although the RC3 is based on optical imaging and is therefore not
optimal in recognising all bars, it is generally reliable in reporting
well-defined bars. We did check the literature and images of
individual galaxies though, and refined the classification in the
following cases. Some galaxies are catalogued as non-barred in the RC3
but do in fact have weak and/or small bars. The first case is NGC~1068, part of our
$\sigma$-drop sample, where a bar has been discovered from near-IR imaging (e.g.,
\cite{SC88}). Other galaxies with a bar discovered on the basis of
modern imaging are the control sample galaxies NGC~3169 (\cite{LA02}), NGC~3593 and NGC~3675 (\cite{LA04}), and NGC~2460, NGC~7331, and NGC~7742 (\cite{LAI02}). Fitting ellipses to a $2MASS$ image we also found strong evidence for a weak bar in NGC~2985.

The bar fractions in our $\sigma$-drop and control samples can thus be
seen in Table~\ref{bars}, where we use Poisson statistics to give an
estimate of the error in the numbers ($\epsilon=\sqrt{f(1-f/N)}$, where
$f$ is the quantity that is measured and $N$ is the sample size in
which the quantity is searched).  The conclusion from the numbers
presented in Table~\ref{bars} is that there are no statistically
significant differences between the two samples. What is evident from this test, however, is that a bar is not
needed to cause a $\sigma$-drop, and that, conversely, the presence of
a bar does not automatically, or even preferentially, lead to a dip in
the central stellar velocity dispersion. As for inner bars, we have found six of them in $\sigma$-drop galaxies (NGC~1068, NGC~1097, NGC~1808, NGC~2460, NGC~4303, and NGC~4725) and three in galaxies of the control sample (namely NGC~3169, NGC~3368, and NGC~7742). All of these inner bars have been reported  independently by Laine et al. (2002) and/or Erwin (2004). 

The difference between inner bar fraction in the $\sigma$-drop and control samples is not very significant, but points in the direction of a link between inner bars and $\sigma$-drops. It is worthy to note here that, except in the case of NGC~3593, $\sigma$-drops have a much smaller scale than outer bars. In the other hand, $\sigma$-drops occur on a size scale similar to that of inner bars (our nine inner bars have a mean size of 350\,pc, compared to  the mean radius of the $\sigma$-drops in the sample, which is 400\,pc).

\begin{table}
\begin{center}
\caption{\label{bars} Bars in the sample galaxies. The data come from the
RC3 except in the cases that are explained in the text.}
\begin{tabular}{l c c c}
\hline
Sample &  SA & SAB & SB \\
\hline
$\sigma$-drop & 7 $(35\% \pm 11\%)$ & 10 $(50\% \pm 11\%)$ & 3 $(15\% \pm 8\%)$\\
control & 8 $(40\% \pm 11\%)$ & 9 $(45\% \pm 11\%)$ & 3 $(15\% \pm 8\%)$\\
\hline
\hline
\end{tabular}
\end{center}
\end{table}

\subsection{Dust classification}

Dust is well traced by structure maps, and we thus used the structure
maps we constructed for our sample galaxies to characterise the
central dust morphology across our samples. The structure maps are
shown in Appendices A and B (on-line only) for the $\sigma$-drop and
control samples, respectively. We draw two ellipses that indicate the
scale of each image, an inner ellipse that indicates $r_{\rm{c}}$, as defined above, in Section~4, 
and an outer one which
indicates the inner kiloparsec in radius. 

We applied the classification criteria developed by \cite{PE06} to the
central area ($r_{\rm{c}}$) of each galaxy. Examples of each
classification class are given in Fig.~2 of \cite{PE06}. The classes are:

\begin{itemize}

\item{\emph{Grand design (GD)}: Two spiral arms with a 180\deg{}
rotational offset between them. At least one arm must reach the
unresolved centre of the galaxy and must be a dominant feature.}

\item{\emph{Tightly wound (TW)}: A coherent spiral over a wide
range of radius, with a pitch angle smaller than 10\deg.}

\item{\emph{Loosely wound (LW)}: A coherent spiral over a wide
range of radius, with a pitch angle greater than 10\deg.}

\item{\emph{Chaotic spiral (CS)}: Evidence of spiral arms with a
unique sense of chirality, but not coherent over a wide radial
range.}

\item{\emph{Chaotic (C)}: No evidence of spiral structure.}

\item{\emph{No structure (N)}: No dust seen.}

\end{itemize}

\begin{figure}
\includegraphics[width=0.45\textwidth]{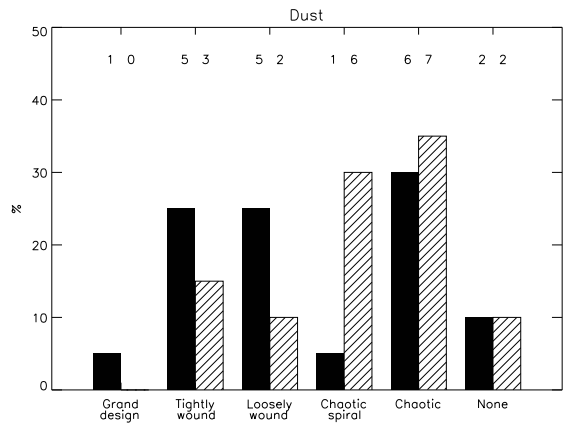}
\caption{\label{dust} Comparison between the dust distribution in the
two samples. Filled bars correspond to the $\sigma$-drop sample and dashed bars
denote the control sample. The numbers at the top of each bin refer to the number of galaxies in each bin.}
\end{figure}

The results of the classification for the $\sigma$-drop and the
control sample are in Tables~\ref{sigmaproperties} and
\ref{controlproperties}, respectively.  One result that is obvious from
the Tables is that there is a correlation between the dust
distribution and the axis ratio. Galaxies with ``chaotic'' dust
distribution tend to have lower axis ratios (they are more highly
inclined), which we interpret as being due to the dust in the disc
obscuring the nucleus and the small-scale dusty features that are in
the same plane as the external discs of dust. To quantify this we can compute the mean $d/D$ for the 13 chaotic galaxies ($\overline{d/D}=0.48$) and for the 40 studied galaxies ($\overline{d/D}=0.65$)

Figure~\ref{dust} summarises the dust morphology for the $\sigma$-drop
and control samples. There is some correlation between the
presence of a $\sigma$-drops and the morphology of the dust
structure. Whereas galaxies with dust spirals (especially LW) are preferentially $\sigma$-drop hosts, the chaotic and chaotic spiral morphological classes have more control sample galaxies.

When using broader bins, i.e., separating galaxies with clear spiral dust structure
(classes LW, TW, and GD) from those without it (N, C, and CS), and thus
to some extent overcoming the low-number statistics, the result is amplified. Eleven of the 20 $\sigma$-drop
galaxies show clear dust spiral structure ($55\% \pm 11\%$), versus only five ($25\% \pm 10\%$) in the control sample .

Does this mean that galaxies without a $\sigma$-drop are preferentially chaotic? To test this idea we first have to remember how the sample was composed. The key matching parameter is morphological type, so axis ratio is not so well matched. This causes the mean and median axis ratios of the control sample to be slightly smaller than those of the $\sigma$ sample. This might explain part of the fraction of control galaxies in the C category whose inner structure is hidden by a dust disc seen edge-on. But CS galaxies clearly show the bulge dust structure and therefore are barely affected by disc dust so the size of the CS class is not overestimated. In summary, chaotic morphologies are indeed favoured by the control sample galaxies.

\subsection{\ha{} classification}

The classification of the \ha{} emission (images shown in Appendices A
and B; on-line only) is based on the scheme developed by
\cite{KN05}. We classify the images separately for two radial regions,
$r_{\rm{c}}$, as defined above, and the inner kpc. In the very central
region ($r_{\rm{c}}$) the \ha\ emission can be dominated by the
following features:

\begin{itemize}

\item{\emph{Peak}: A peaked emission region centred on the nucleus of
the galaxy.}

\item{\emph{Amorphous}: Clear \ha\ emission, but without any
significant peak.}

\item{\emph{Peak+UCNR}: A peaked emission region centred on the
nucleus of the galaxy surrounded by an \ha-emitting Ultra-Compact Nuclear Ring
(UCNR; Comer\'on et al. 2008) of a few tens of parsec in radius.}

\item{\emph{Peak+Bipolar Structure+UCNR}: A peak in the centre
accompanied by both a bipolar structure and an UCNR. This appears only
in NGC~4579 where the inner peak and the bipolar structure are
surrounded by an extremely fuzzy UCNR ring of 130\,pc in radius.  Its
existence has been confirmed from UV images (\cite{CO08}).}

\item{\emph{Peak+Nuclear spiral}: A peak in the centre surrounded by a short \ha{} spiral that does not reach $r_{\rm{c}}$. NGC~3031 is the only example of this kind of galaxy in our samples.}

\end{itemize}

In the zone between $r_{\rm{c}}$ and the inner kpc radius, we used the following classification
criteria:

\begin{itemize}

\item{\emph{Patchy}: Individual regions of \ha{} emission, not
centrally peaked, looking like the \hii{} regions that can be found in
the discs of galaxies.}

\item{\emph{Ring}: \ha{} emission organised in a well-defined ring of
star formation bigger than a UCNR, which means that it measures, at
least, two hundred parsecs in radius. We are classifying only the inner
kpc in radius, so rings with bigger radii are not taken into
account.}

\item{\emph{Diffuse}: \ha{} emission is present but cannot be ascribed
to individual and well-defined \hii{} regions.}

\item{\emph{Filaments}: \ha{} emission distributed in long filaments
of \hii{} regions.}

\item{\emph{None}: Less than 1\% of the area of the inner kiloparsec
region has any \ha{} emission.}

\end{itemize} 

\begin{figure}
\includegraphics[width=0.45\textwidth]{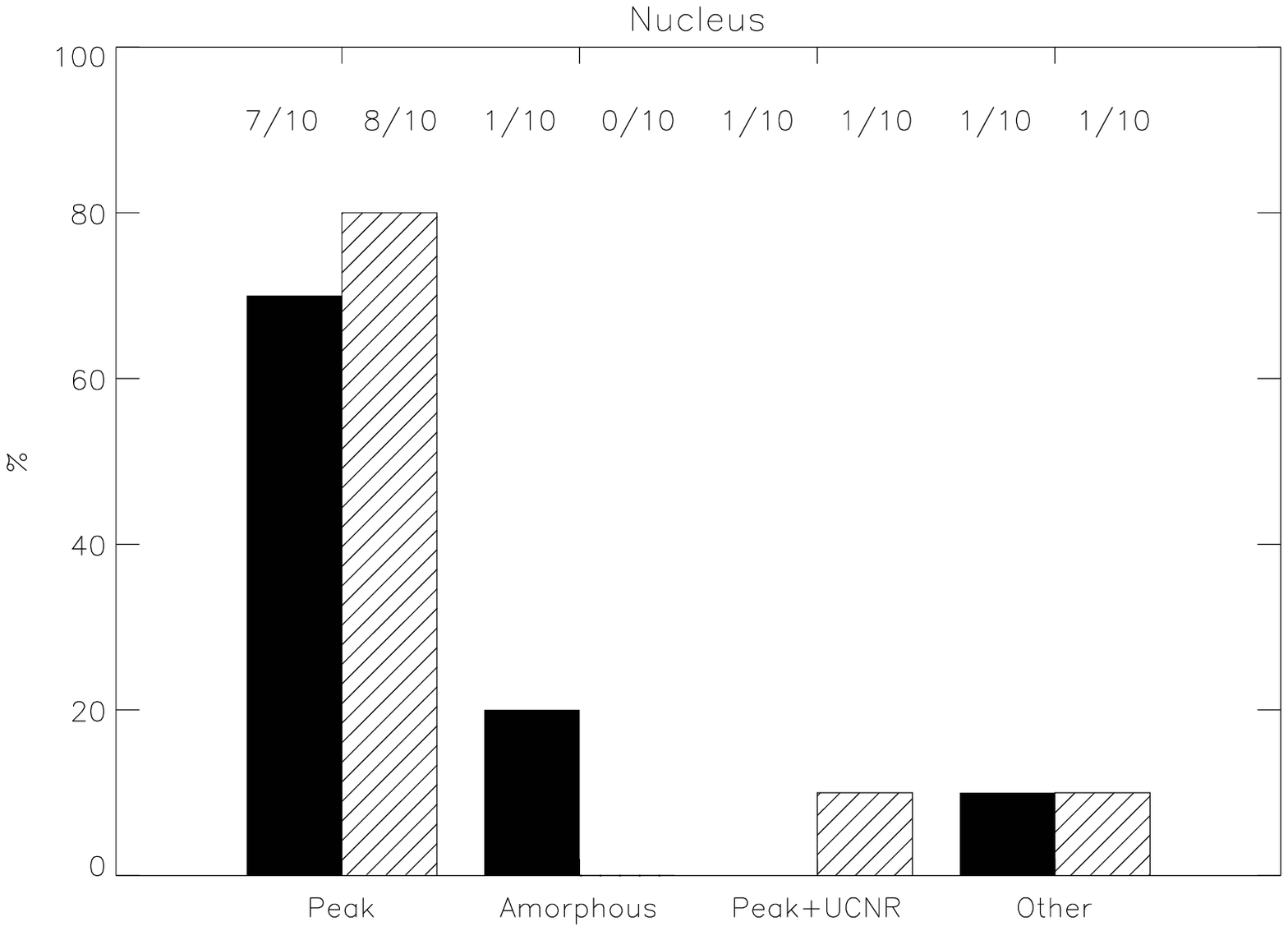}\\
\includegraphics[width=0.45\textwidth]{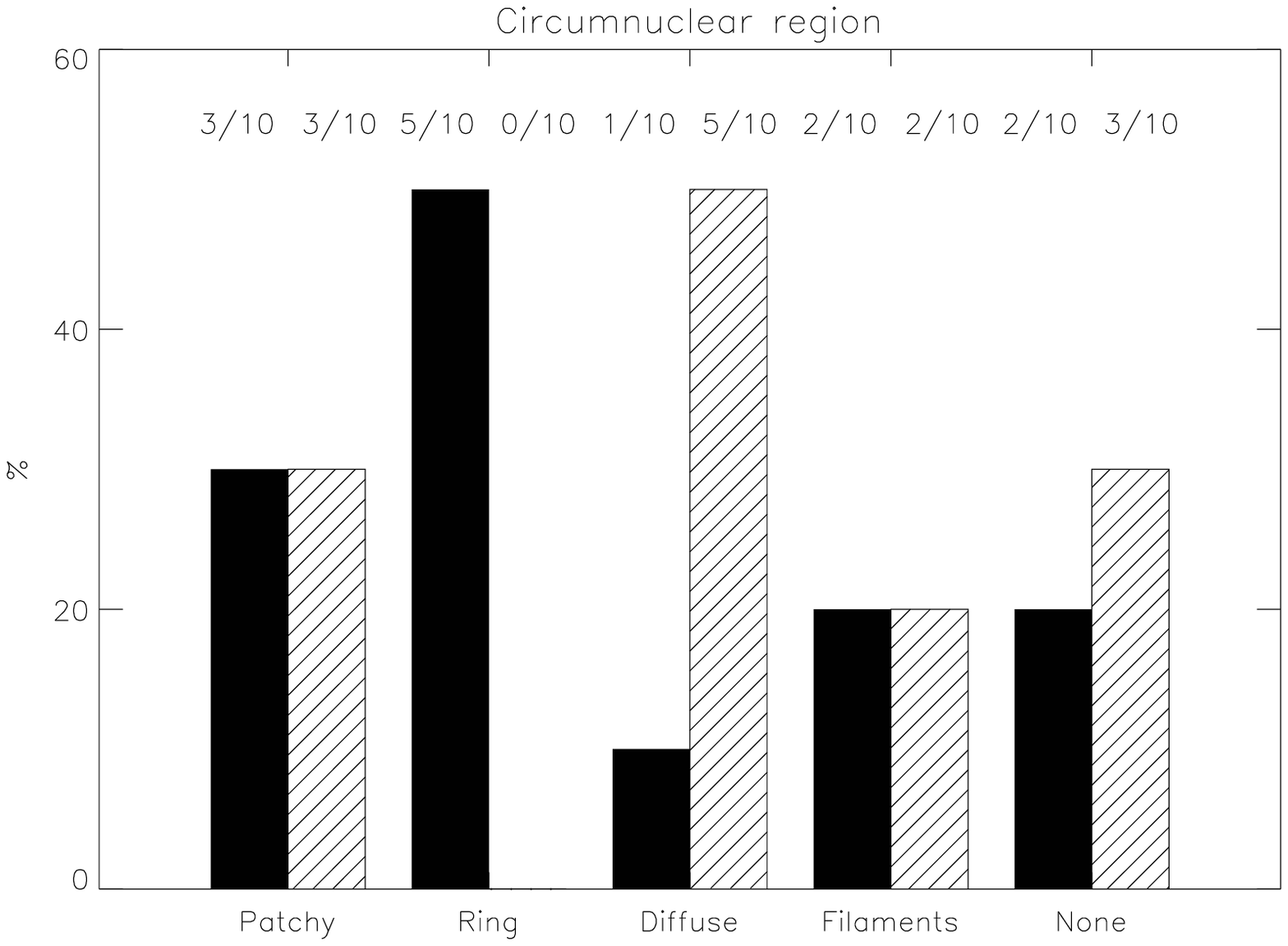}\\
\caption{\label{halfa} Comparison between the \ha{} distribution in
the two samples. Filled
bars correspond to the $\sigma$-drop sample and dashed bars denote the control
sample. In the histogram that shows the nuclei classification the
\textquoteleft other' bars refer to the \textquoteleft
Peak+Bipolar Structure', and \textquoteleft Peak + Bipolar Structure +
UCNR'. The numbers at the top of each bin refer to the number of galaxies in the bin and the number of galaxies in the subsample.}
\end{figure}

This classification has been applied to the 10 $\sigma$-drop galaxies and the 10 control sample galaxies
for which we have \ha{} images. Results for the $\sigma$-drop sample are listed in
Table~\ref{sigmaproperties}, and for the control sample in
Table~\ref{controlproperties}, and are summarised as histograms in
Fig.~\ref{halfa}.

There are no significant differences at all in the very central \ha\
emission morphology. For the inner 1\,kpc
region, however, we notice that star-forming activity is much higher in $\sigma$-drop galaxies than in control-sample galaxies. Patches and filaments in $\sigma$-drop host galaxies are more extended (covering a larger surface area) and more obvious in the images than in the control sample galaxies. Most of the control galaxies are devoid of \ha{} formation or have a weak diffuse emission. Galaxies in the  $\sigma$-drop sample often show strong starbursts with bright \ha{} rings and patches.

The differences between these two subsamples could be due to the fact that they are not as well matched as the two samples of 20 galaxies. The mean axis ratio for the $\sigma$-drop galaxies is 0.56 while it is 0.62 for the control galaxies. Four $\sigma$-drop galaxies have axis ratios smaller than 0.4. Two of these galaxies (NGC~1808 and NGC~3593) show clear evidence of nuclear star-burst rings, which means that we are seeing the inner region without strong extinction. In the other two cases (NGC~3623 and NGC~3627) there is no \ha{} emission, possibly due to dust obscuration. In the control sample dust seems to hide the inner regions in NGC~3169 ($d/D=0.62$ and diffuse circumnuclear emission) and in NGC~3718 ($d/D=0.49$ with no circumnuclear emission). It is difficult to explain using only obscuration the virtual absence of circumnuclear emission in the two other galaxies, NGC~3031 which shows a tiny nuclear spiral and the nearly face-on galaxy NGC~6340.

The other difference between the two subsamples is that the mean morphological type of the control sample is earlier than that of the $\sigma$-drop sample (mean Hubble morphological type 1.8 and 2.6, respectively). Both subsamples have galaxies with types ranging between zero and six. In Knapen et al.~(2006) galaxies with no circumnuclear emission tend to be early type and patchy emission tends to be in later-type galaxies which is coherent with what we have found in the two subsamples. Diffuse emission is also more prevalent in the control sample. As the diffuse emission is an intermediate case between no emission and patchy emission, the overabundance of this type in the control sample is probably due to the fact that it contains earlier-type galaxies. But in Knapen et al.~(2006) the nuclear starburst ring distribution peaks around $T=1$, {\it so the control sample should have a higher number of nuclear rings, which is not the case}. According to the statistics of Knapen et al.~(2006), corrected for the nuclear ring fractions reported there for each morphological type of the host galaxy, four of our $\sigma$-drop galaxies and six control galaxies should have a circumnuclear ring---the actually observed numbers are five and zero. Therefore it is reasonable to think that the higher proportion of circumnuclear star-forming rings in $\sigma$-drop galaxies is a true effect correlated with the $\sigma$-drop.

\section{Nuclear activity}

From Tables~\ref{sigmaproperties} and \ref{controlproperties} we can see that the proportion of galaxies in the $\sigma$-drop sample that have a Seyfert nucleus is $65\%\pm11\%$ while in the control sample it is $30\%\pm10\%$. On the other hand, LINERs make up $25\%\pm10\%$ of the $\sigma$-drop sample but $60\%\pm11\%$ of the control sample. This points to a link between $\sigma$-drops and Seyferts, and links the other galaxies with a more reduced level of activity.

If this is indeed the case, assuming that Seyfert nuclei are more active than LINERs, we might intuitively think that more energetic phenomena are related to a bigger inflow of gas. This gas could have been partly spent in a generation of stars in the inner volume of the galaxies. The signature of this recent star formation would be a $\sigma$-drop. This result has to be handled with care because in both samples we have a proportion of AGN of around 80\%, which is far higher than the proportion of active galaxies that is found in a randomly taken sample. A direct explanation might be that $HST$ is usually employed to observe `interesting' objects, which implies an enhanced fraction of AGN in the galaxies that have been observed with the space telescope.

One possibility is that the inflow of gas giving rise to a $\sigma$-drop feeds the energetic phenomena detected as nuclear activity. When the inflow stops the activity fades, and the $\sigma$-drop gradually relaxes. As the timescale for the former is shorter than that for the latter, any correlation will be only partial.

A way to test if this correlation is indeed true is to study a bigger sample, for example all the galaxies from the papers from which we selected the control sample. That yields 85 $\sigma$-drop galaxies and 95 control galaxies. There are around 40 galaxies that are difficult to classify and therefore are not included in this test. We found that of the $\sigma$-drop galaxies $32\%\pm5\%$ are Seyfert hosts and $19\%\pm4\%$ are LINER hosts. For the control sample we find that $16\%\pm4\%$ are Seyfert hosts and $24\%\pm4\%$ are LINER hosts. These data must be taken with caution because they do not come from matched samples, so the effects that we see may be due to other effects such as differences in morphological type. More than a half of the galaxies used for this test (as performed on the total sample of $85+95$ galaxies, defined above) come from the papers from H\'eraudeau \& Simien (1998), H\'eraudeau et al.~(1999), and Simien \& Prugniel (1997abc, 1998, 2000, and 2002), which have dispersion profiles from galaxies that are so distant that in many cases they have not been studied in any survey of nuclear activity.

From this discussion we may tentatively conclude that $\sigma$-drop are preferably accompanied by Seyfert nuclei, and that both phenomena can be traced to increased gas inflow. Given the different spatial and time scales involved, exactly why and how this is so is slightly puzzling. We note that a similar, and similarly puzzling, relation has been reported in the literature between the presence of nuclear activity and that of nuclear rings (see Knapen 2005, in particular the discussion in Section~6 of that paper). 

\section{Luminosity profiles}

As a way to check another possible connection between $\sigma$-drops
and their host galaxies, we searched for a correlation between the
presence of a $\sigma$-drop and the shape of the luminosity profiles
of their host galaxies. Our aim here was to derive luminosity
profiles from the nucleus to the outer parts of the galaxies or to use
profiles obtained by other authors. We obtained Sloan Digital Sky Survey release 5
(SDSS5) (\cite{AD07}) $r$-band images to make the profiles. In several cases we had
to join two or three SDSS frames to get a complete image of a
galaxy. Each pixel in the SDSS images subtends 0.396\,arcsec,
obviously not nearly small enough to do an analysis at the same scale as we
did for dust and \ha, so we concentrate on the host disc here.

We obtained 13 profiles for the $\sigma$-drop sample and 13 more for the control subsample. These subsamples are very well matched in morphological type. There are not so well matched in axis ratio but this is only marginally important because ellipse fitting and posterior deprojection allow good photometry. The biggest difference is in the galaxies' magnitudes. The $\sigma$-drop subsample is on average 0.23~magnitudes brighter than the control one.

To produce the luminosity profile, we start by subtracting the sky
background contribution from the images. That was done using the
average of sky measurements over outer parts of an image where the disc of
the galaxy is well below the noise level. The sky measures were done by
averaging circles of 10 pixels in radius. Once the sky had been
subtracted, we fitted ellipses for increasing radius, which yielded,
first of all, the position angle (PA) and the ellipticity of the outer
regions of the galaxy. The radial profiles were usually featureless
across the outer regions. We used this PA and ellipticity to
measure surface photometry at increasing radii from the
centre of the galaxy until the disc `fades out'. More details about the
method can be found in \cite{ER08}.  The results are
shown as radial surface brightness profiles for all the galaxies in
Appendices C and D (on-line only), for the $\sigma$-drop and control
samples, respectively. The fitted exponentials are overplotted. The
vertical dashed-dotted lines correspond to the position of the end of the bar. When
there are two vertical lines, they indicate a lower and an upper
limit estimate of the end of the bar.

We classified each profile following \cite{ER08}, and thus
divide the galaxies in three groups:

\begin{itemize}

\item{\emph{Type~I}: The radial variation in disc surface brightness
is described by a single exponential, allowing some irregularities due
to structure (e.g., arms, big star-forming regions). This type of profile
was established by \cite{FR70}.}

\item{\emph{Type~II}: Galaxies with a break in the surface brightness
  characterised by a convex, steeper, outer slope ({\it
  truncation}). This type of profile was also established by
  \cite{FR70}. In barred galaxies the Type~II class can be subdivided
  into two subclasses, namely \emph{Type~II.i}, where the break is
  inside the bar radius, and \emph{Type~II.o}, where it is outside,
  as explained by \cite{ER08}. All the \emph{Type~II} galaxies in this paper have
  the break outside the bar radius, so we ignored this subdivision.}

\item{\emph{Type~III}: Galaxies with a concave break followed by a
  less steep slope ({\it antitruncation}; Erwin et al. 2005).}

\end{itemize}

In some cases, antitruncations and truncations appear in the same
galaxy at different radii. That is the reason that we can have, for
example, II+III and II+III+II galaxies. We fitted exponentials to
every section of the surface brightness profile that was roughly
straight in the plots, except in the case of features directly related
to bars or dust rings, as judged from a visible comparison with the
images.

\begin{table}
\begin{center}
\caption{\label{profiles}Luminosity profiles of the studied galaxies.}
\begin{tabular}{l c | l c} 
\hline
\multicolumn{2}{c |}{$\sigma$-drop sample} & \multicolumn{2}{c}{Control sample}\\
Galaxy & Type & Galaxy & Type\\
\hline
NGC~1068 & II & NGC~2964 & I\\
NGC~2460 & II & NGC~2985 & II+III+II\\
NGC~2639 & I & NGC~3031 & II\\
NGC~2775 & III & NGC~3169 & III+II\\
NGC~3021 & II+III & NGC~3198 & II+III\\
NGC~3593 & III & NGC~3227 & I\\
NGC~3623 & I & NGC~3368 & II\\
NGC~3627 & II+III & NGC~3675 & III+II\\
NGC~4030 & III & NGC~3718 & III\\
NGC~4477 & II+III & NGC~4102 & I\\
NGC~4579 & II+III & NGC~4151 & II\\
NGC~4725 & I & NGC~4459 & I\\
NGC~4826 & III & NGC~5055 & III\\
\hline
\hline
\end{tabular}
\end{center}
\end{table}

Table~\ref{profiles} summarises the classified shapes of the surface
brightness profiles for all the galaxies of our samples that are
either in the SDSS survey, or in Pohlen \& Trujillo (2006; NGC~1068 and
NGC~5806), in Erwin et al. (2008; NGC~3368 and NGC~4102), or in L. Guti\'errez et al. (2008, in preparation; NGC~2460, NGC~2775, and NGC~2985). NGC~4151 also appears in \cite{ER08} but we have performed deeper photometry and have shifted the classification of this galaxy from Type~I to Type~II. In
the latter two papers, the procedure for constructing the luminosity profiles
is the same as that used in the current paper.

The data in Table~\ref{profiles}  show that there is no apparent correlation between the
presence of a $\sigma$-drop  and the shape of the disc surface
brightness profile.

We have also fitted a sersic+exponential profile to the 2MASS H-band profiles in order to find excess or lack of light in the circumnuclear zone. This excess or lack could be either related to a recent star-burst or to a high dust extinction. We find no differences between the two samples.

\section{Analysis of radio data}

The spatial resolution of radio observations is not generally good
enough to allow us to see details in the centres of galaxies. It
seems, nevertheless, reasonable to think that there may be a link between
the total quantity of gas  in a galaxy, as measured through their 21\,cm in \hi, or the millimetre emission in CO, and the quantity of gas available in the central region.
We thus investigate whether there are any statistical differences
between the gas content of the $\sigma$-drop and the control galaxies.

\subsection{Analysis of CO data}

\begin{table}
\begin{center}
\caption{\label{gas}Emission in CO and \hi{} by the sample
galaxies. Both the fluxes and the errors are in units of $10^{54}$
\,W\,km\,(Hz\,s)$^{-1}$ for CO, and of
$10^{52}$\,W\,km\,(Hz\,s)$^{-1}$ for
\hi. The error column corresponds to the photometric error and does
not take into account the uncertainty in the distance. In some cases,
no error estimate is given for the \hi{} emission because this was not
given in the literature. A hyphen (--) means that no data are
available.}
\begin{tabular}{l c c c c}
\hline
ID & \multicolumn{2}{c}{CO}& \multicolumn{2}{c}{\hi{}}\\
   & Flux & Error & Flux & Error\\
\hline
\multicolumn{5}{c}{$\sigma$-drop sample}\\
\hline
NGC~1068 & 114.9 &  6.1  &   71.0 &  4.0 \\
NGC~1097 &  --   &  --   &  256.9 & 10.3 \\
NGC~1808 &  --   &  --   &   80.3 &  4.8 \\
NGC~2460 &  --   &  --   &  235.0 &  9.4 \\
NGC~2639 &  --   &  --   &Very little& --\\
NGC~2775 &  24.3 &  4.4  &   20.4 &  1.8 \\
NGC~3021 &  --   &  --   &   79.2 &  3.2 \\
NGC~3593 &  10.7 &  1.9  &    8.7 &  0.7 \\
NGC~3623 &$<16.7$&  8.4  &   26.5 &  1.4 \\
NGC~3627 &  83.5 &  8.4  &   78.8 &  3.5 \\
NGC~4030 &  55.9 & 11.2  &  202.6 &  8.1 \\
NGC~4303 & 155.0 &  9.6  &  433.7 & 15.6 \\
NGC~4579 &  57.6 & 12.7  &   58.3 &  4.7 \\
NGC~4725 &  40.7 & 14.6  &  139.0 &  6.7 \\
NGC~4826 &  12.4 &  1.5  &   26.6 &  1.3 \\
NGC~6503 &   3.3 &  1.1  &   23.9 &  1.1 \\
NGC~6814 &  16.5 &  3.1  &  162.3 &  6.5 \\
NGC~6951 & 102.6 & 21.4  &  172.5 &  5.5 \\
\hline
\hline
\multicolumn{5}{c}{Control sample}\\
\hline
NGC~488  &  66.6 & 16.0  &  190.1 &  10.7\\
NGC~864  &  29.0 &  7.4  &  358.9 &  11.5\\
NGC~1566 &   --  &  --   &  312.7 &  17.5\\
NGC~2964 &  17.3 &  3.1  &   86.6 &   3.1\\
NGC~2985 &  39.0 & 13.2  &  361.6 &  28.9\\
NGC~3031 &   --  &  --   &  133.8 &   8.6\\
NGC~3169 &  48.7 &  9.8  &  127.6 &  13.3\\
NGC~3198 &   --  &  --   &  123.8 &   5.9\\
NGC~3227 &   --  &  --   &   81.6 &   4.9\\
NGC~3368 &   --  &  --   &   73.8 &   3.8\\
NGC~3675 &  32.8 &  6.4  &   89.1 &   3.2\\
NGC~3718 & $<8.3$&  4.4  &  381.1 &  15.2\\
NGC~4102 &  27.3 &  6.2  &   38.1 &   2.3\\
NGC~4151 &  13.6 &  3.2  &  122.5 &   5.4\\
NGC~4459 & $<5.0$&  2.5  &   --   &  --  \\
NGC~5055 & 124.2 &  9.0  &  693.4 &  36.1\\
NGC~6340 &   --  &  --   &   55.3 &   4.4\\
NGC~6501 &   --  &  --   &  103.5 &  17.4\\
NGC~7331 &  75.4 &  9.7  &  366.4 &  10.3\\
NGC~7742 &   --  &  --   &   59.4 &   2.1\\
\hline
\hline
\end{tabular}
\end{center}
\end{table}

We have used data from two CO surveys (\cite{YO95} and \cite{HE03}) to
search for a correlation between CO emission and the presence of
$\sigma$-drops. Not all the galaxies in our samples were included in
the surveys; the CO fluxes and the uncertainties of those that do
appear are listed in Table \ref{gas}. Using the distance to each
galaxy, we converted the observed flux values into the
total emission of the galaxy in the CO line. We obtained CO emission data for 13 galaxies for the $\sigma$-drop sample and 12 more for the control sample. These two subsamples are well matched in axis ratio, morphological type, and brightness. They are not so well matched in distance which is not really important because we are interested in an integrated CO luminosity.

\begin{table}
\begin{center}
\caption{\label{costats}Statistics of CO emission in the $\sigma$-drop and
control samples. All data are in units of $10^{54}$\,W\,km\,(Hz\,s)$^{-1}$.}
\begin{tabular}{l c c c }
\hline
sample & mean & median & std. deviation\\
\hline
$\sigma$-drop & 52.1 & 40.7 & 48.88\\
control & 39.5 & 30.9 & 35.5 \\
\hline
\hline
\end{tabular}
\end{center}
\end{table}

The statistical results for the CO are summarised in
Table~\ref{costats}, which shows that there are no significant
differences between the two samples. This is relevant because CO is a
good tracer of molecular gas and $\sigma$-drops are thought to be
related to accumulations of gas in the nuclear region. The absence of
observed differences between the amount of gas in
galaxies with and without $\sigma$-drops could thus be seen as evidence
against this assumption, although our current data do not allow us to say anything about the level of concentration of the CO, and thus of the amount of gas in their nuclear regions.

\subsection{Analysis of \hi{} data}

We obtained integrated \hi{} emission for our sample galaxies from LEDA, except for NGC~2639 where the data come from \cite{SP05}. As in the case of
the CO emission, we used the distance to each galaxy to derive the
total emission. The results are listed in Table \ref{gas}. We found data for 18 galaxies from the $\sigma$-drop sample and 19 for the control sample. The two subsamples of galaxies for which we have \hi{} data are well matched.

\begin{table}
\begin{center}
\caption{\label{histats}Statistics of \hi{} emission in $\sigma$-drop
and control samples. All data are in units of
$10^{52}$\,W\,km\,(Hz\,s)$^{-1}$.}
\begin{tabular}{l c c c }
\hline
sample & mean & median & std. deviation\\
\hline
$\sigma$-drop & 115.3 & 79.0 & 112.8\\
control & 197.8 & 123.8 & 170.1 \\
\hline
\hline
\end{tabular}
\end{center}
\end{table}

The statistical results for the integral \hi{} emission are summarised in
Table~\ref{histats}. Although the \hi{} emission of the $\sigma$-drop sample is more than two times lower than that of the control sample, this difference is not statistically significant. The conclusion from
our analysis of both CO and \hi\ emission must be that there is no
compelling evidence for either more or less gas in the $\sigma$-drop galaxies as
compared to those without $\sigma$-drops.

\section{Discussion and conclusions}

In this paper we have presented an analysis of the morphology of dust
and \ha{} emission on scales smaller than 0.1\,arcsec and within the central
1\,kpc radius, of the disc surface brightness profiles, and of the CO
and \hi{} emission of a sample of $\sigma$-drop host galaxies, and
compared the results to those obtained for a matched control sample.

Our control sample is not guaranteed to be  completely free of $\sigma$-drop galaxies,
but we estimate (Sect.~2) that less than 20\% of our control
sample galaxies can be expected to host a $\sigma$-drop. This possible contamination somewhat weakens the significance of our results. Another factor that can weaken our results is that, even though this is the biggest $\sigma$-drop galaxy sample studied in such a consistent way, it is not big enough to avoid problems with small number statistics. In addition, to study correlations for several parameters we had to work with subsamples that were not as well matched.

Our overall conclusion is that $\sigma$-drops are related to nuclear dust spirals and nuclear star-forming rings. We find that the $\sigma$-drop radius is independent of the other tested parameters of the host galaxy. We note the possibility that Seyfert galaxies are related to $\sigma$-drops and find that LINERs are found more often in galaxies without this feature. We also find that bars, at least
strong bars, are not likely to be a necessary feature to cause a
$\sigma$-drop in a galaxy.

One of the theories for the creation of $\sigma$-drops is based on a
massive and concentrated dark matter halo which can remove kinetic
energy from the stellar component of galactic nuclei
(\cite{AT02}). Such a dark matter halo would, however, have other
effects on a galaxy's morphology and kinematics, such as the
stimulation of a strong bar (\cite{AT02}). But in our $\sigma$-drop
sample there are only three strongly barred and ten weakly barred galaxies,
and these numbers are not significantly different for the control
sample. This indicates that massive and compact dark matter halos are
not likely to be the only $\sigma$-drop formation mechanism, and may well not affect this phenomenon at all.

There are other theories for $\sigma$-drops. Dressler \& Richstone (1990) propose that $\sigma$-drops may be due to the lack of a central super-massive black hole and van der Marel (1994) suggests that a $\sigma$-drop may be due to the presence of a super-massive black hole that broadens emission lines and makes `traditional' line fitting overestimate the true velocity. These hypotheses can now be completely discarded, mainly because the presence of a $\sigma$-drop has been found to be a widespread phenomenon (in the literature cited in the introduction they are found in 30\%-50\% of the disc galaxies) and because $\sigma$-drops are not found only in galaxies whose velocity profiles indicate the presence of an anomalous super-massive black hole.

Another theory for $\sigma$-drop formation points to cold gas accumulation and a posterior starburst in the inner parts of the galaxy (e.g., simulations by \cite{WO03}). Such a concentration of gas would start to create stars at a given density threshold. This critical density would probably only be related to local gas properties so the $\sigma$-drop size is independent of macroscopic galactic parameters. This is confirmed by our results. The huge dispersion in $\sigma$-drop sizes can be related to a reduction of their size when they become old or to the different extents of the cool gas volume at the moment it reached the critical density.

As we have shown in Section 4.2, $\sigma$-drop hosts have in more than 50\% of the cases a clear inner dust system of spiral arms that probably trace the path of the inflowing material. In contrast, less than 20\% of the control sample galaxies show such a clear spiral pattern. As there are no significant differences in bar characteristics between the two samples, it is safe to say that there is some correlation between $\sigma$-drops and inner dust spiral arms. This dust could trace the inflowing gas fuelling the $\sigma$-drop, or it could be the product of relatively recent star-formation.

We have also found that $\sigma$-drop galaxies often (in 50\% of the cases) have a nuclear star-forming ring visible in \ha{} with a size between 400\,pc and 1\,kpc and that there are no star-forming rings at those scales in the control sample. It thus seems reasonable to say that the rings are a manifestation of the same phenomena causing the $\sigma$-drops. The stars that compose the $\sigma$-drops can be formed in such nuclear starburst rings, usually due to shock focusing of gas near the location of one or more inner Lindblad resonances (ILRs; Schwarz 1984; Athanassoula 1992; Knapen et al.~1995, Heller \& Shlosman 1996; see also Shlosman 1999, Knapen 2005 for reviews). The ILRs are thought  to be related to bars or interactions between close galaxies. Four of our five galaxies with star-forming rings are barred. The fifth, NGC~6503 is so edge-on that a bar may be difficult to detect.

We have found that $\sigma$-drop galaxies are more often Seyfert than LINER hosts. On the other hand, control galaxies more often have LINER emission than Seyfert emission. Seyfert galaxies are more energetic phenomena, which probably implies a more continuous or more recent inflow of material to maintain the central activity. The dust spiral arms that are frequent in $\sigma$-drop galaxies are a plausible method to feed the central black hole. It is therefore possible to postulate that both $\sigma$-drops and Seyferts are due to a more efficient feeding of the inner parts of galaxies. As Seyferts are shorter lived than $\sigma$-drops (few $10^{8}$\,yrs versus 1\,Gyr) the correlation can only hold partially.

We thus suggest the following model for the creation of a $\sigma$-drop. Gas is driven inwards by spiral arms and maybe by bars. The gas is focused by an ILR into a dynamically cold ring where the density increases until it reaches a critical value when star formation starts. These stars are dynamically cold and start to create a $\sigma$-drop. Alternatively, if there is no ILR, gas can be focused in a cold nuclear disc that also starts to create stars when it reaches the critical density. In both cases friction causes some gas to lose enough angular momentum to drift inwards to feed the super-massive black hole. This gas can be traced with the dust related to the relatively recent star formation. It is entirely possible that $\sigma$-drops are exclusively created in rings that disappear due to the transient nature of the bars and/or the inflow. As a $\sigma$-drop is a long-lived feature (as shown by modelling by \cite{WO06}, and as indicated by the high fraction of galaxies that host them) it would survive long after the ring has disappeared. These conclusions should be strengthened in future work by improving the statistics and by further numerical modelling.

\begin{acknowledgements}

We thank Leonel Guti\'errez and Rebeca Aladro for their help with some
of the luminosity profiles used in this study, Peter Erwin for his help
in the classification of luminosity profiles, Paul Martini for
letting us peruse his structure map IDL scripts, and the referee, Eric Emsellem, for comments that helped improve the paper. This work was
supported by projects AYA 2004-08251-CO2-01 of the Ministerio de
Educaci\'on y Ciencia and P3/86 and 3I2407 of the Instituto de
Astrof\'isica de Canarias. Based on observations made with the
NASA/ESA {\it Hubble Space Telescope}, obtained from the data archive
at the Space Telescope Science Institute. STScI is operated by the
Association of Universities for Research in Astronomy, Inc. under NASA
contract NAS 5-26555. This work is also based in SDDS release 5.

\end{acknowledgements}

\clearpage

\appendix
\onecolumn
\section{Structure maps and \ha{} images of  the $\sigma$-drop sample
  galaxies}
  
Where two images for one galaxy are given on a row, the structure map is shown on the left, and the continuum-subtracted
\ha\ image on the right; where only one image is shown there is no
\ha\ image available. The orientation of the images is shown with
pairs of arrows, where the longer indicates north and the shorter
one east. The small ellipse indicates $r_{\rm c}$, as defined in the
text, and the large one is at 1\,kpc radius. Where one ellipse is drawn
it indicates $r_{\rm c}$---in those cases the image scale is such
that the 1\,kpc radius falls outside the image. Grey-scales are such
that dark indicates more dust (less emission) in the structure maps, and
more \ha\ emission in the \ha\ images. Each image has at the top left the name of the galaxy, the morphological classification given in NED and the activity, also given in NED.

\clearpage
 
\begin{figure}[h]
\begin{center}
\begin{tabular}{c}
\includegraphics[width=0.4\textwidth]{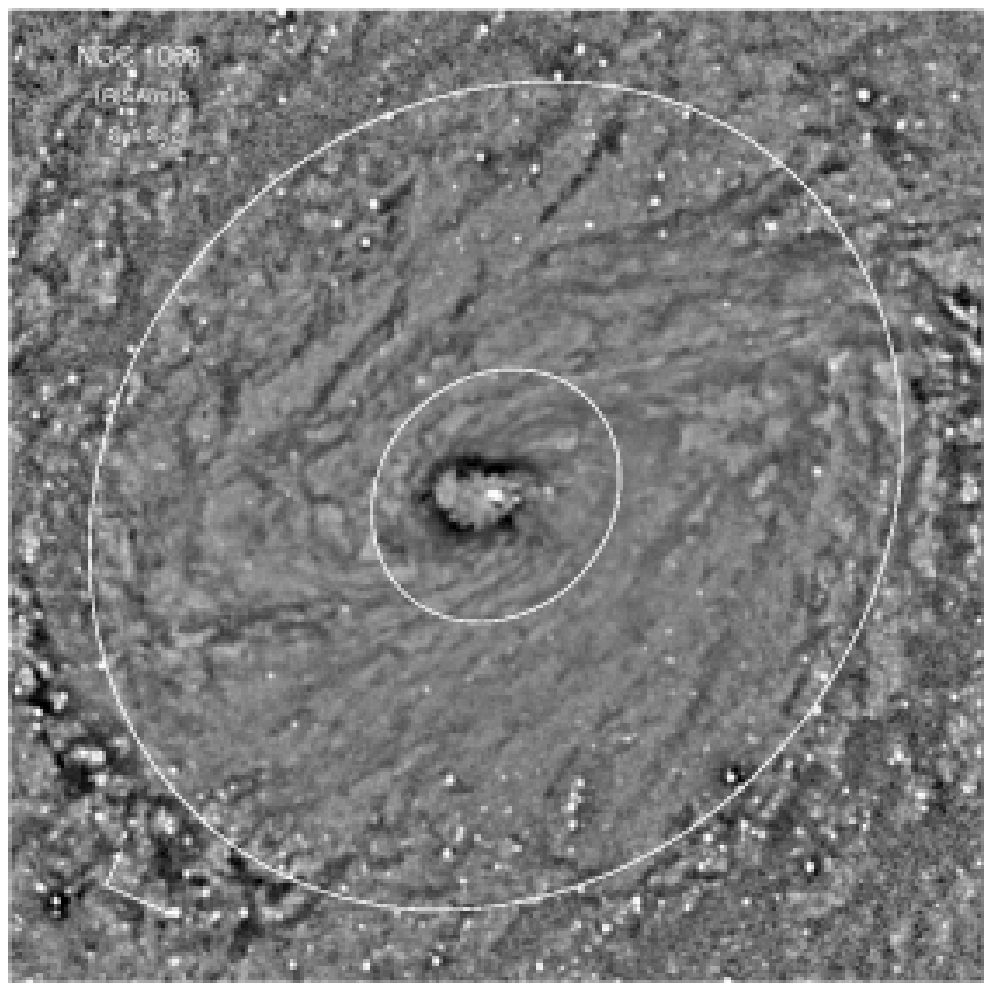}
\includegraphics[width=0.4\textwidth]{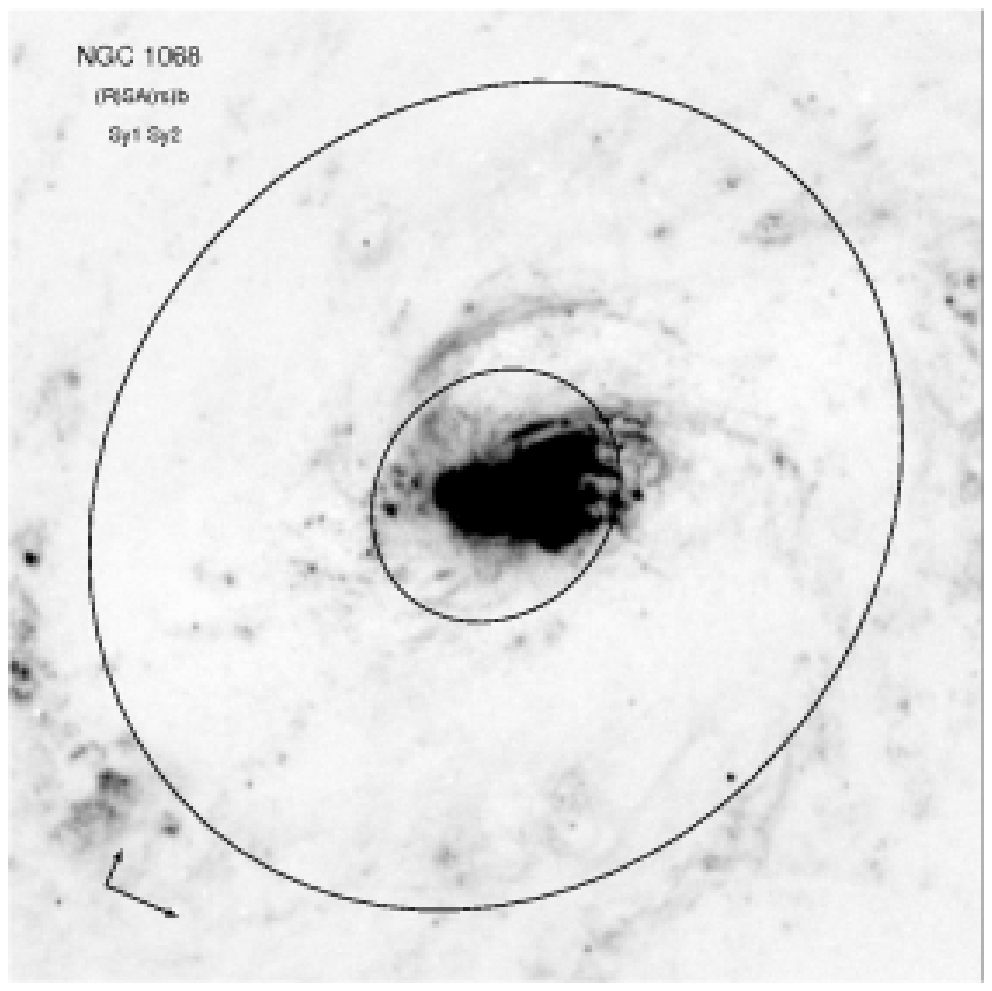}\\
\includegraphics[width=0.4\textwidth]{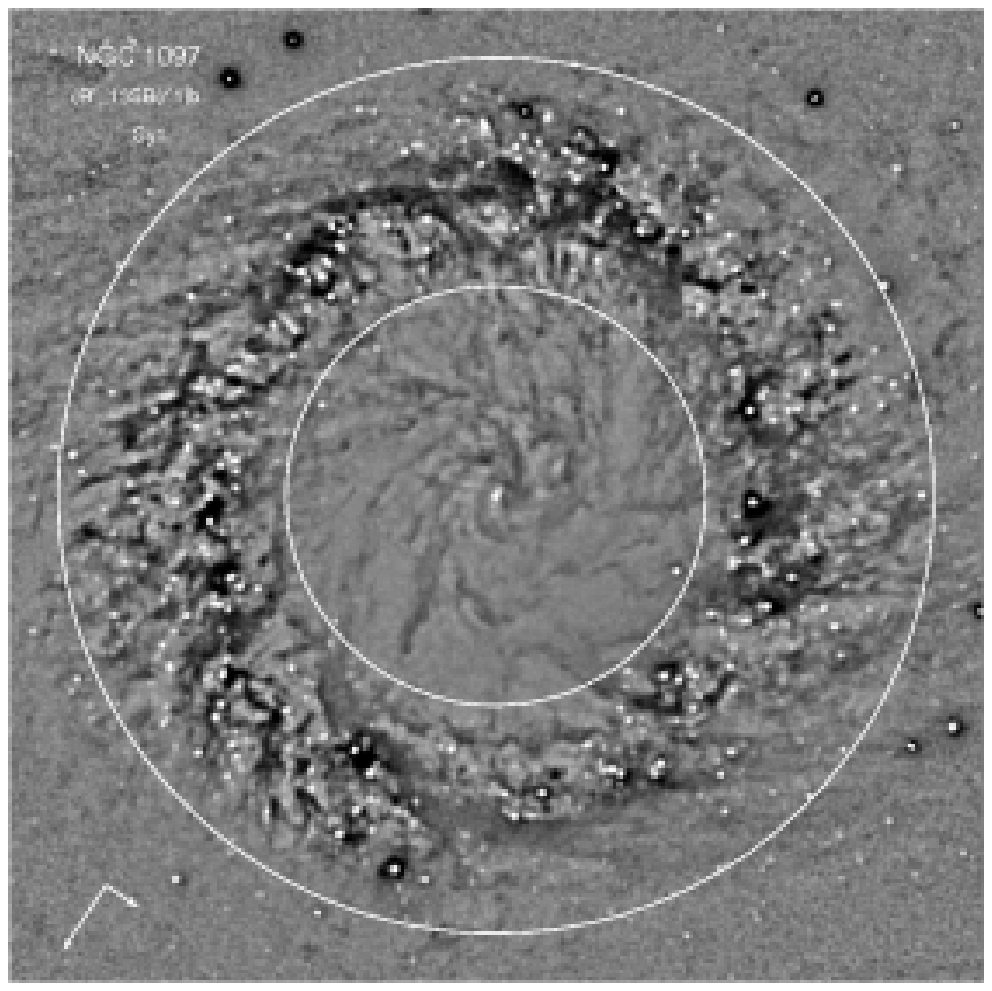}
\includegraphics[width=0.4\textwidth]{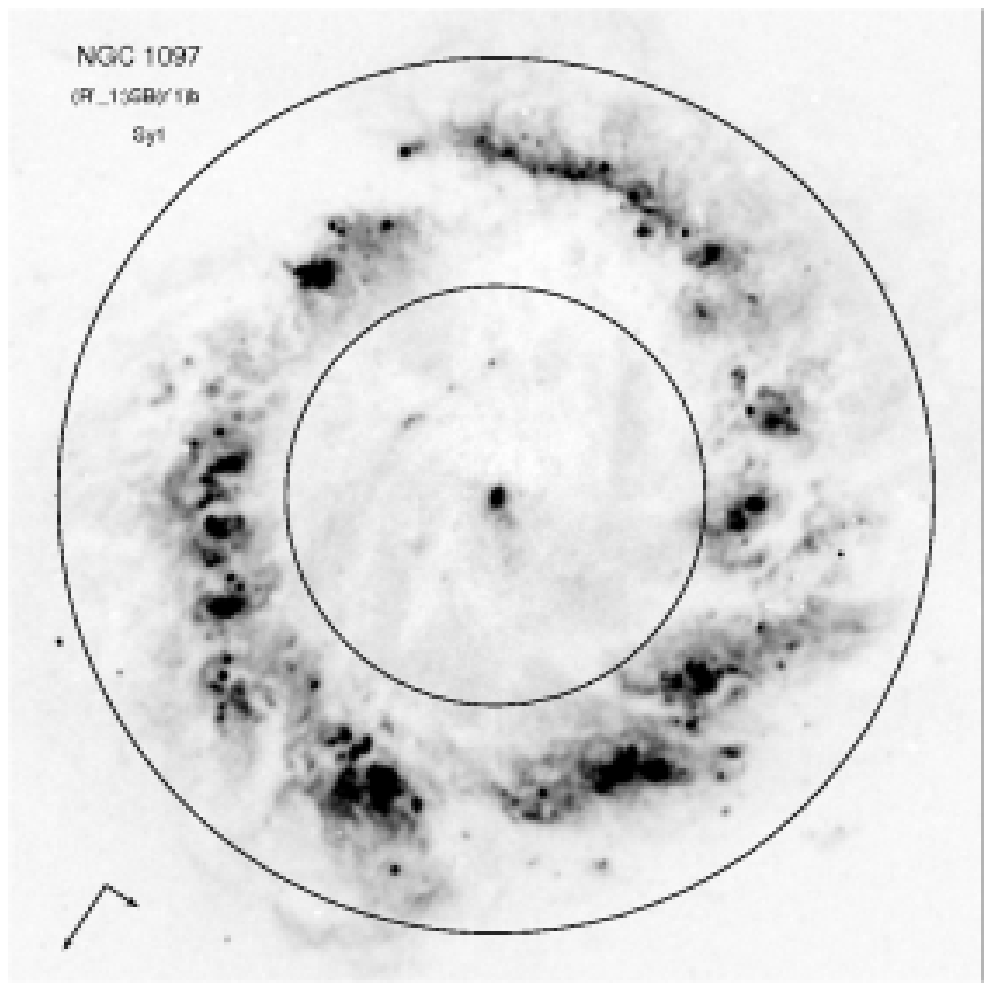}\\
\includegraphics[width=0.4\textwidth]{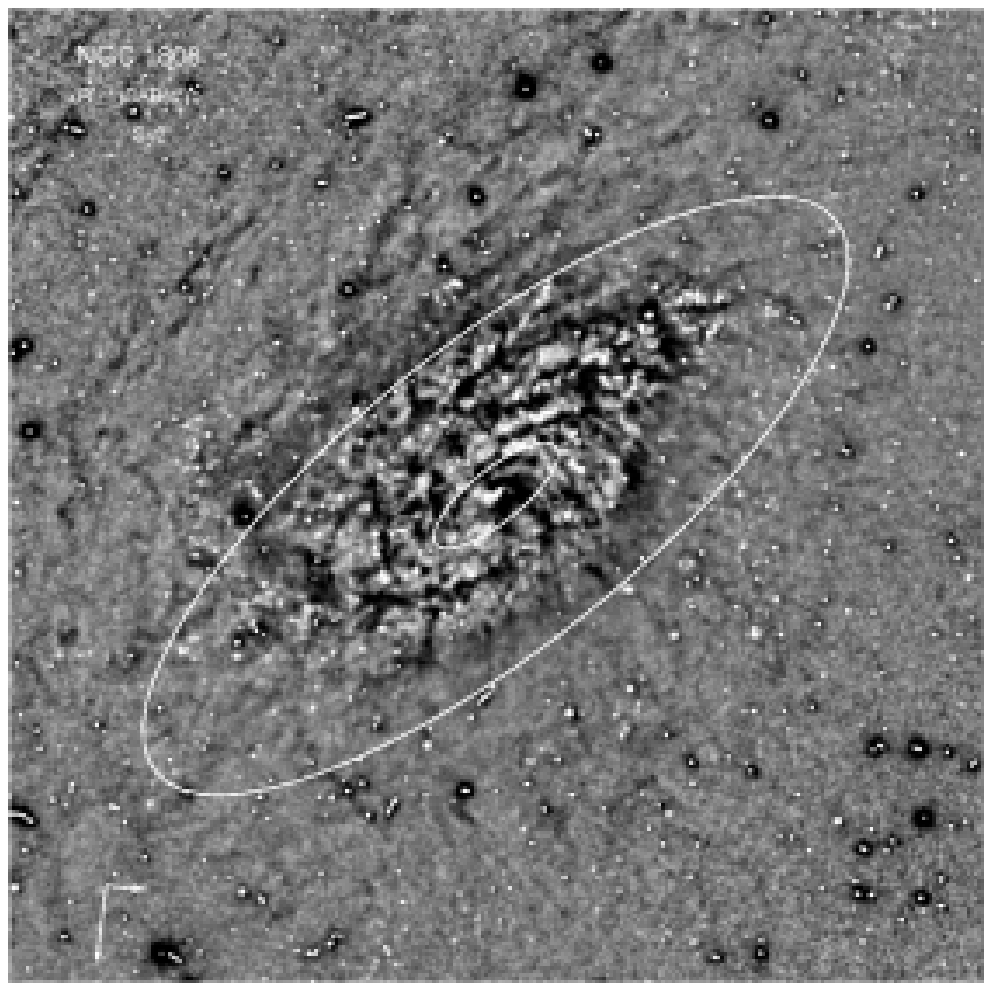}
\includegraphics[width=0.4\textwidth]{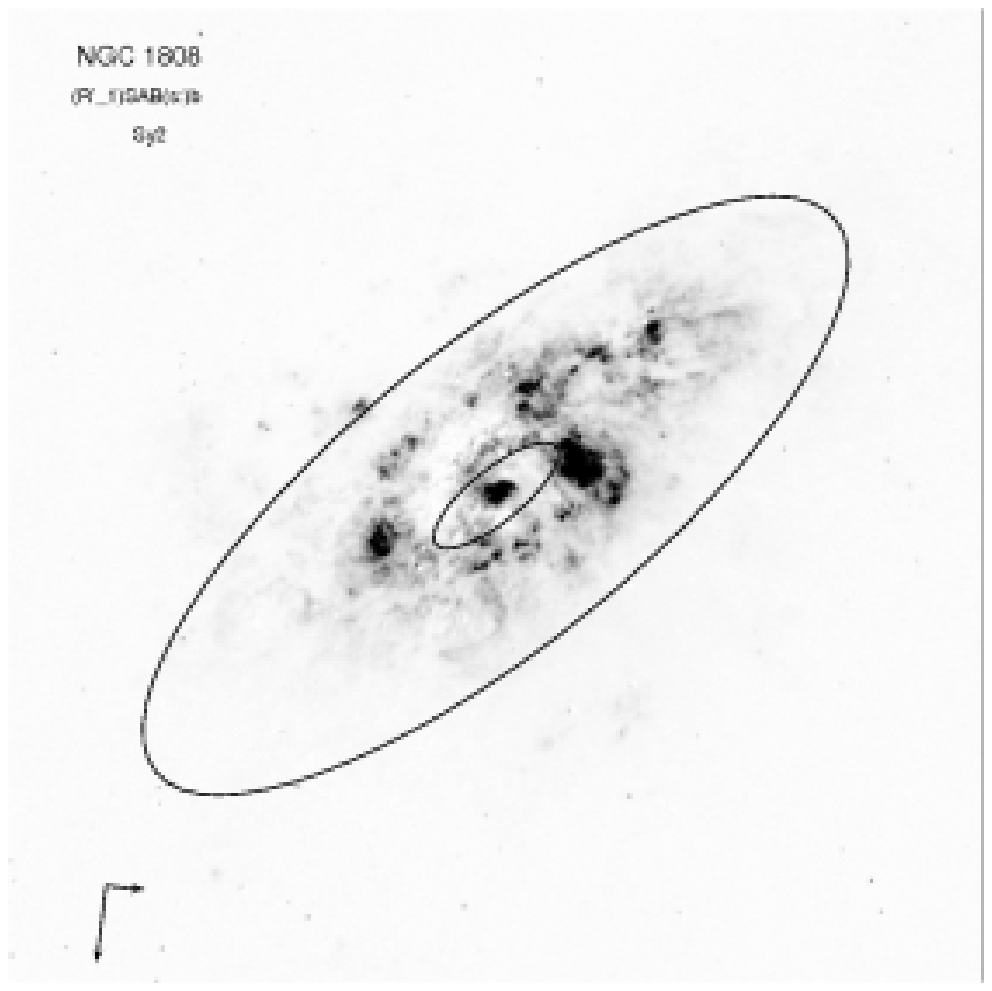}\\
\end{tabular}
\end{center}
\end{figure}

\begin{figure}
\begin{center}
\begin{tabular}{c}
\includegraphics[width=0.4\textwidth]{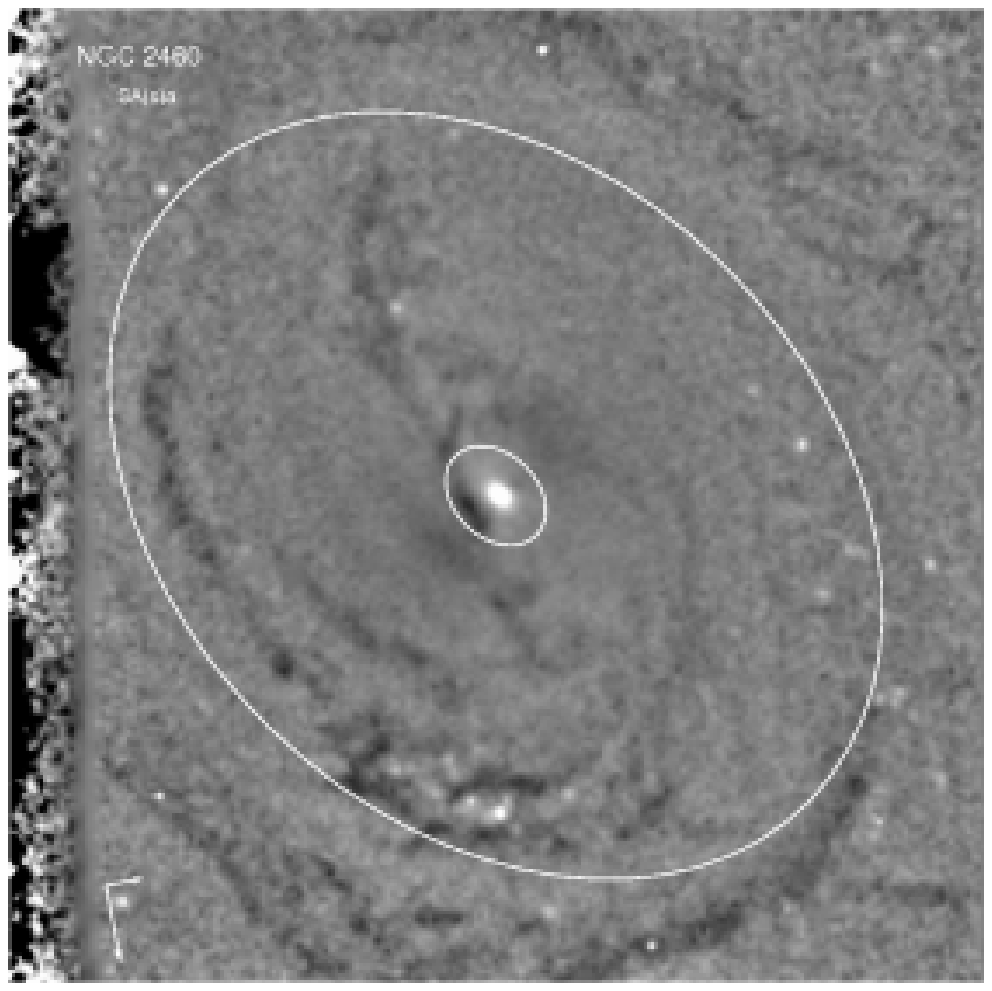}\\
\includegraphics[width=0.4\textwidth]{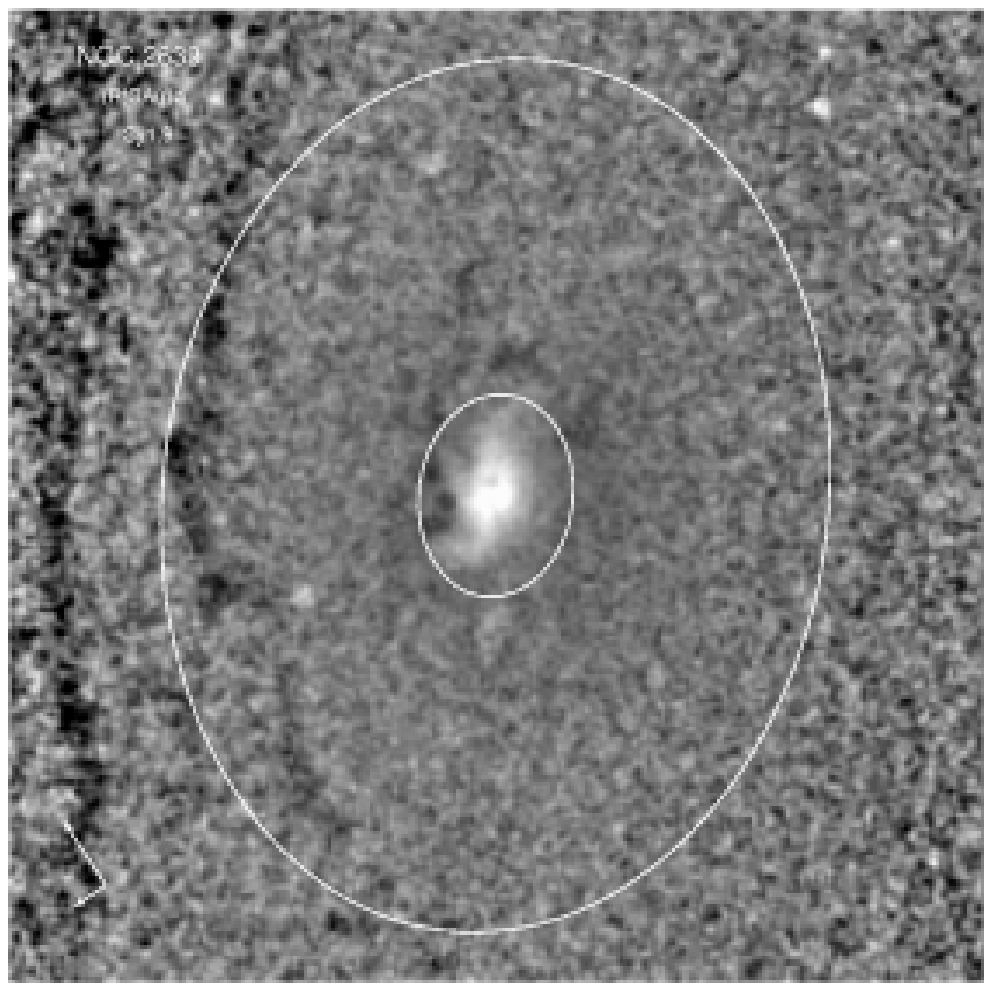}\\
\includegraphics[width=0.4\textwidth]{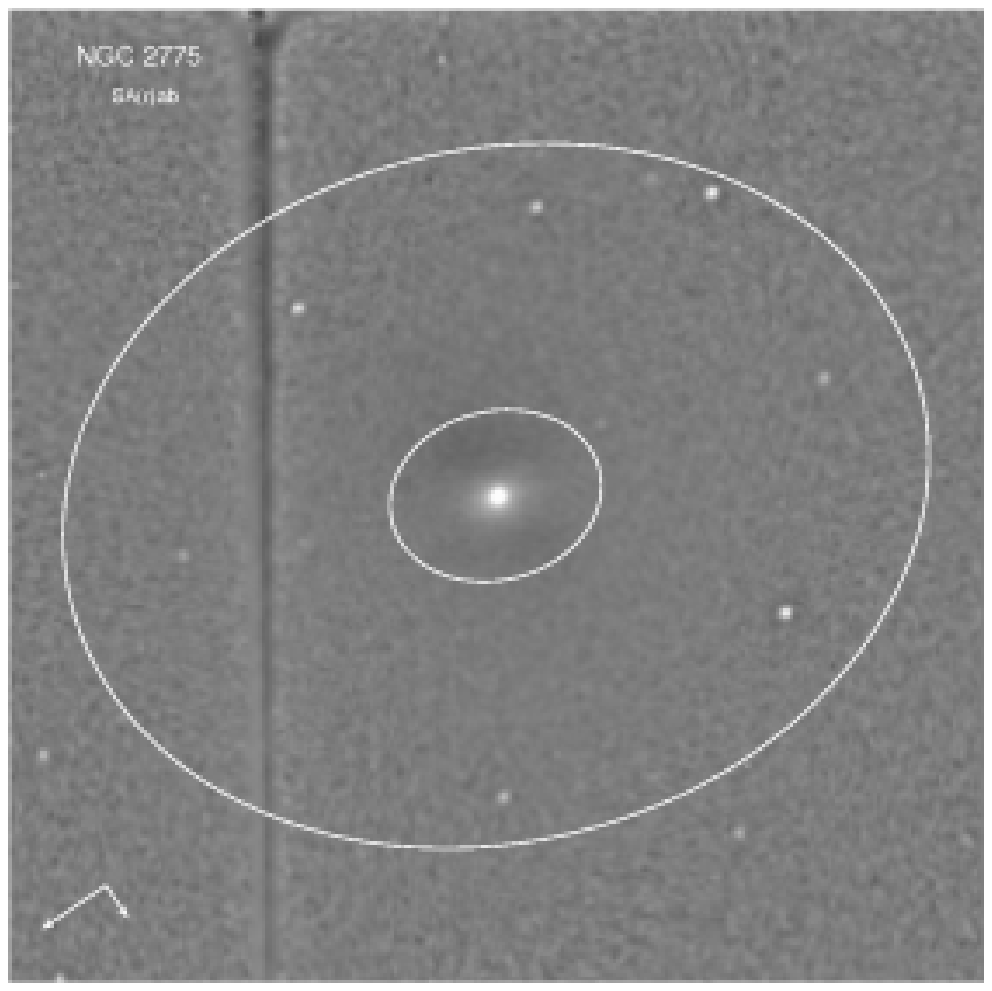}\\
\end{tabular}
\end{center}
\end{figure}

\begin{figure}
\begin{center}
\begin{tabular}{c}
\includegraphics[width=0.4\textwidth]{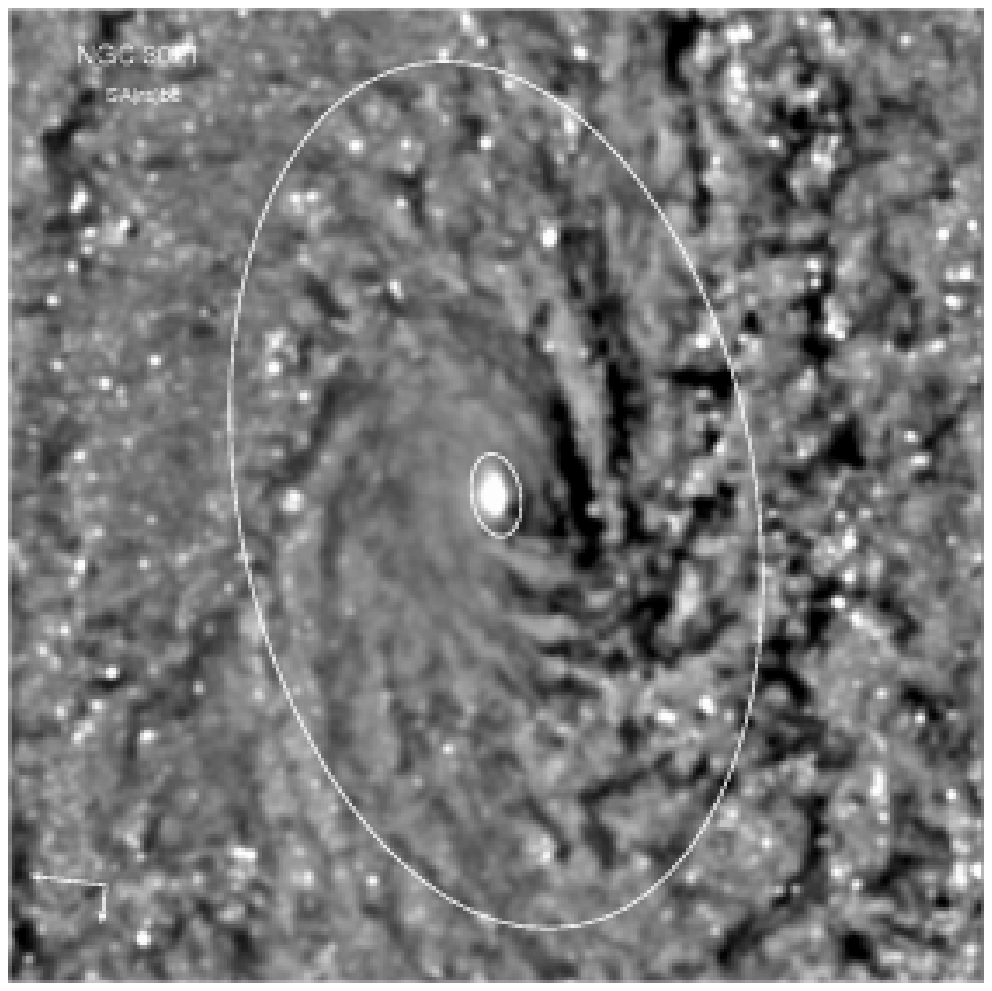}\\
\includegraphics[width=0.4\textwidth]{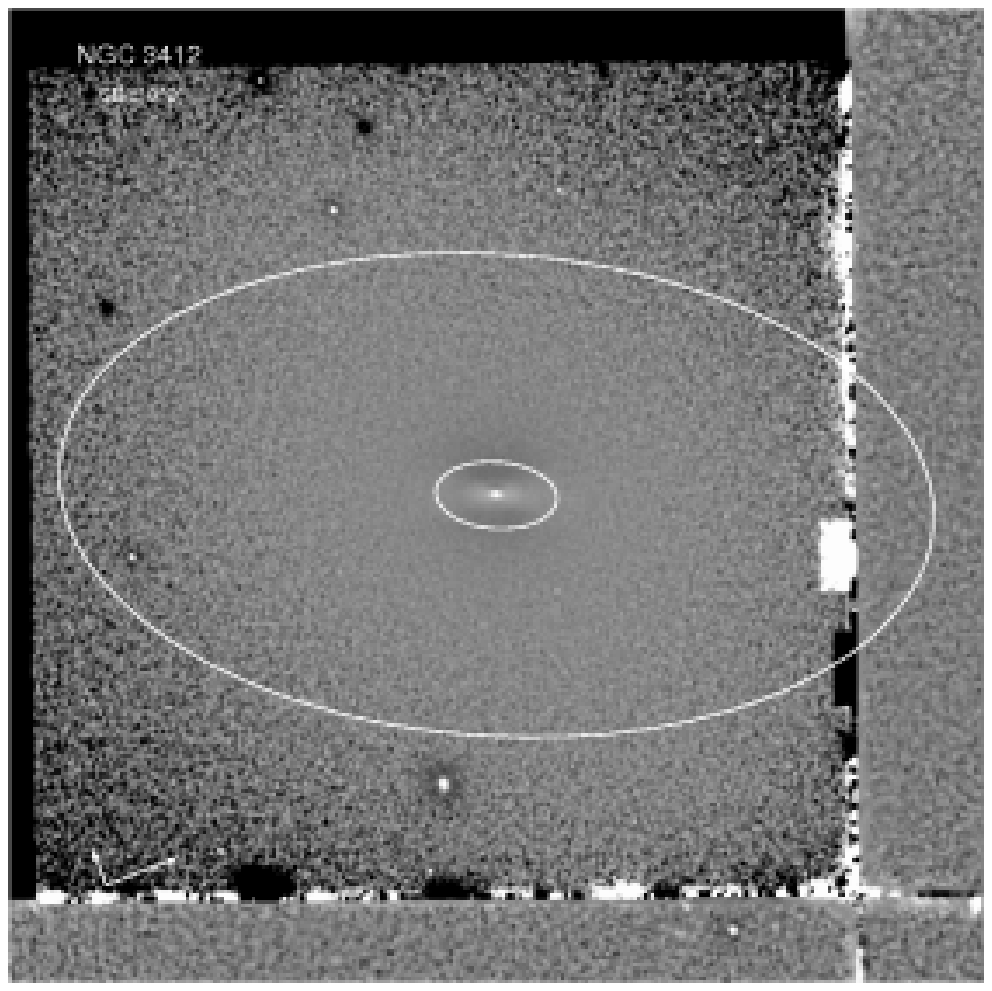}\\
\includegraphics[width=0.4\textwidth]{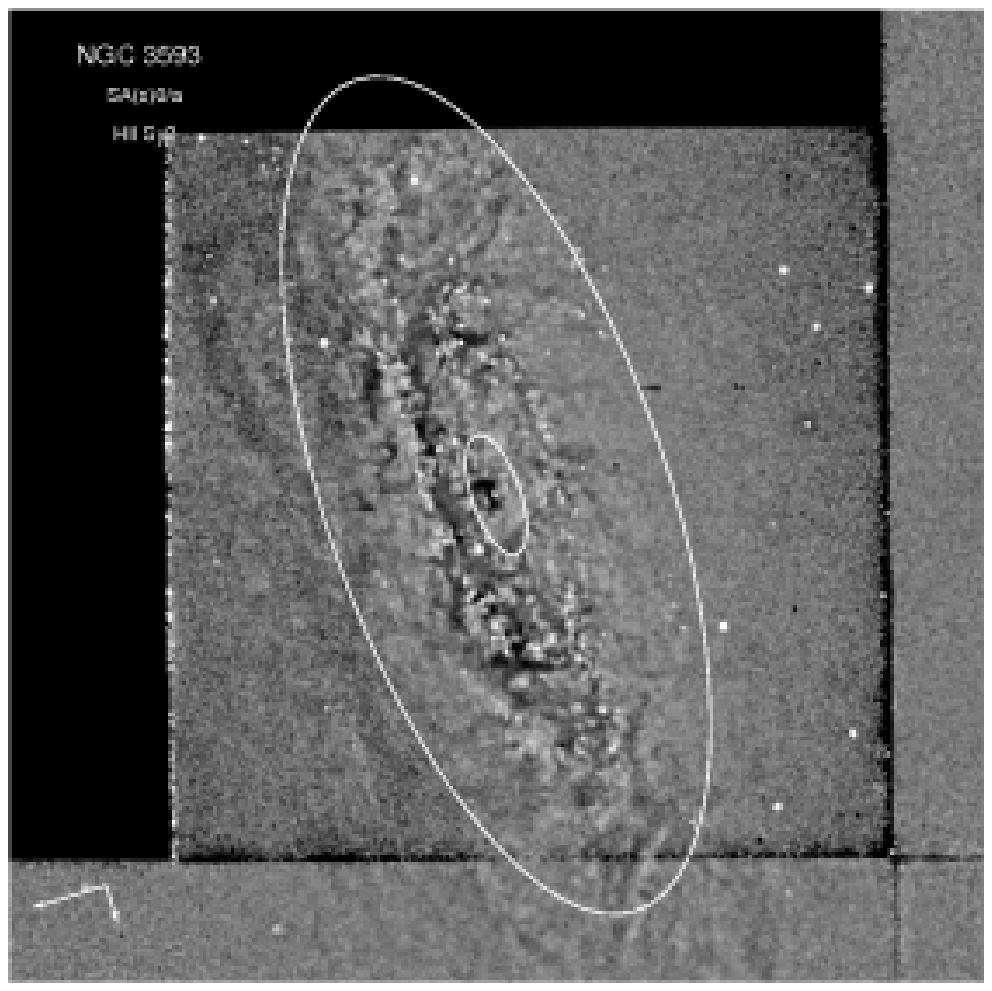}
\includegraphics[width=0.4\textwidth]{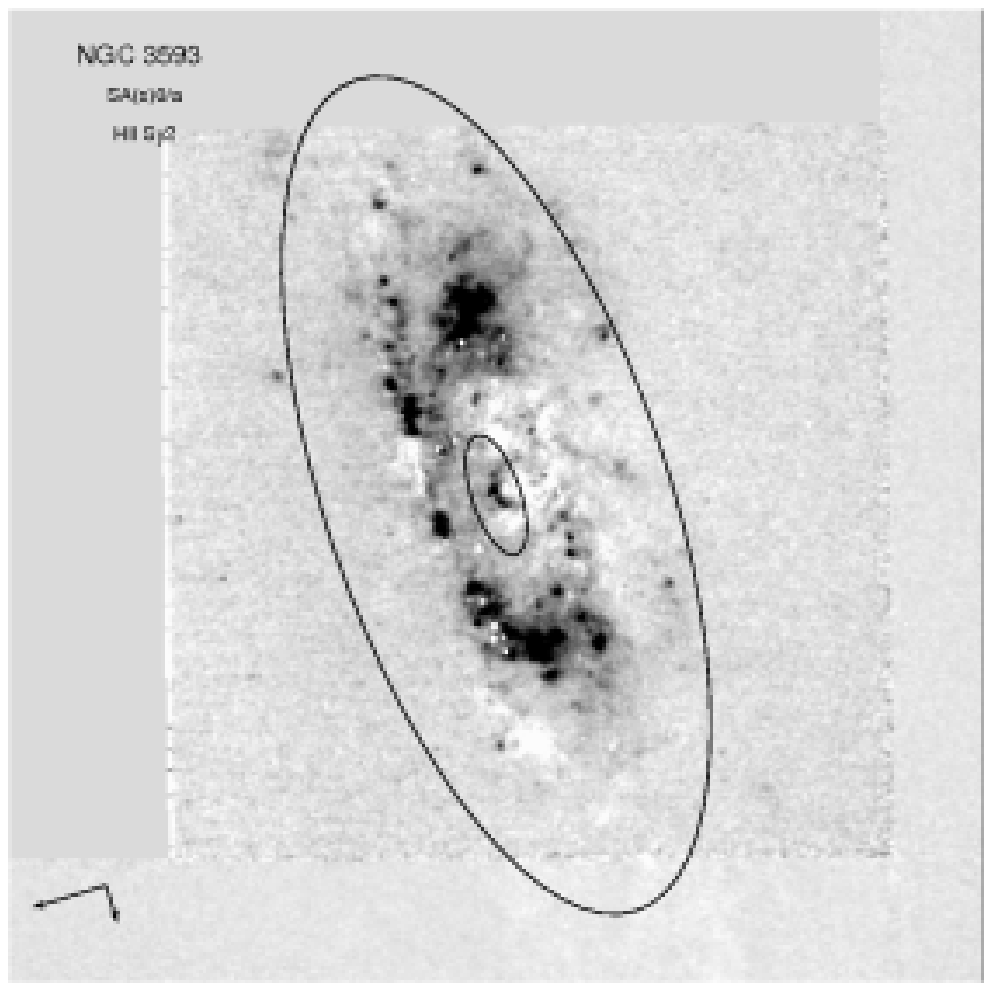}\\
\end{tabular}
\end{center}
\end{figure}

\begin{figure}
\begin{center}
\begin{tabular}{c}
\includegraphics[width=0.4\textwidth]{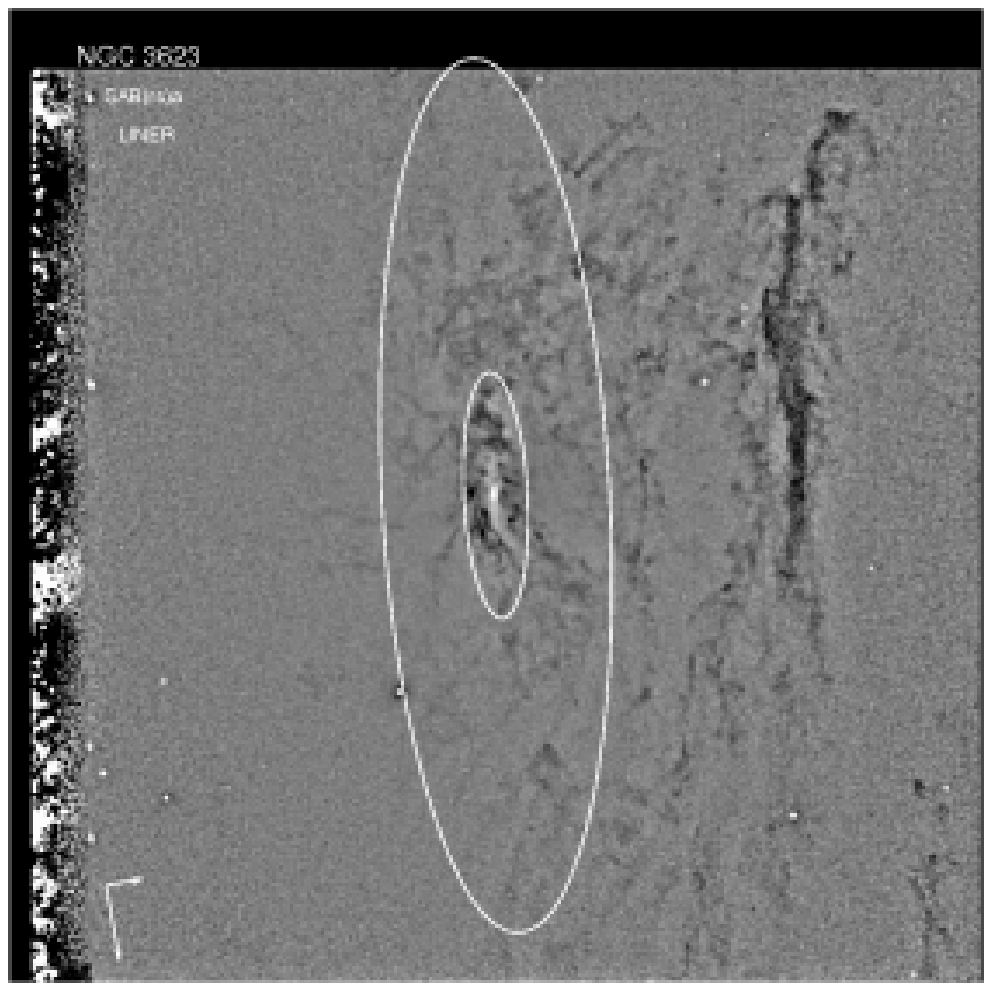}
\includegraphics[width=0.4\textwidth]{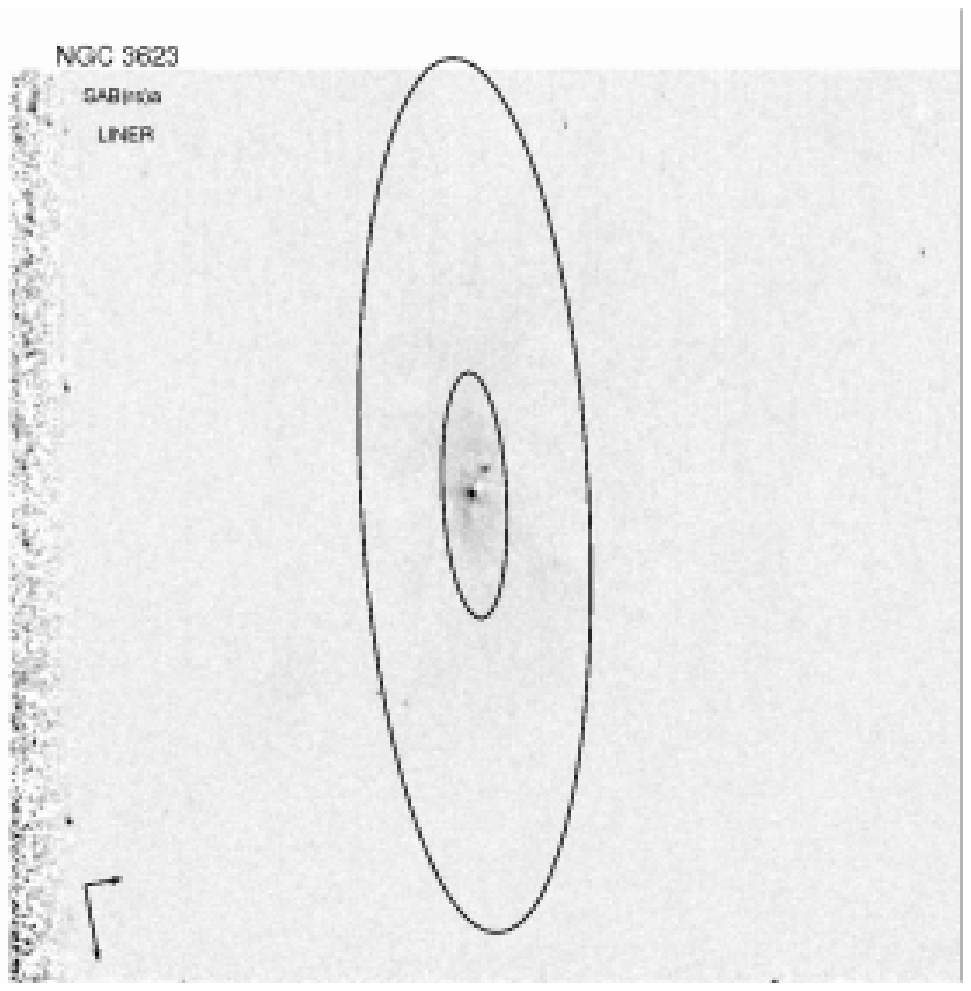}\\
\includegraphics[width=0.4\textwidth]{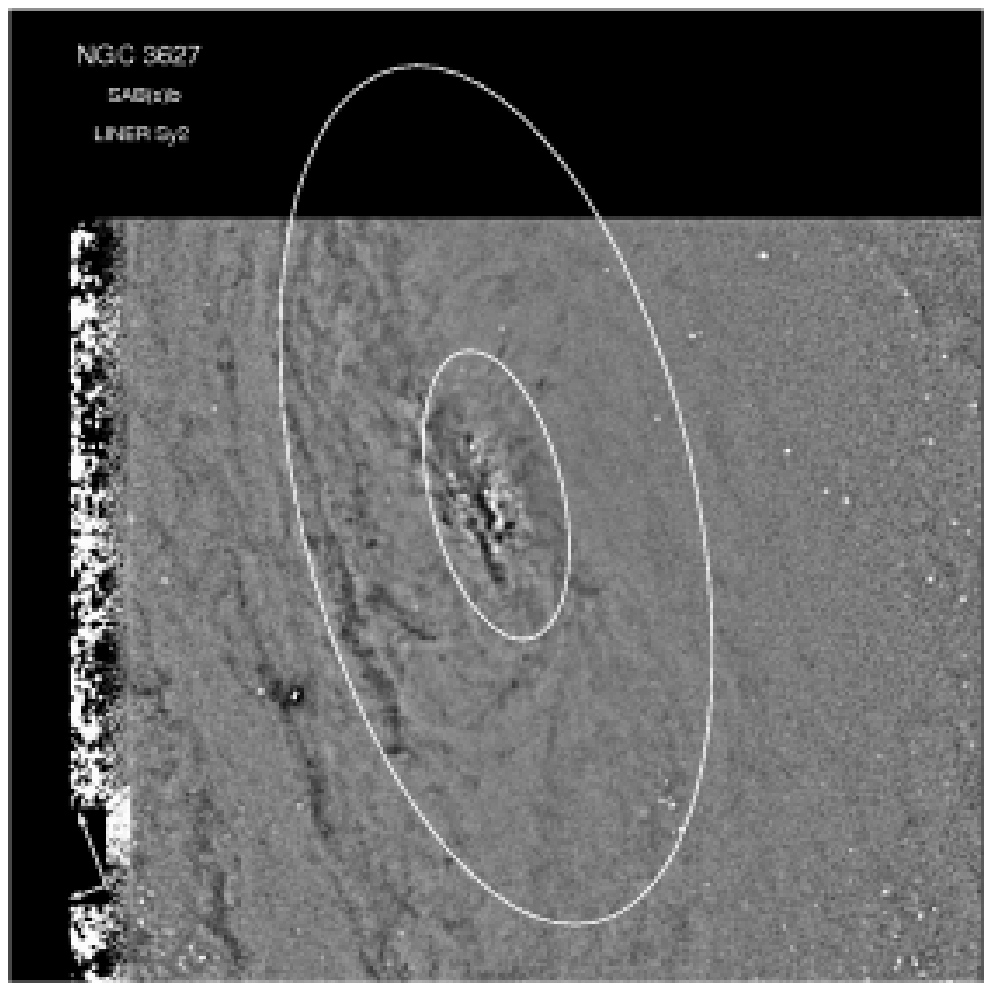}
\includegraphics[width=0.4\textwidth]{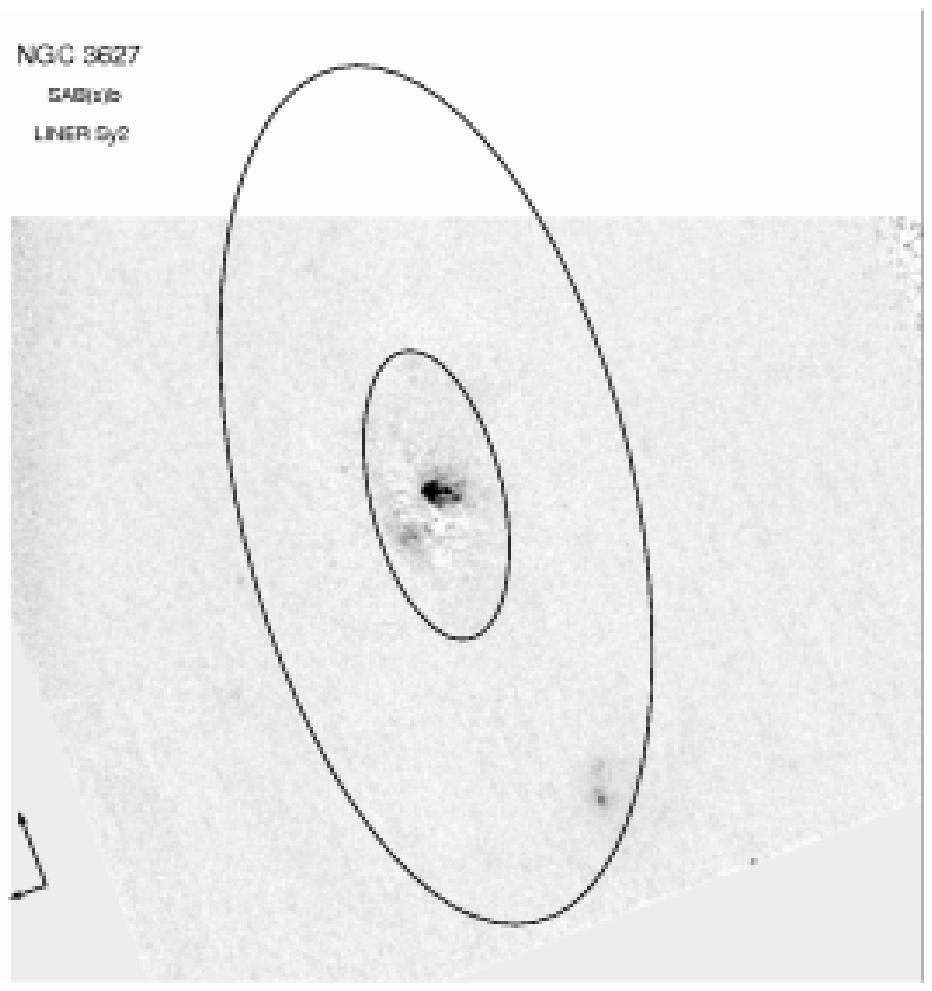}\\
\includegraphics[width=0.4\textwidth]{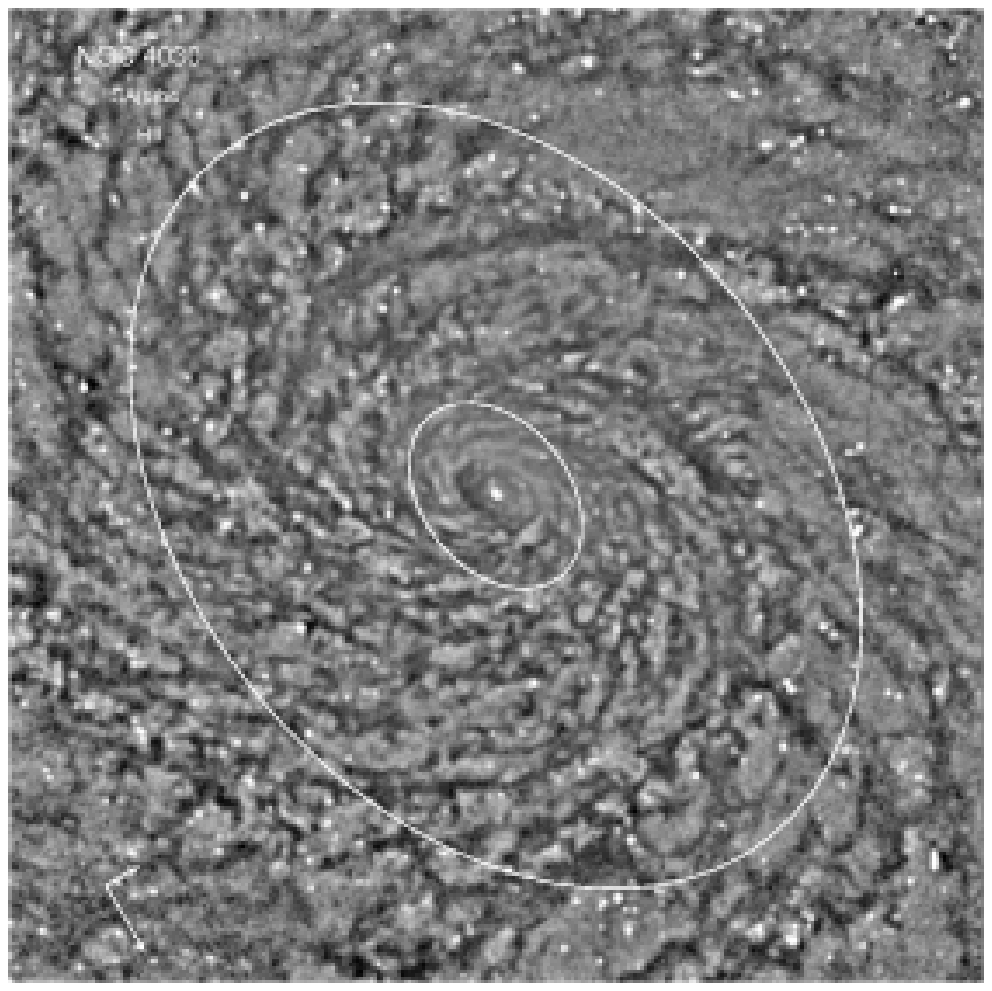}\\
\end{tabular}
\end{center}
\end{figure}

\begin{figure}
\begin{center}
\begin{tabular}{c}
\includegraphics[width=0.4\textwidth]{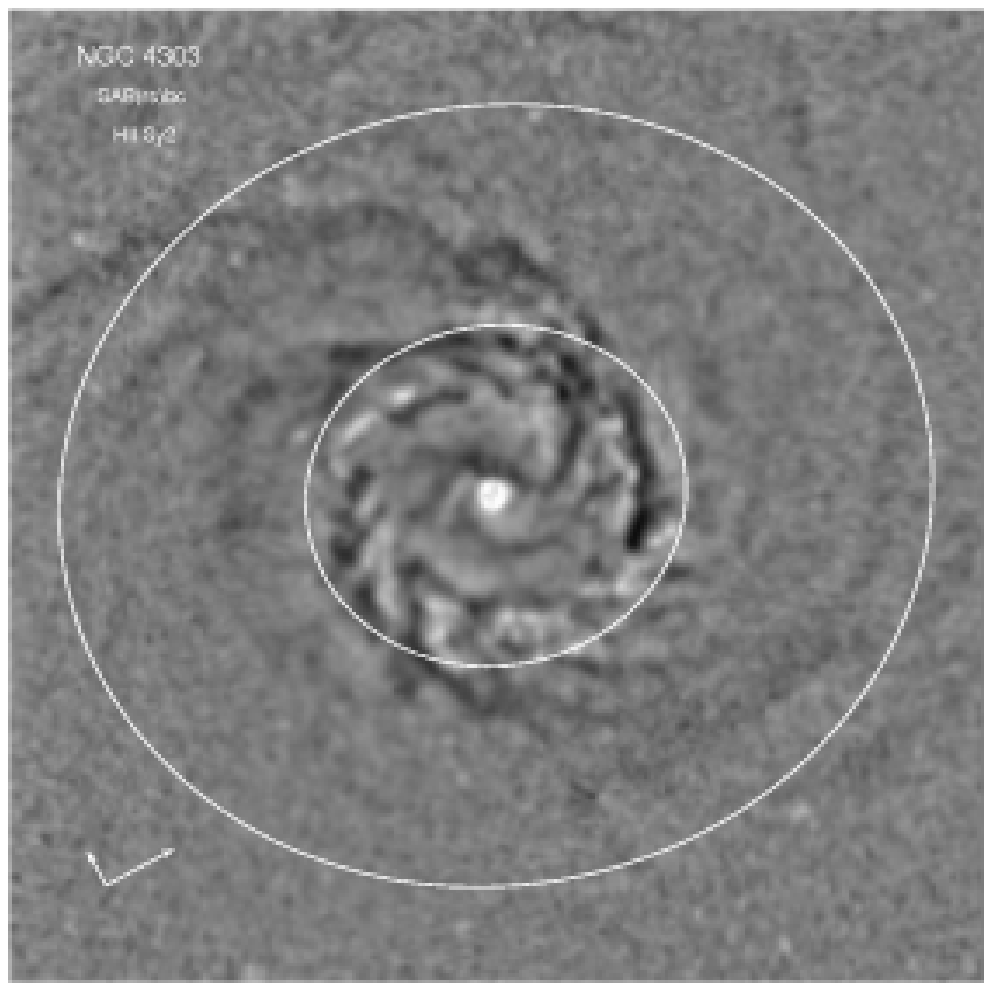}\\
\includegraphics[width=0.4\textwidth]{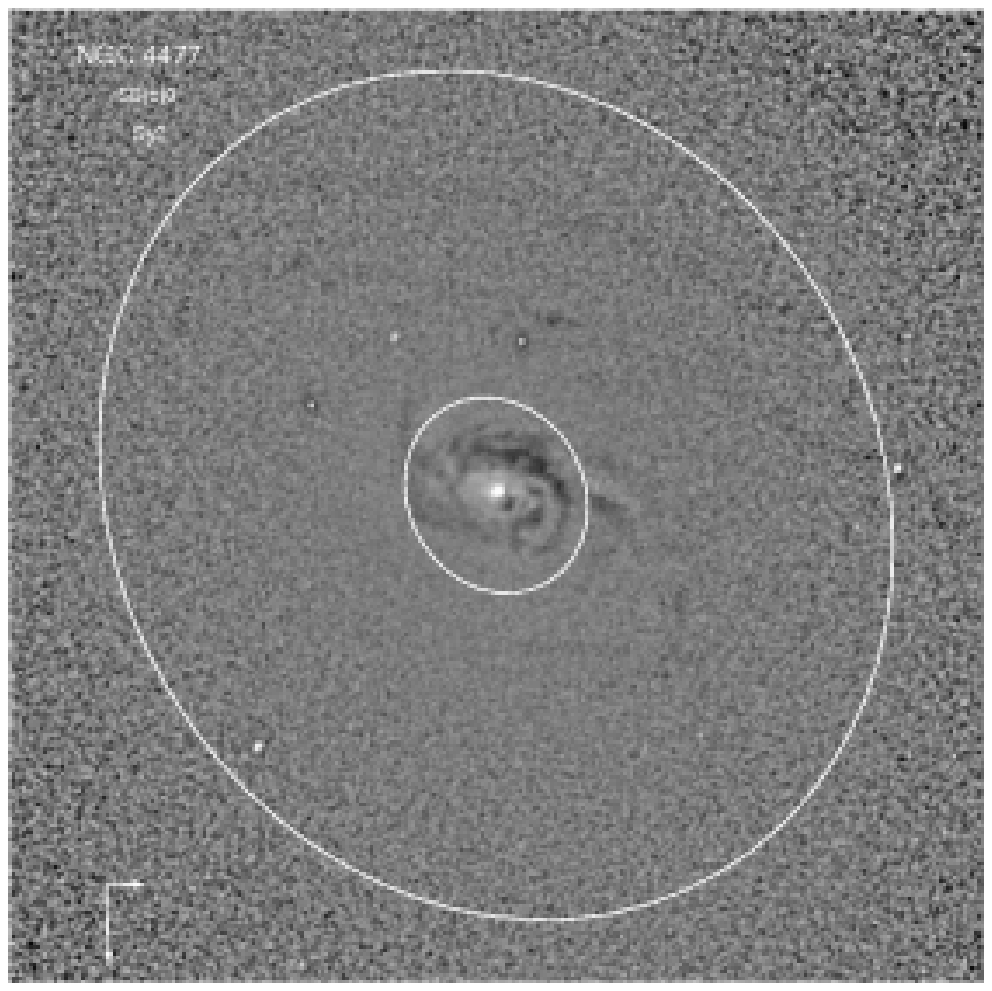}\\
\includegraphics[width=0.4\textwidth]{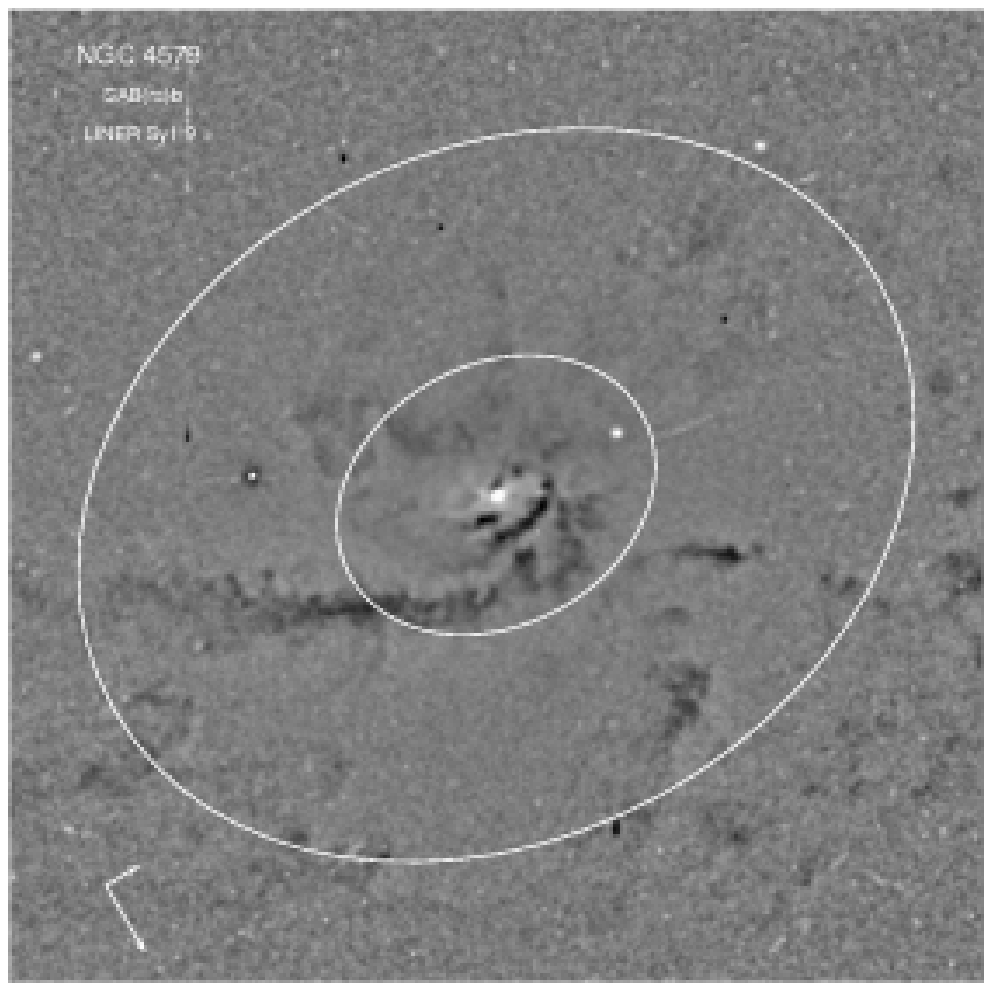}
\includegraphics[width=0.4\textwidth]{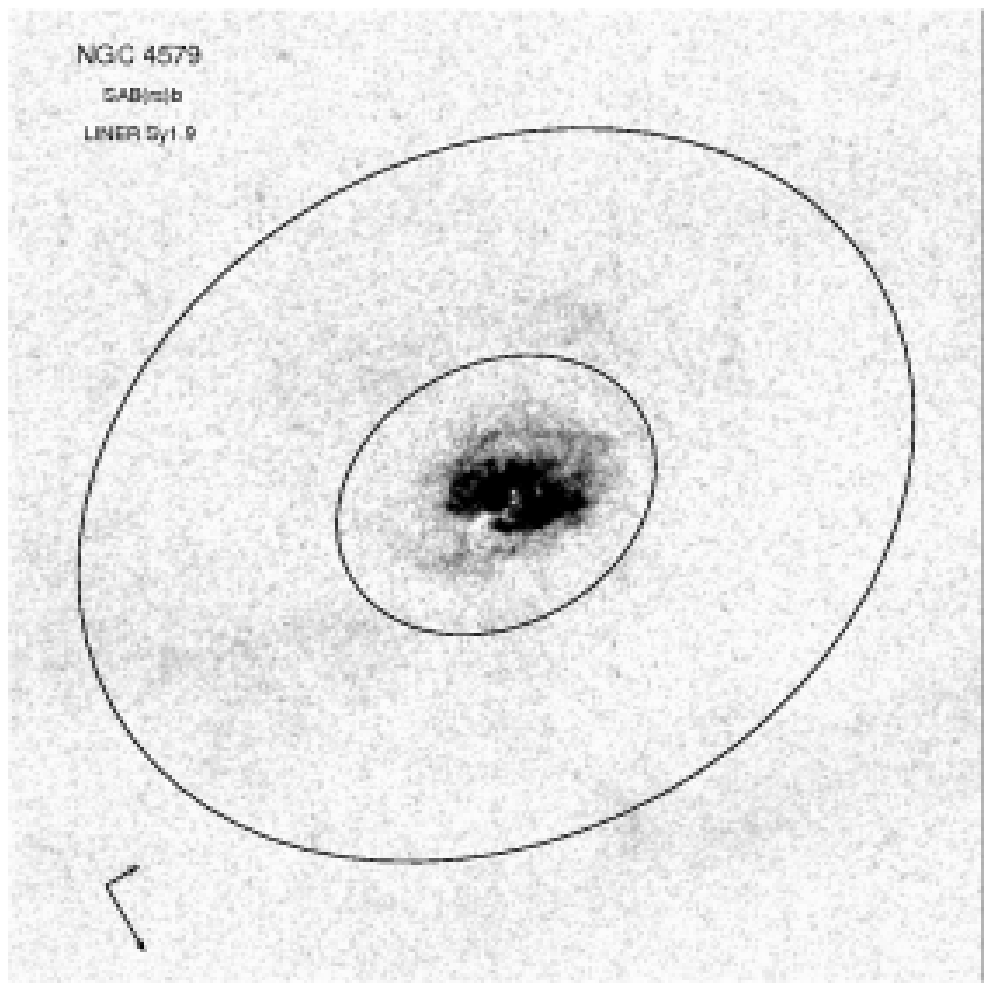}\\
\end{tabular}
\end{center}
\end{figure}

\begin{figure}
\begin{center}
\begin{tabular}{c}
\includegraphics[width=0.4\textwidth]{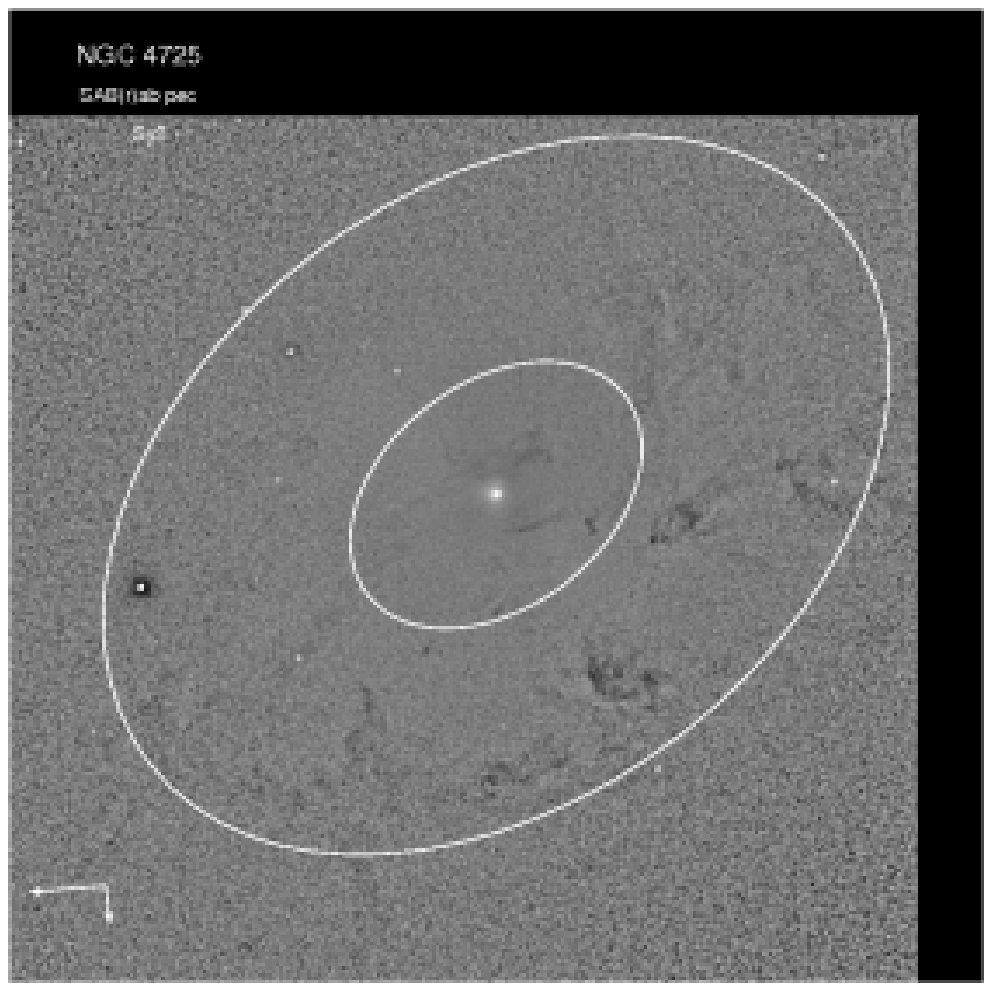}\\
\includegraphics[width=0.4\textwidth]{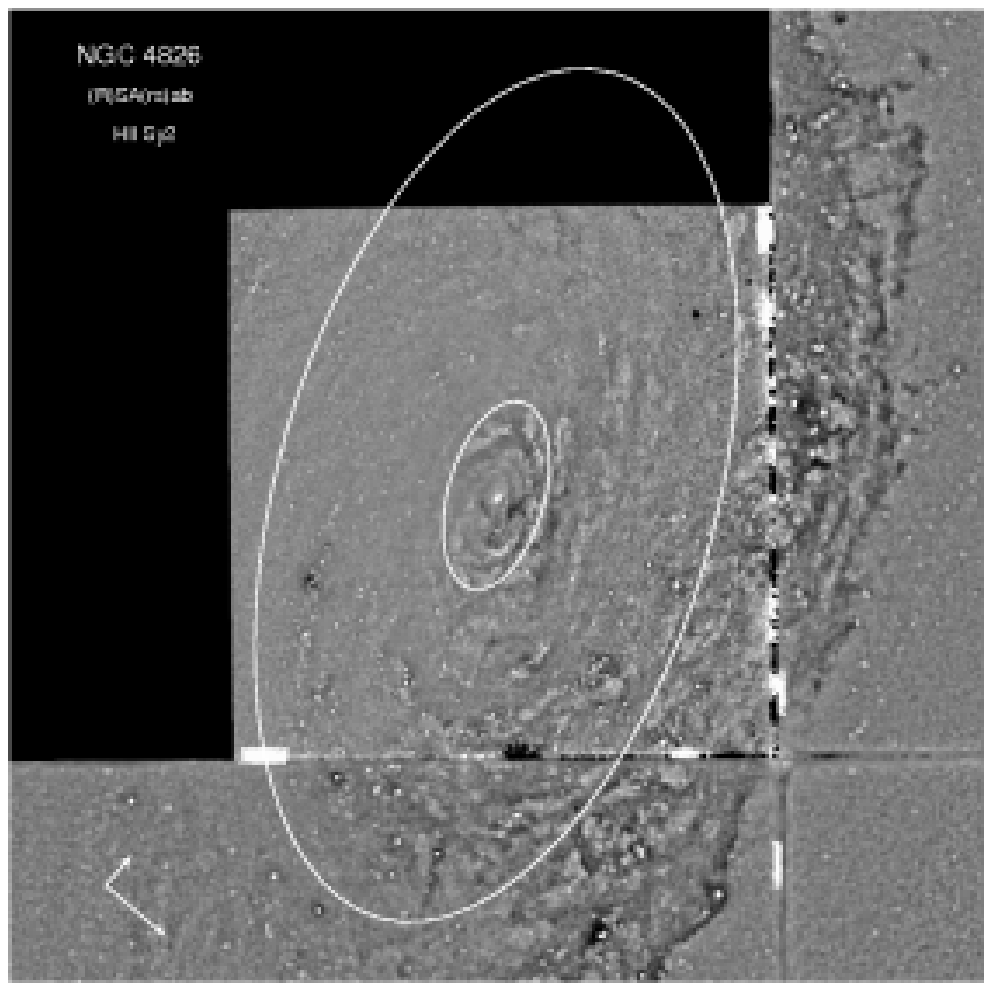}
\includegraphics[width=0.4\textwidth]{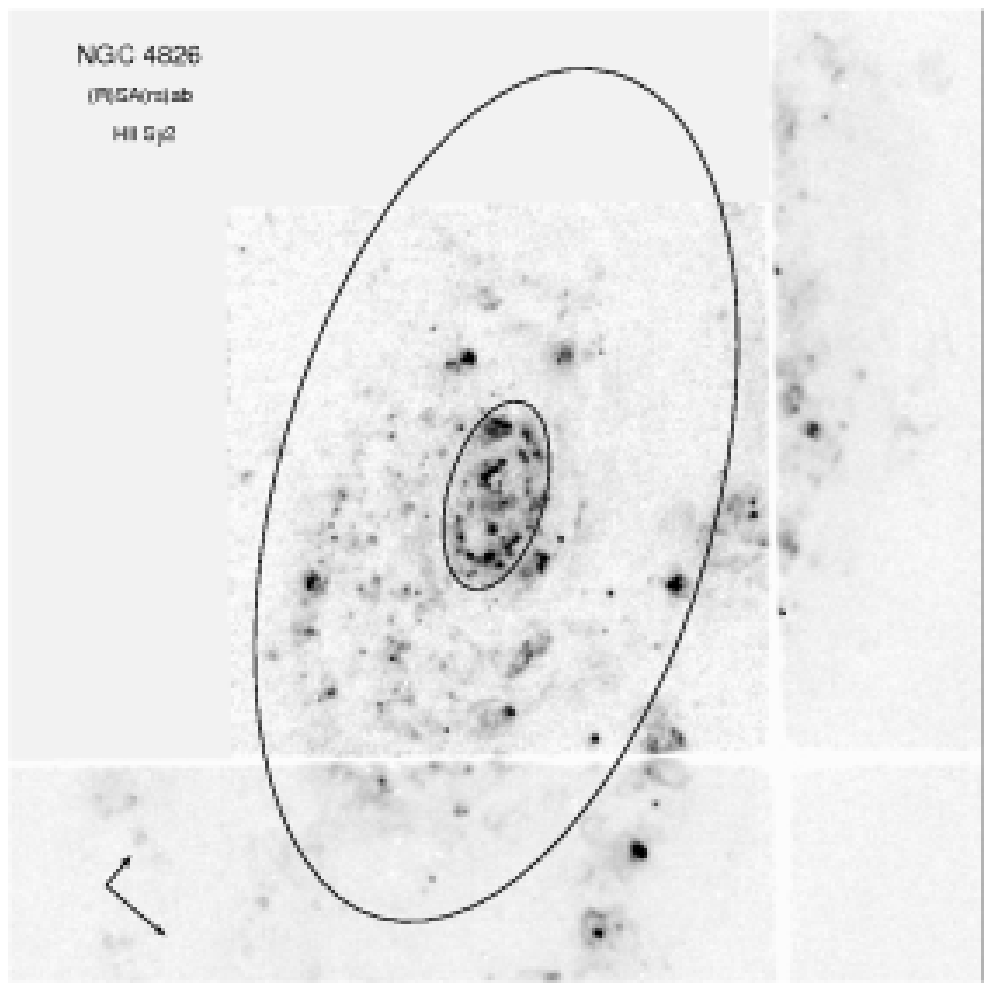}\\
\includegraphics[width=0.4\textwidth]{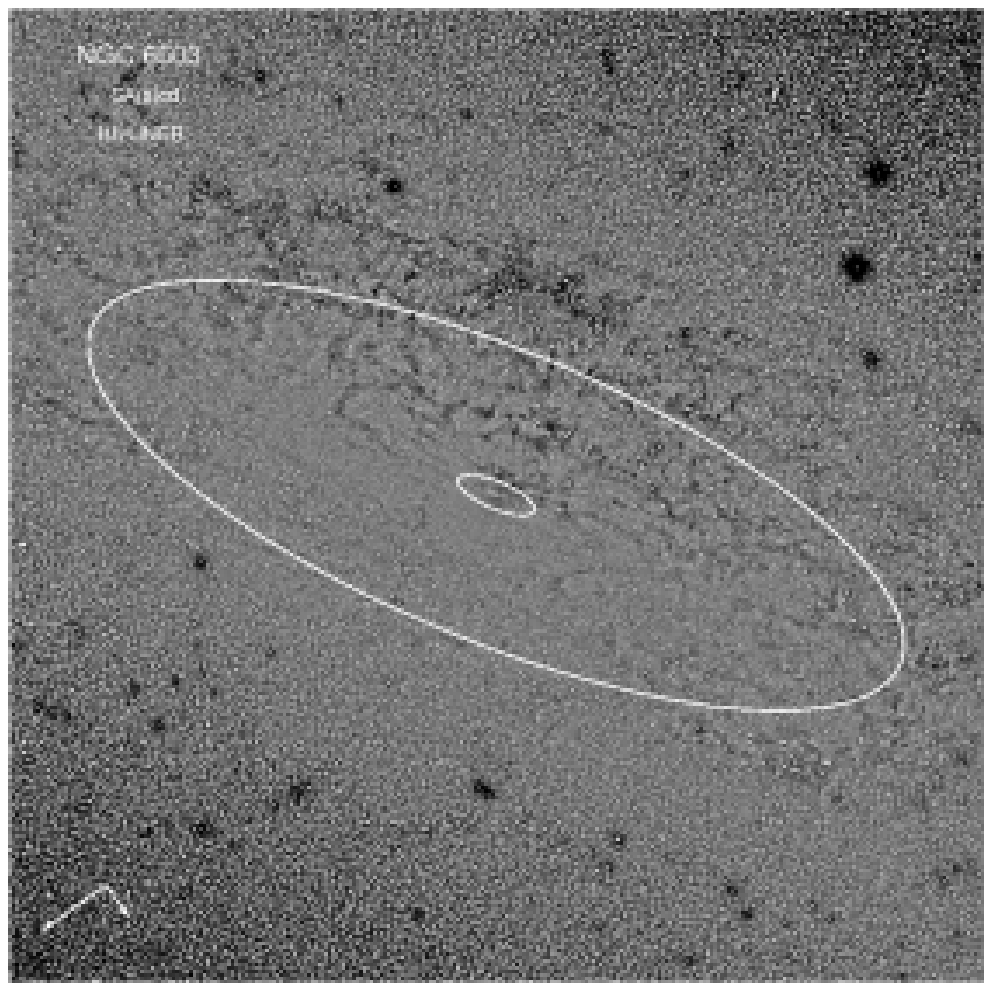}
\includegraphics[width=0.4\textwidth]{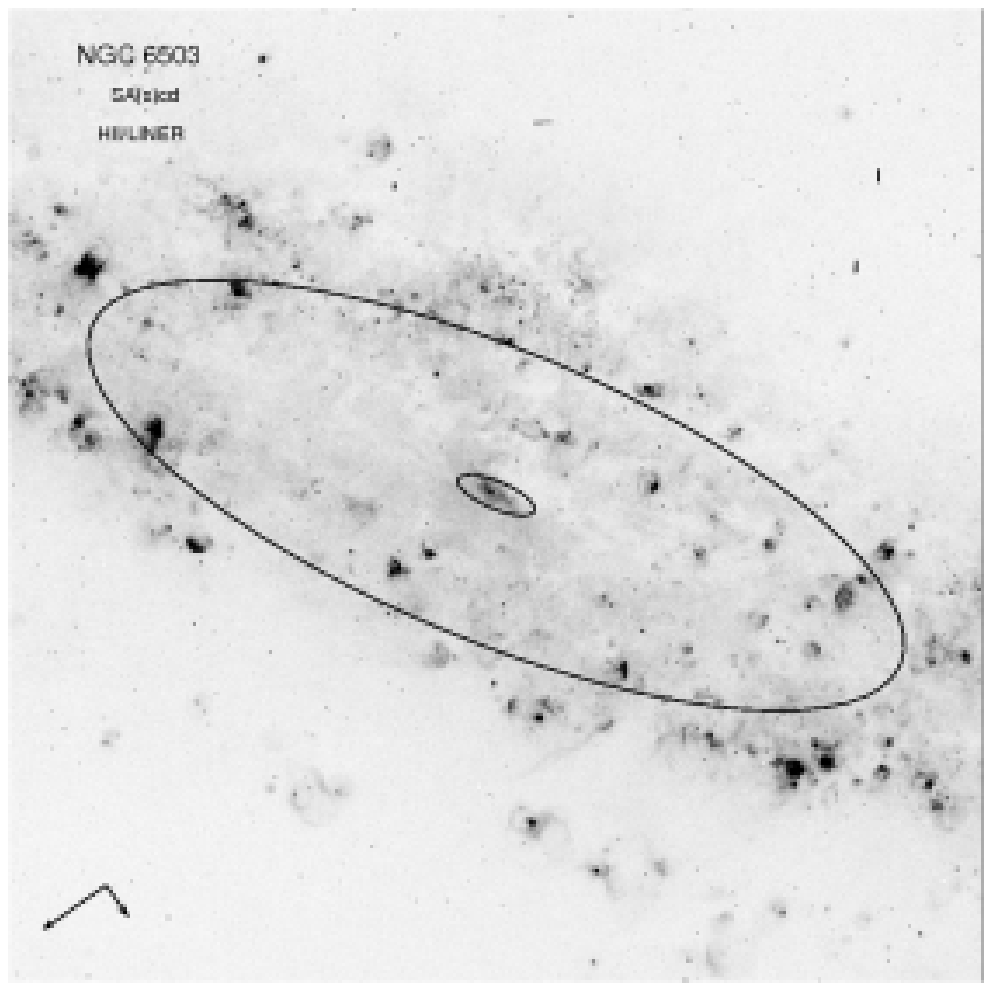}\\
\end{tabular}
\end{center}
\end{figure}

\begin{figure}
\begin{center}
\begin{tabular}{c}
\includegraphics[width=0.4\textwidth]{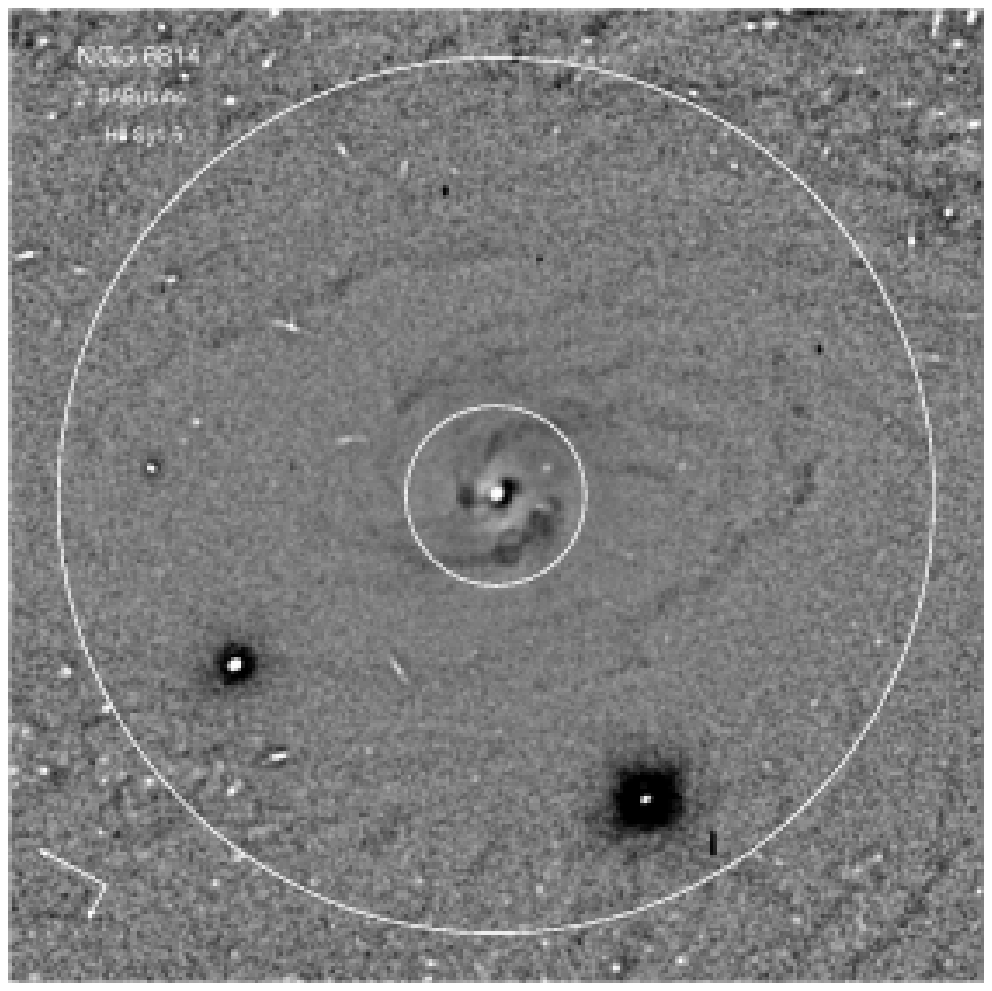}\\
\includegraphics[width=0.4\textwidth]{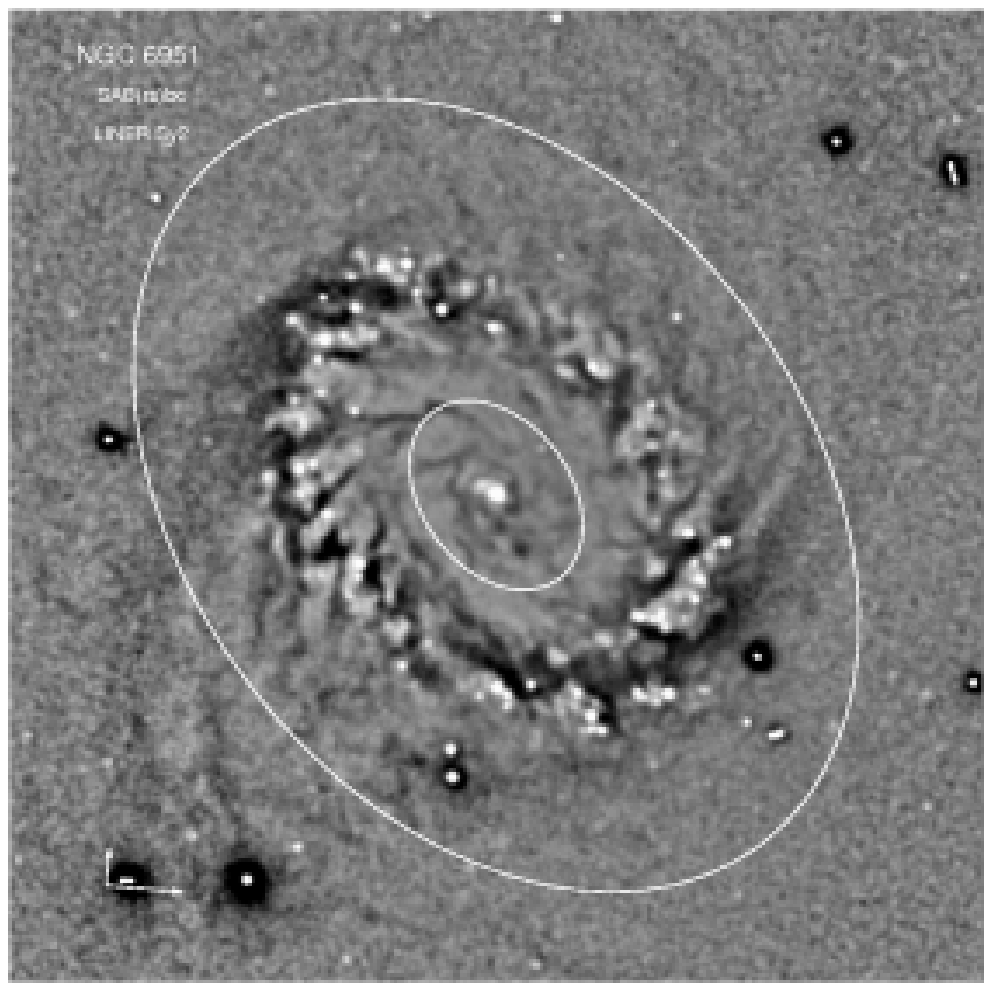}
\includegraphics[width=0.4\textwidth]{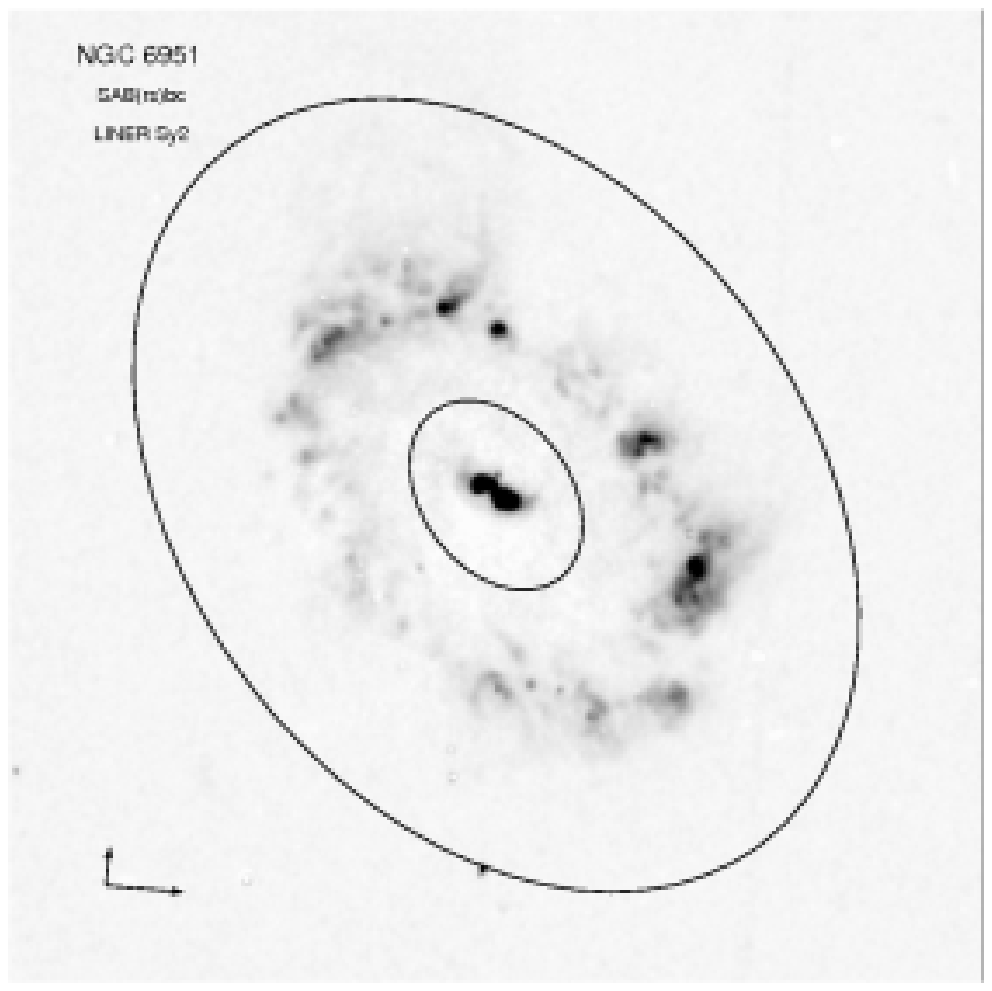}\\
\end{tabular}
\end{center}
\end{figure}

\clearpage

\section{As Appendix~A, now for the control sample galaxies.}

\begin{figure}[h]
\begin{center}
\begin{tabular}{c}
\includegraphics[width=0.4\textwidth]{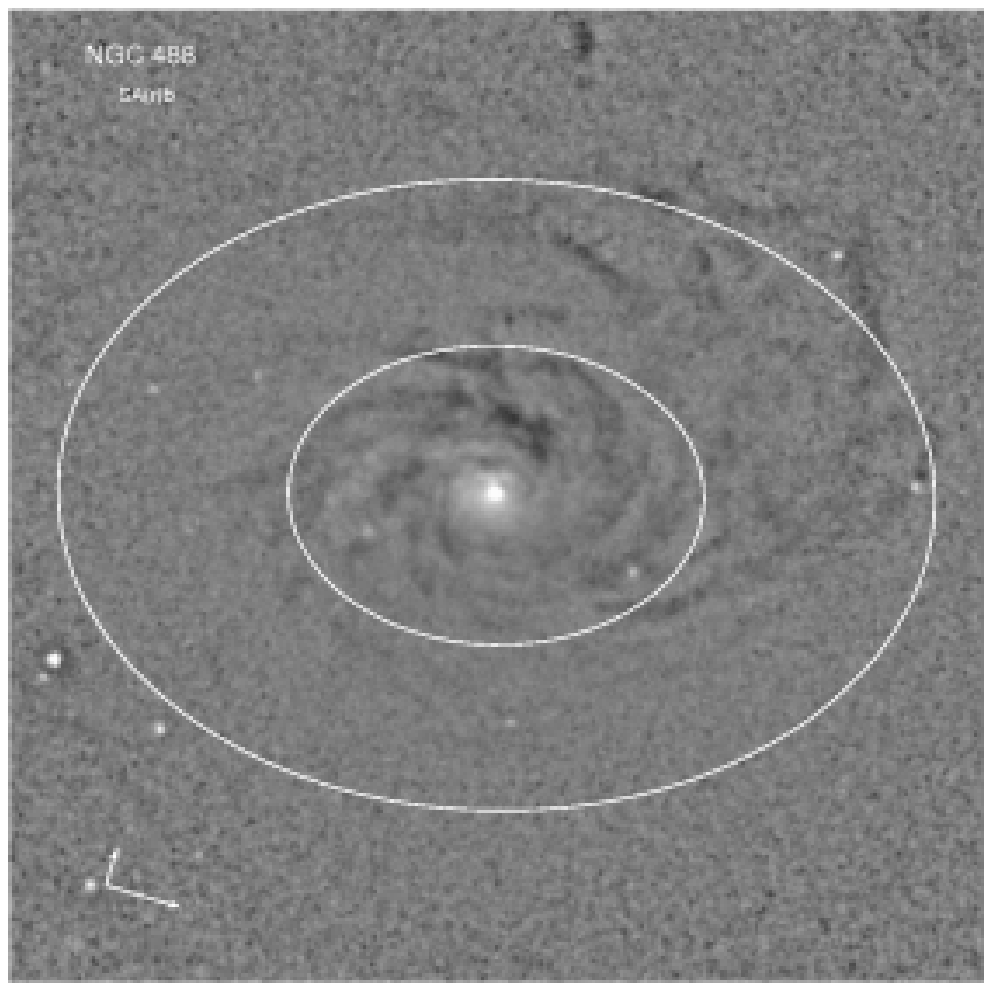}\\
\includegraphics[width=0.4\textwidth]{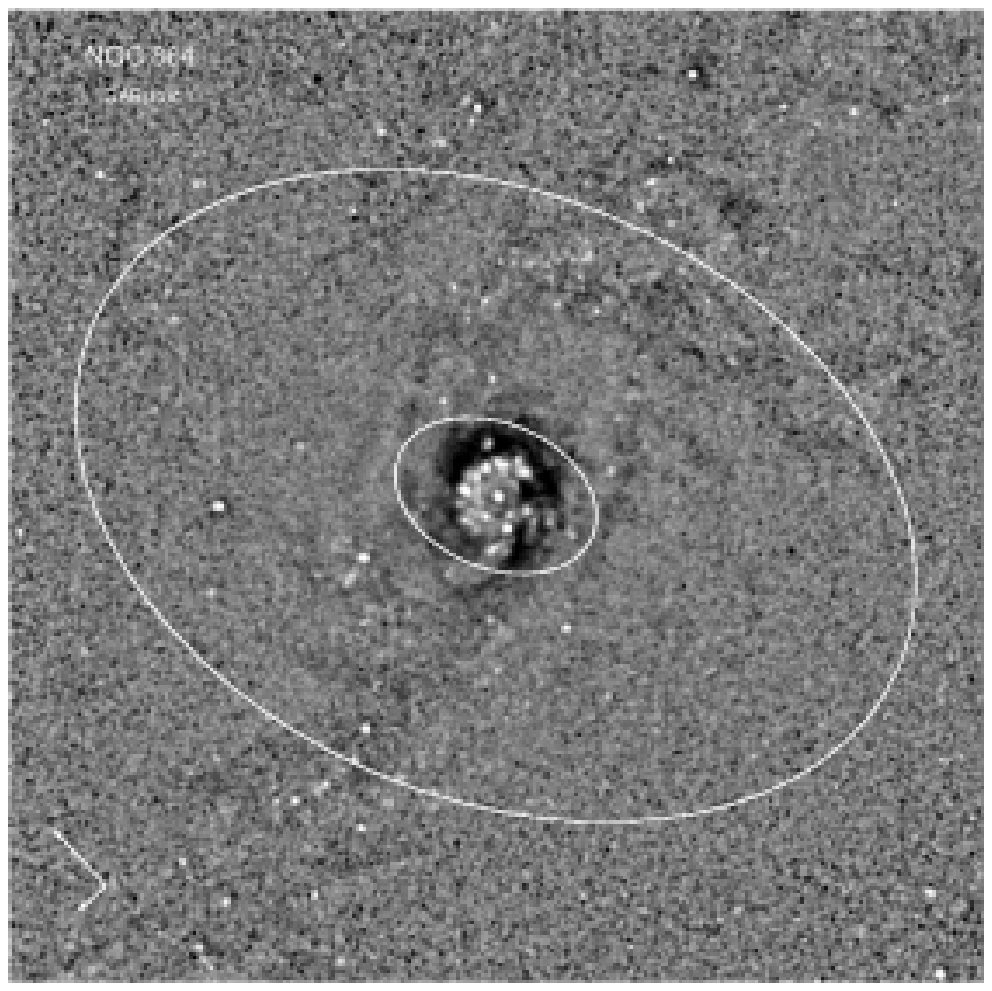}\\
\includegraphics[width=0.4\textwidth]{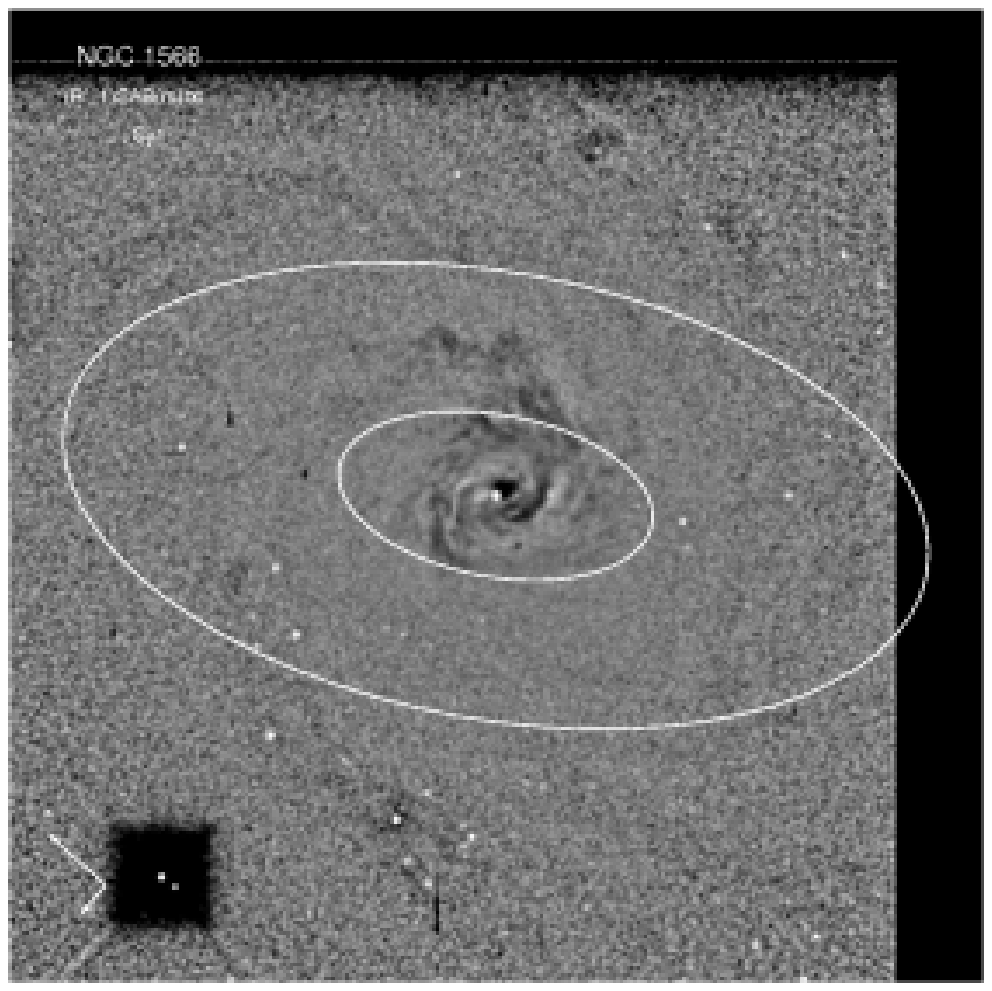}\\
\end{tabular}
\end{center}
\end{figure}

\begin{figure}[h]
\begin{center}
\begin{tabular}{c}
\includegraphics[width=0.4\textwidth]{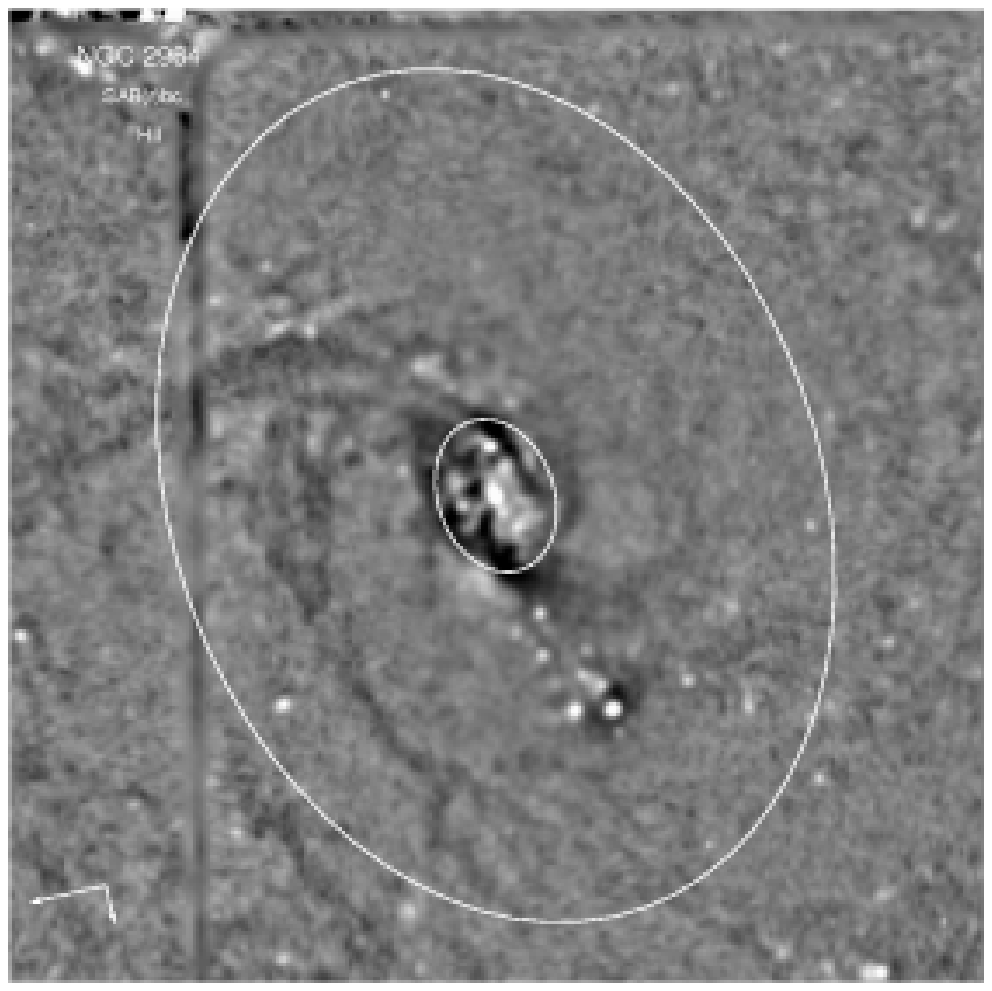}\\
\includegraphics[width=0.4\textwidth]{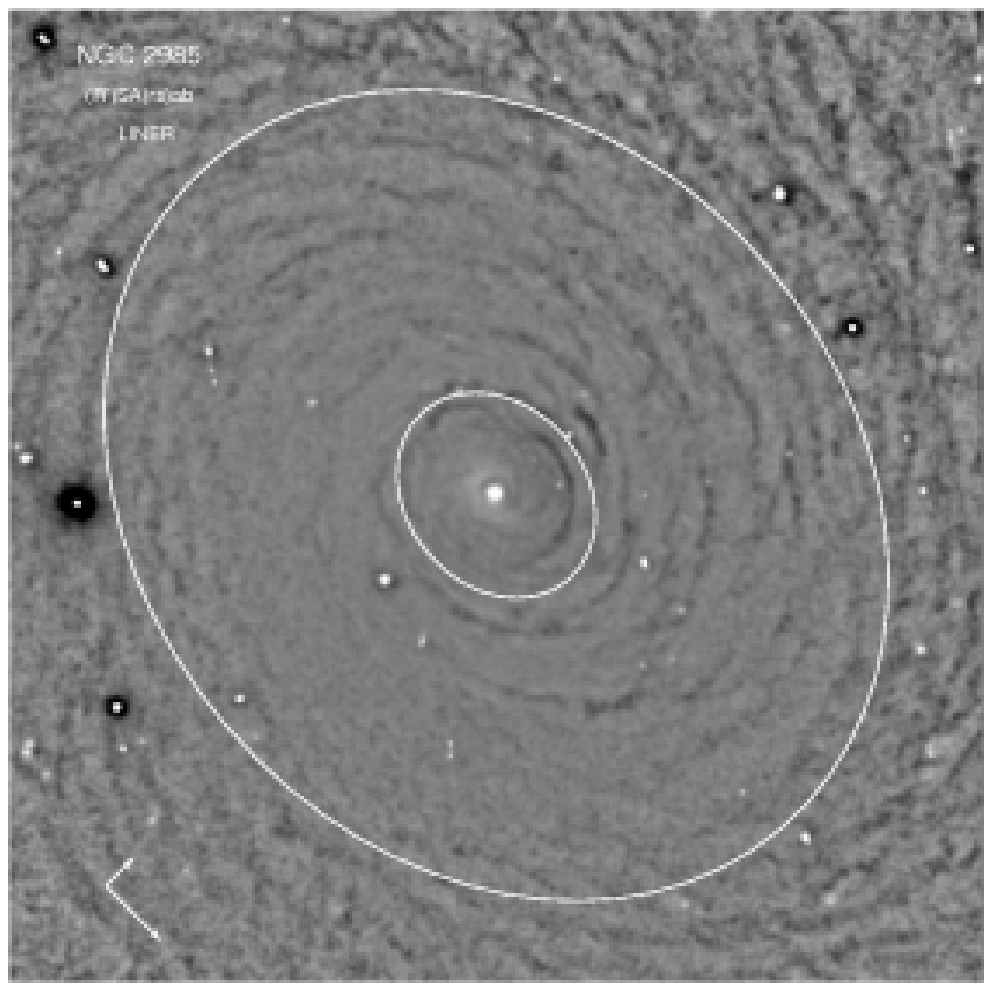}
\includegraphics[width=0.4\textwidth]{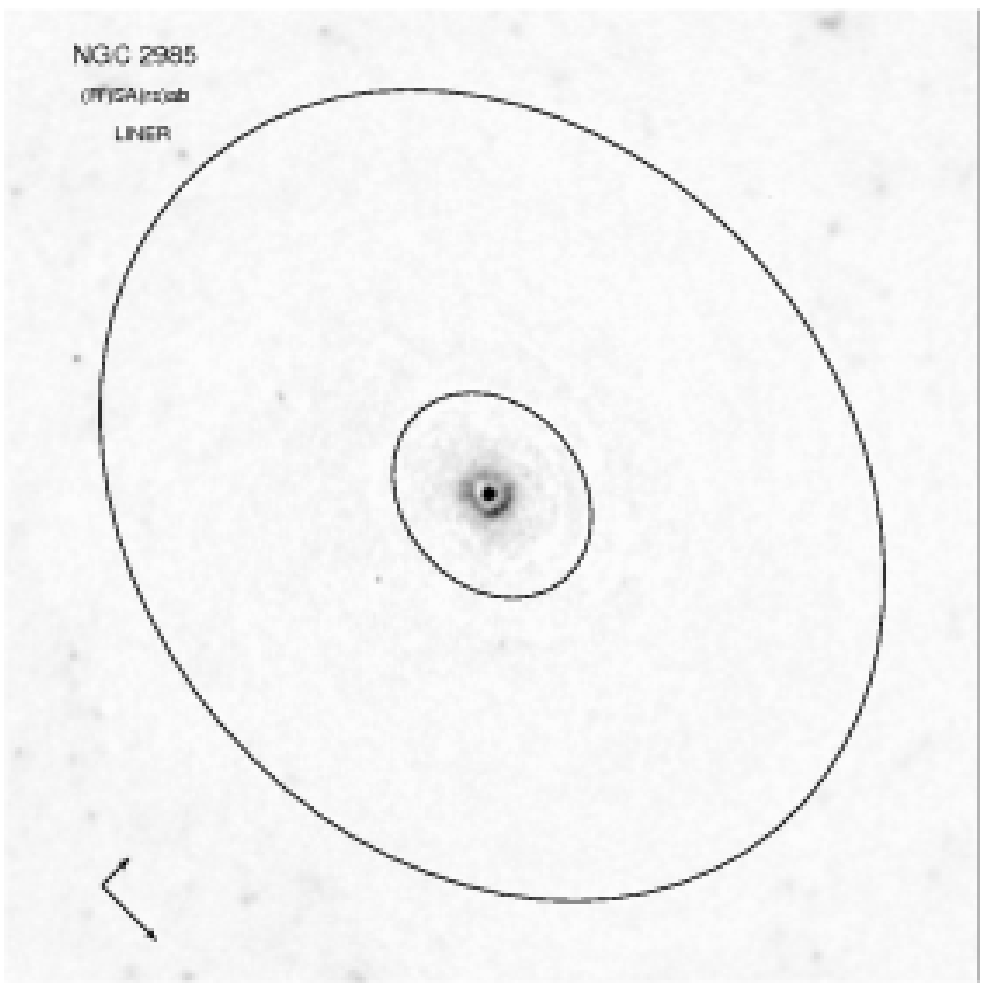}\\
\includegraphics[width=0.4\textwidth]{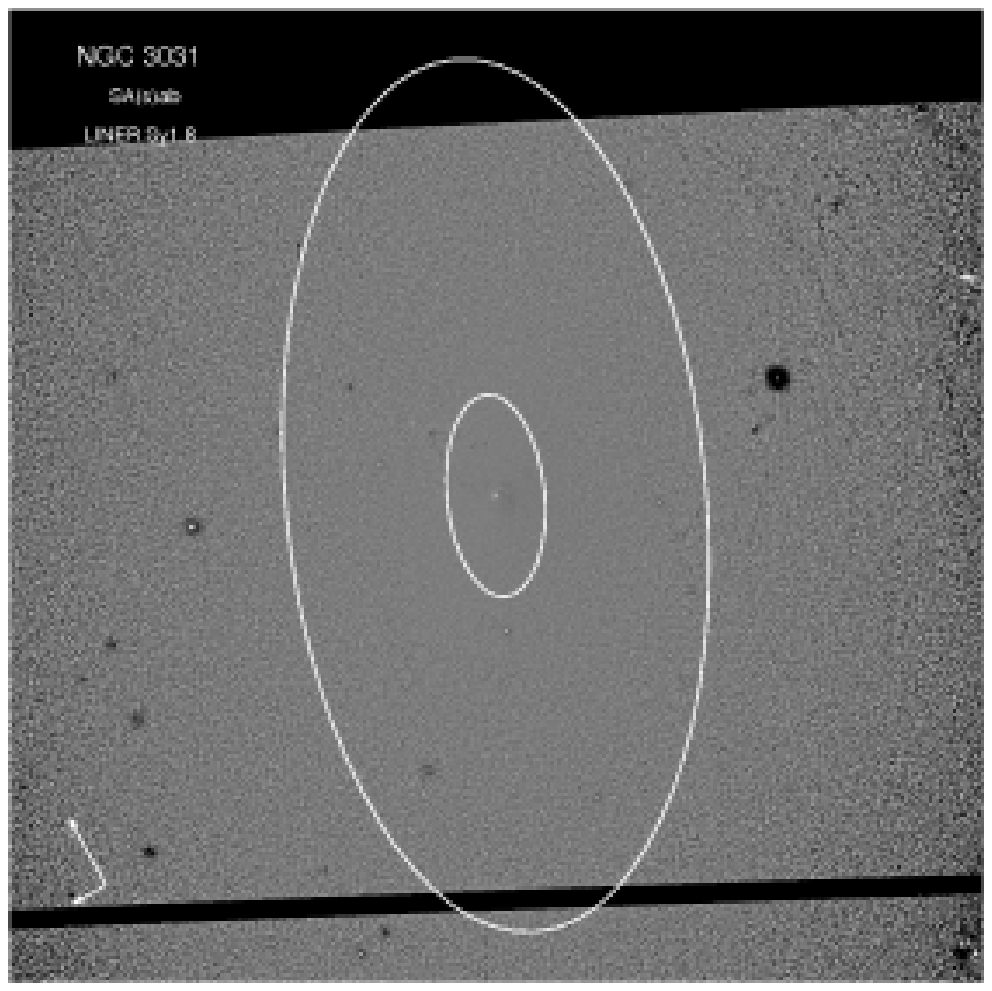}
\includegraphics[width=0.4\textwidth]{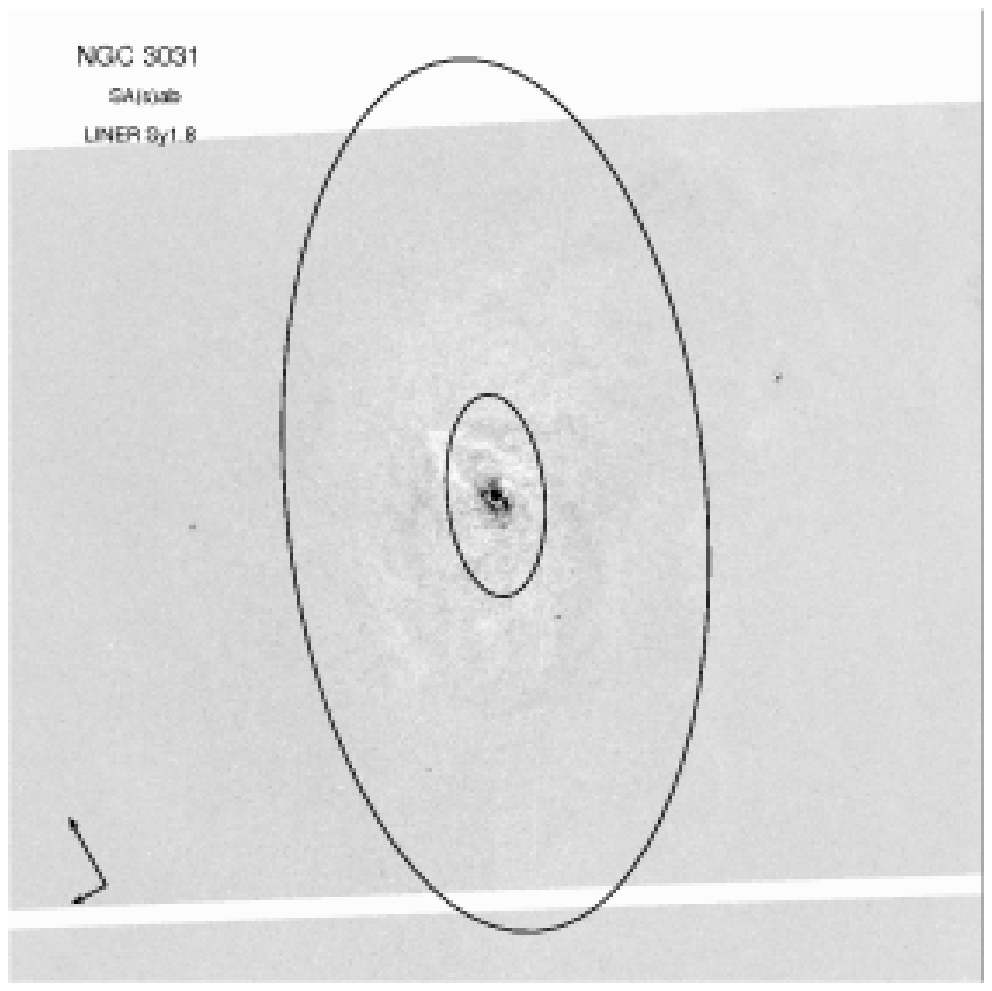}\\
\end{tabular}
\end{center}
\end{figure}

\begin{figure}[h]
\begin{center}
\begin{tabular}{c}
\includegraphics[width=0.4\textwidth]{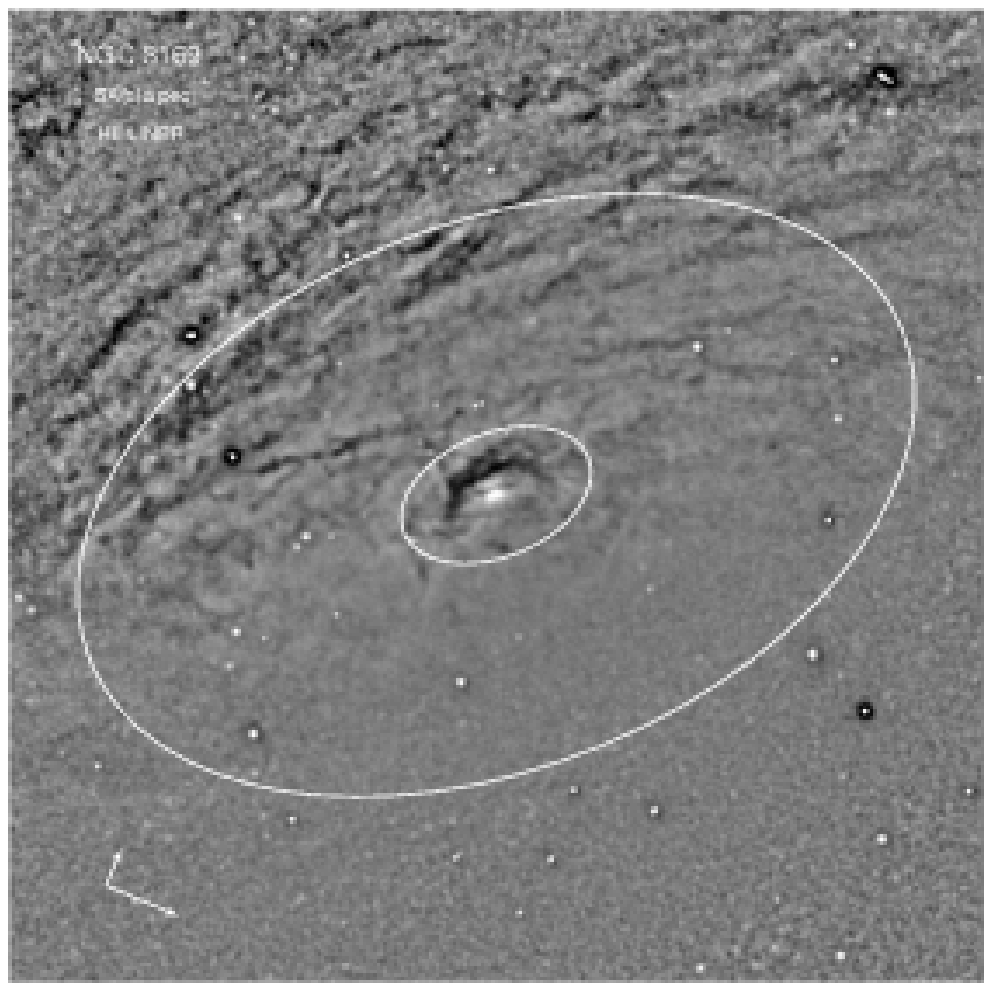}
\includegraphics[width=0.4\textwidth]{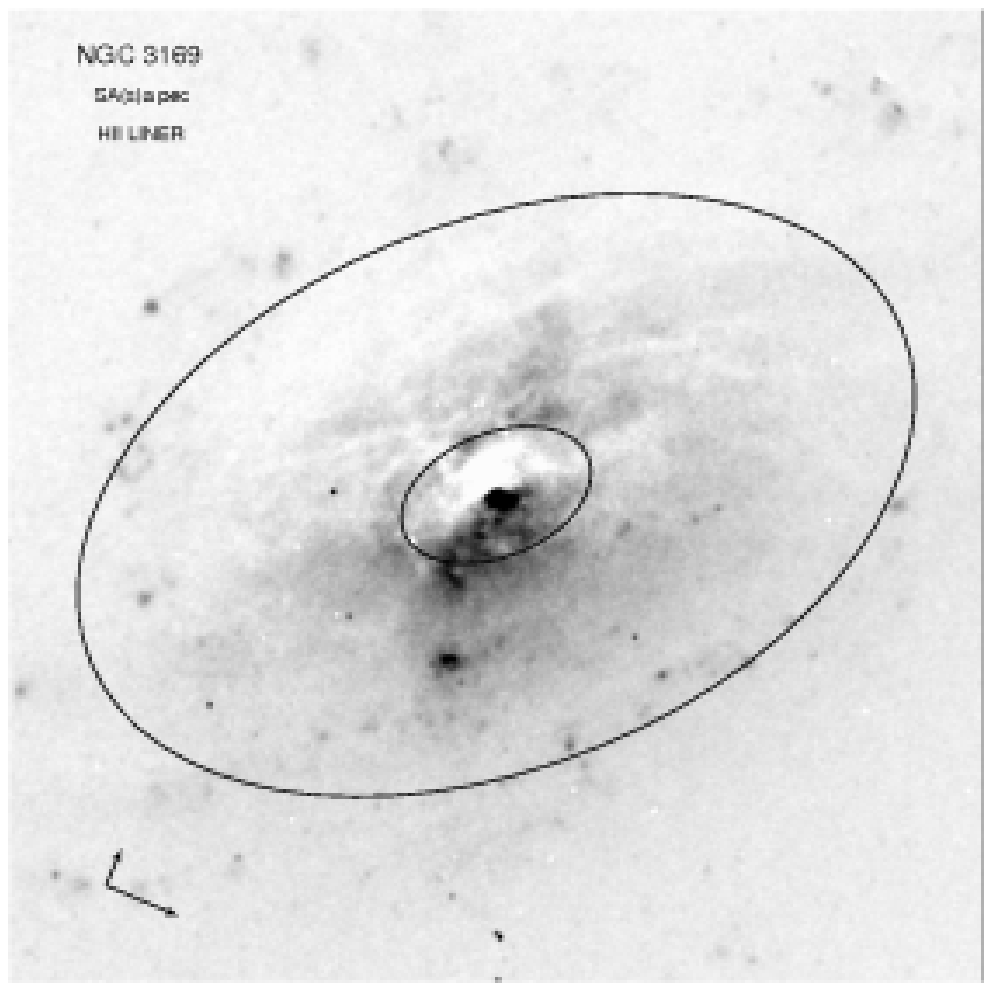}\\
\includegraphics[width=0.4\textwidth]{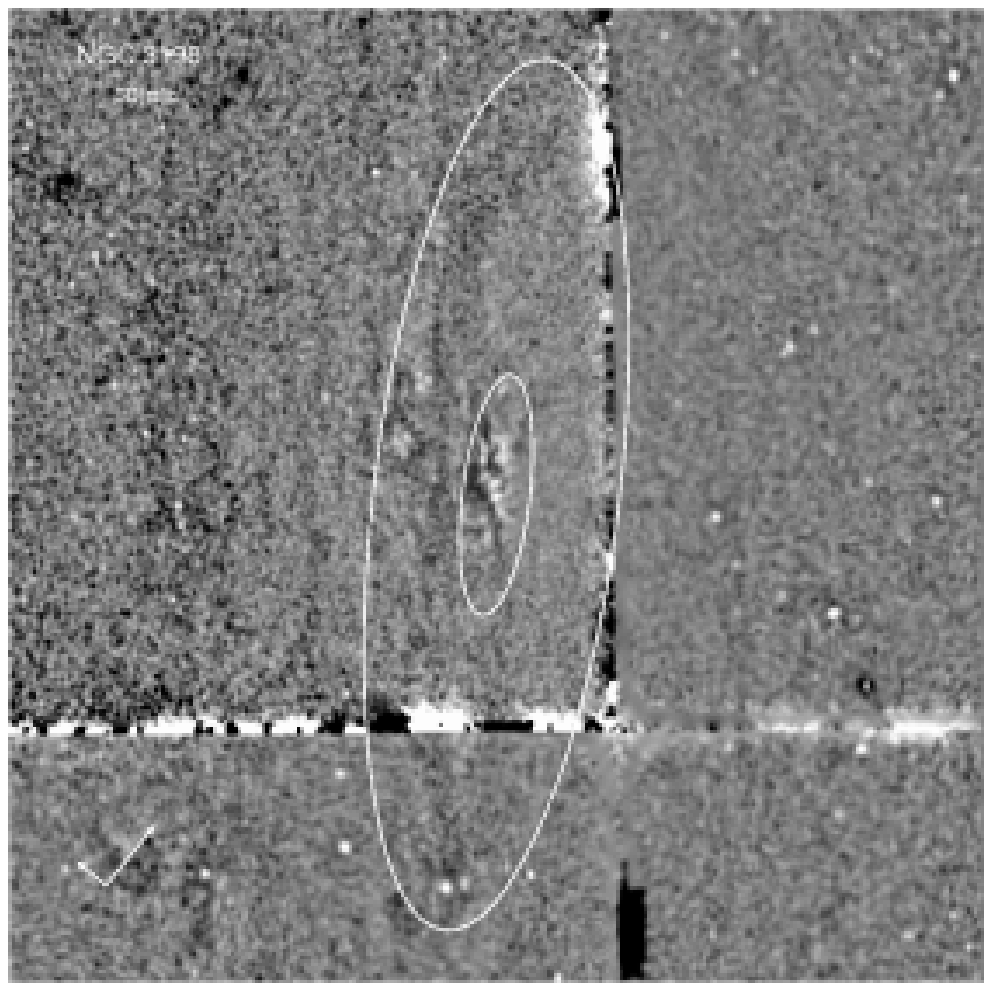}\\
\includegraphics[width=0.4\textwidth]{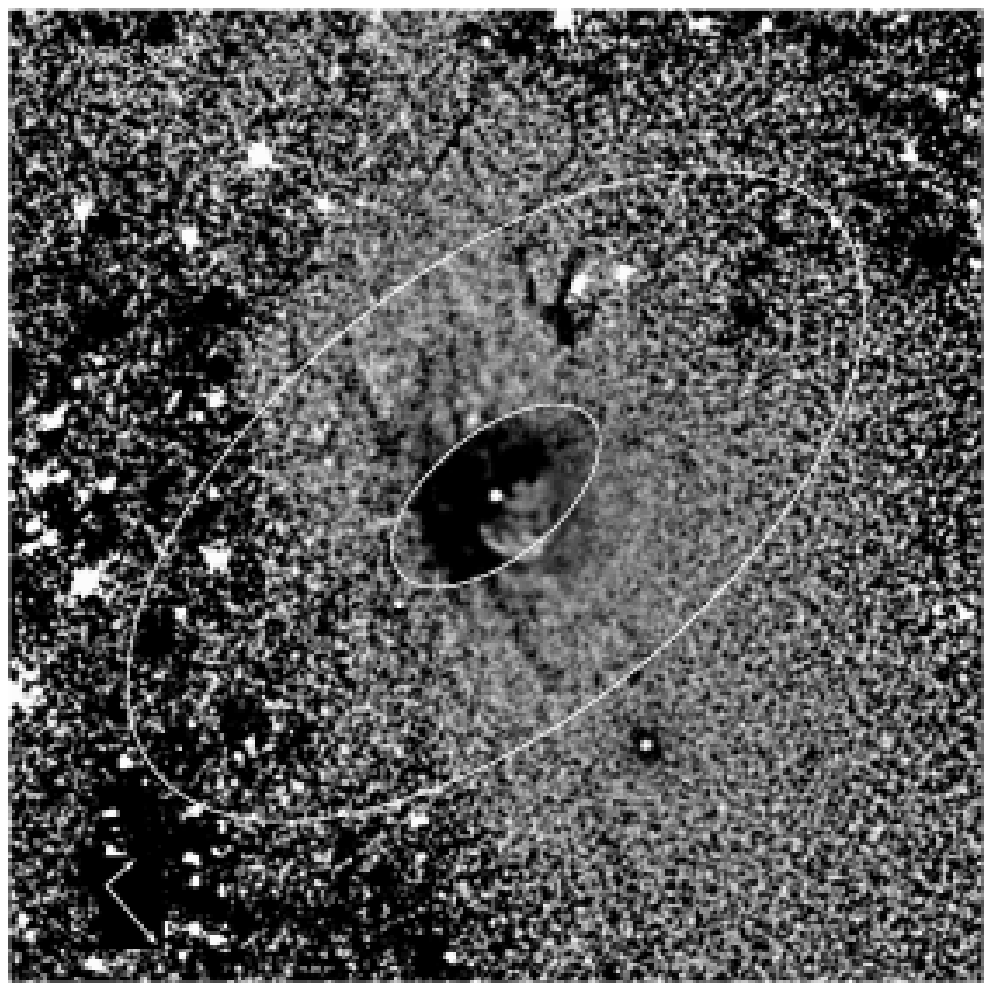}
\includegraphics[width=0.4\textwidth]{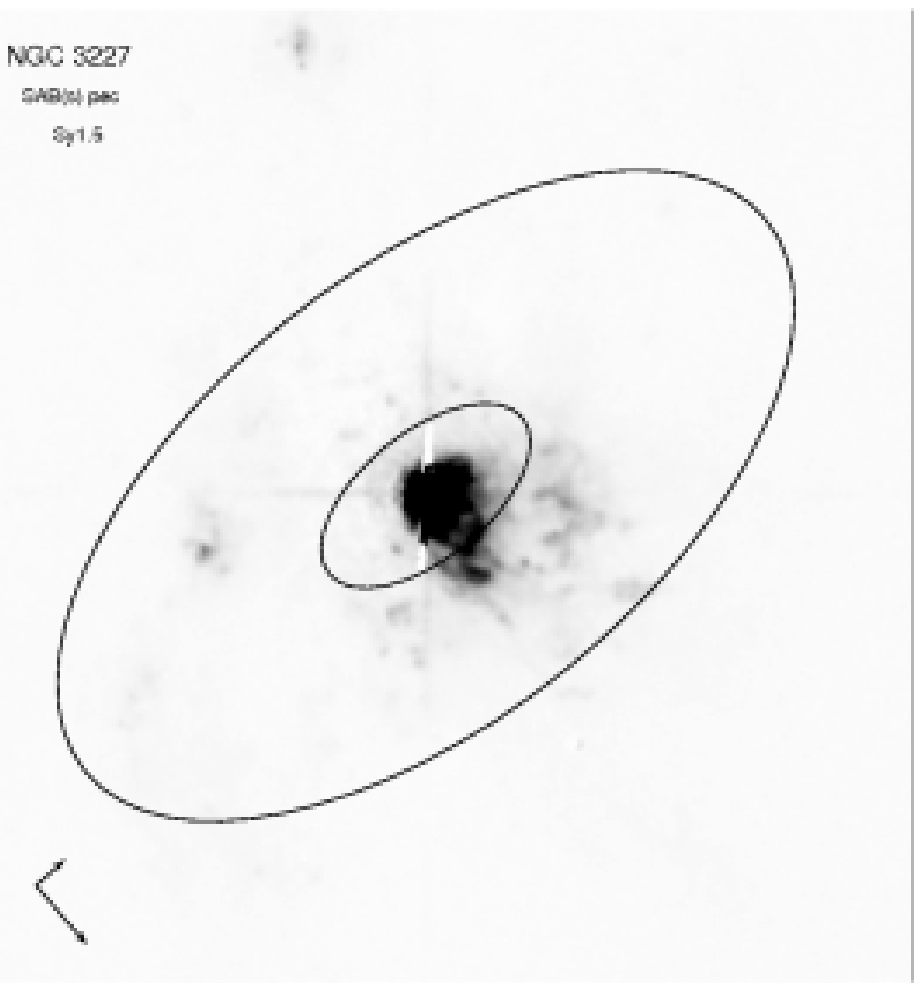}\\
\end{tabular}
\end{center}
\end{figure}

\begin{figure}[h]
\begin{center}
\begin{tabular}{c}
\includegraphics[width=0.4\textwidth]{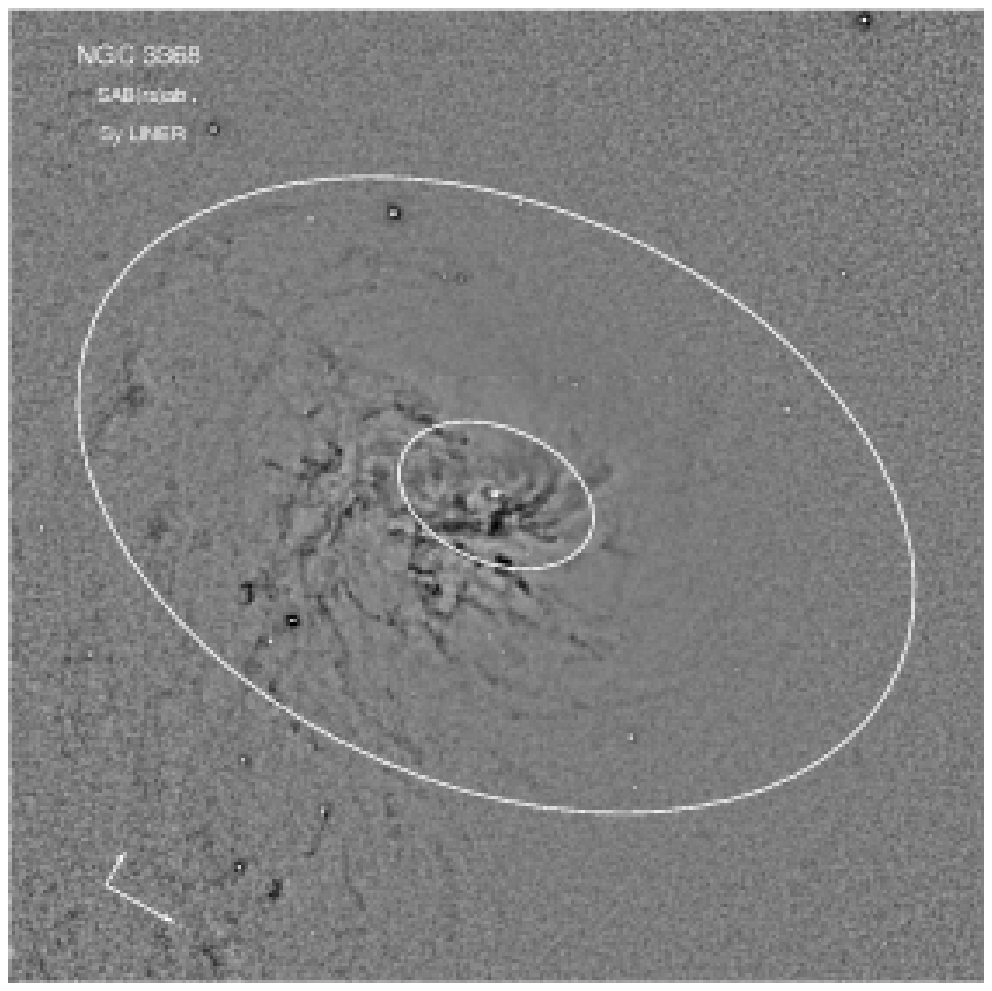}
\includegraphics[width=0.4\textwidth]{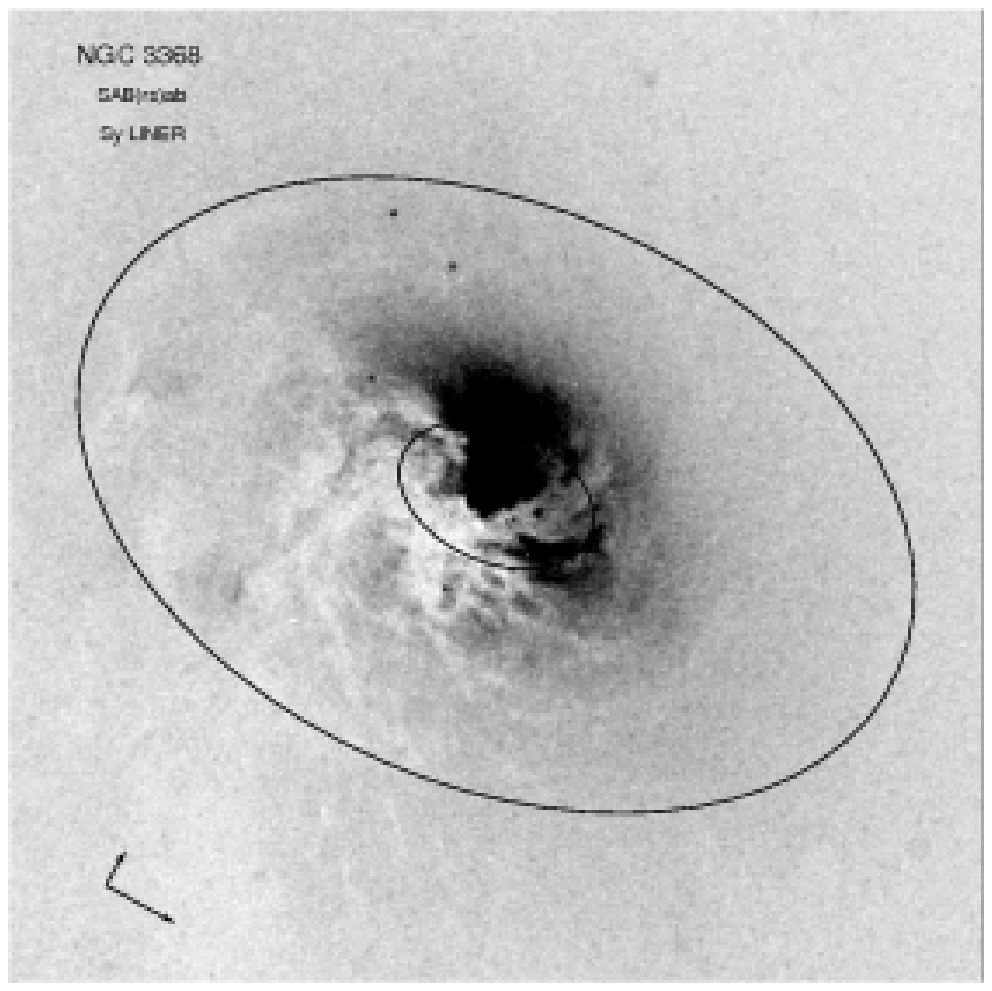}\\
\includegraphics[width=0.4\textwidth]{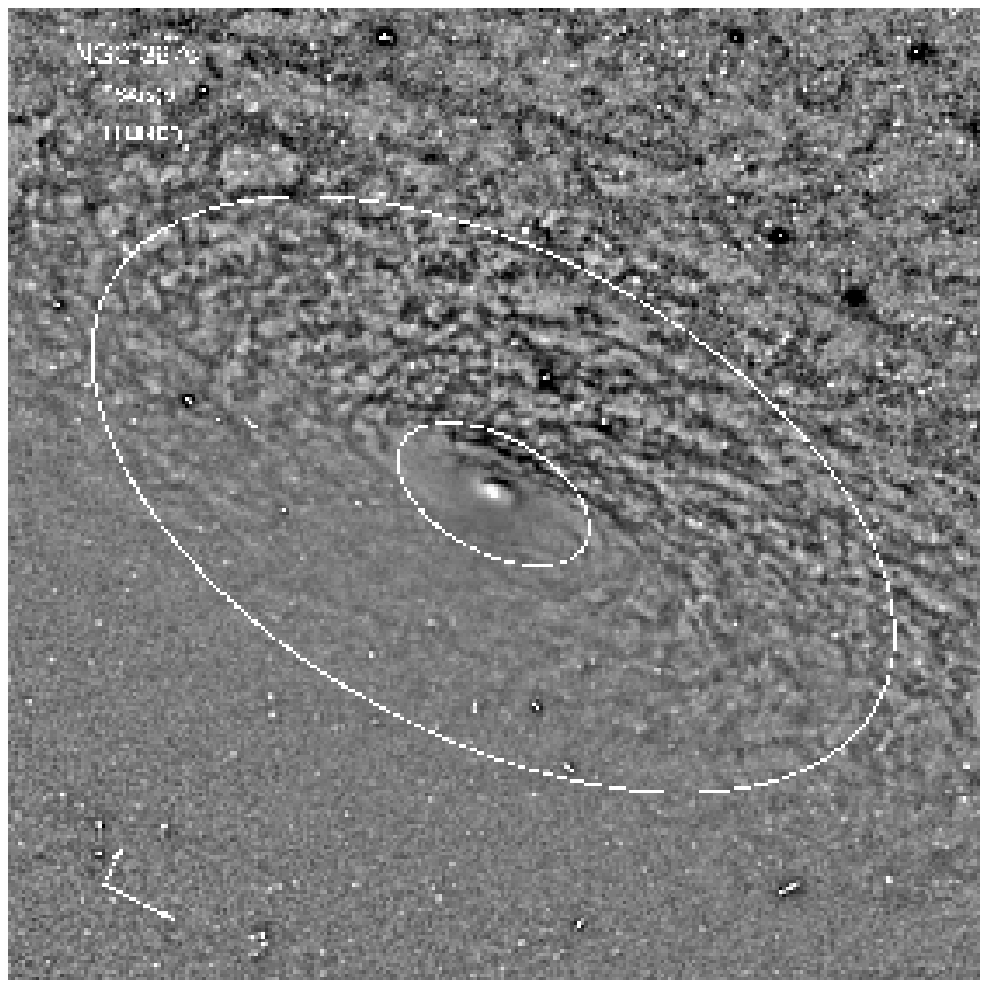}
\includegraphics[width=0.4\textwidth]{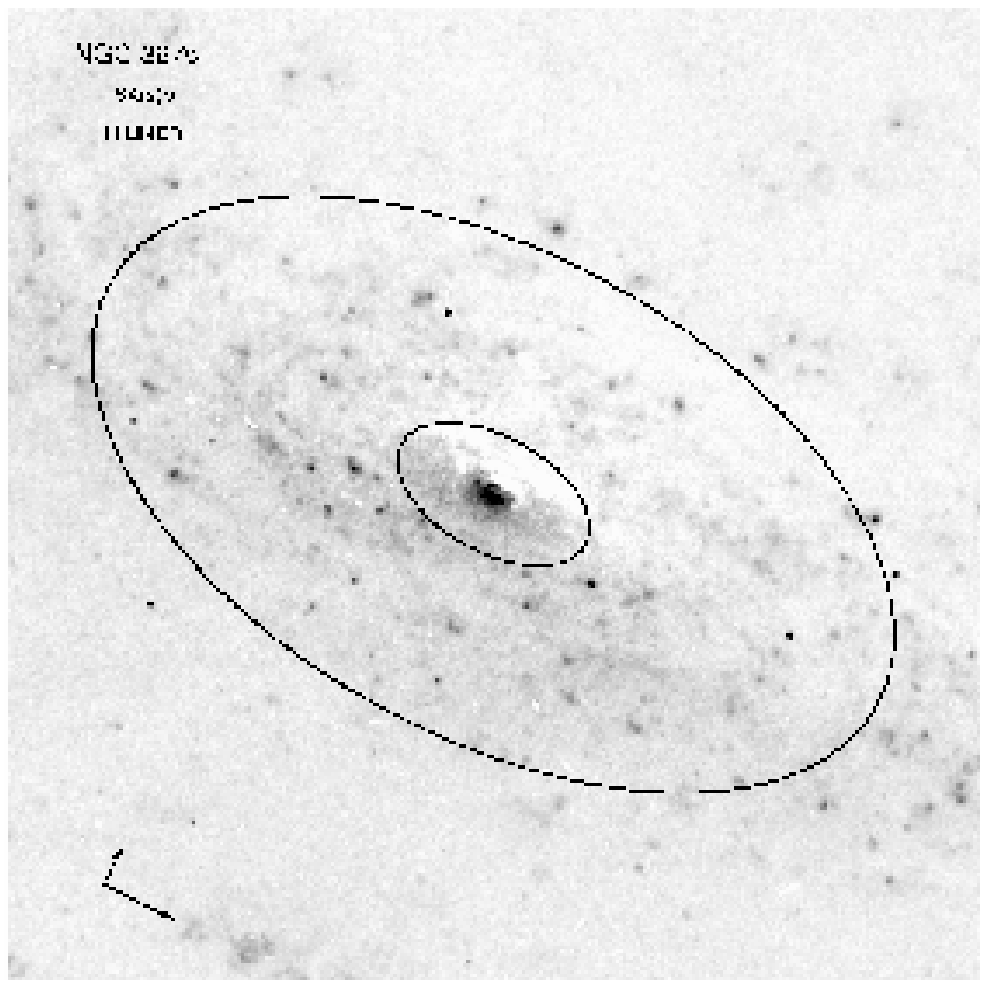}\\
\includegraphics[width=0.4\textwidth]{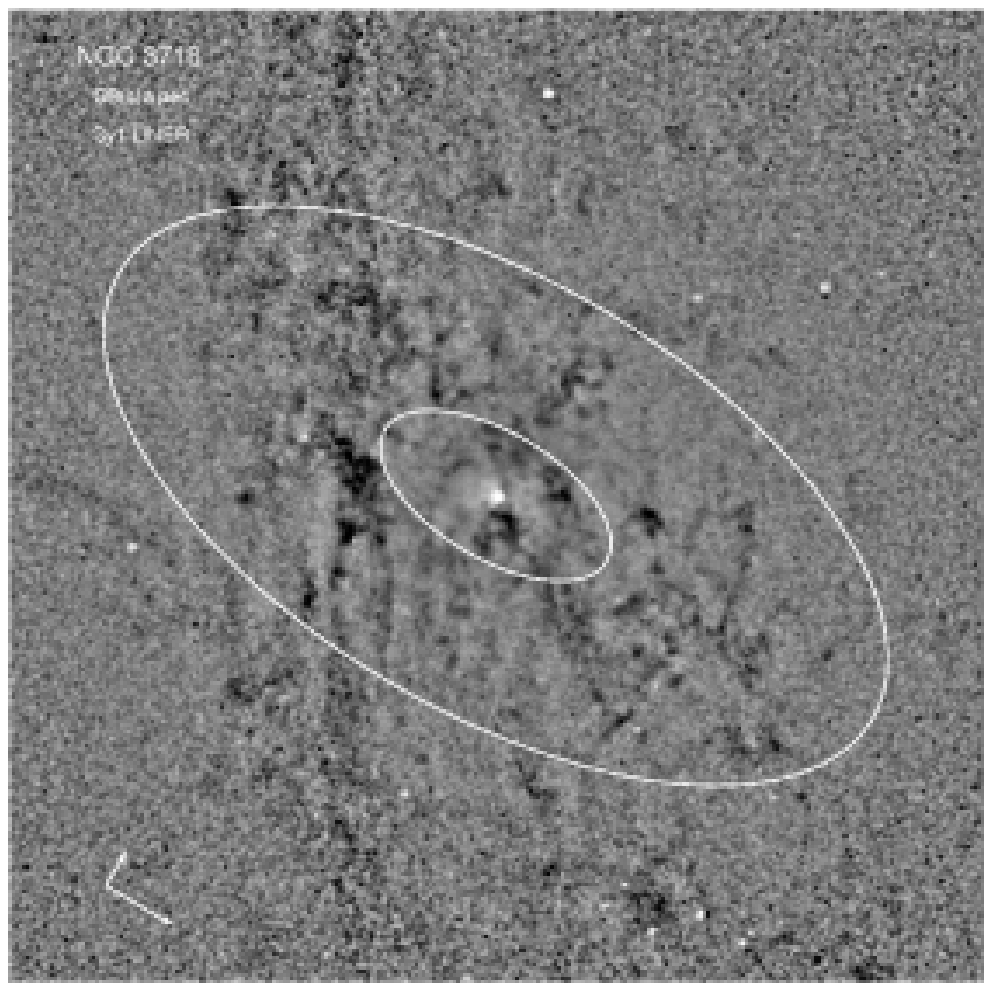}
\includegraphics[width=0.4\textwidth]{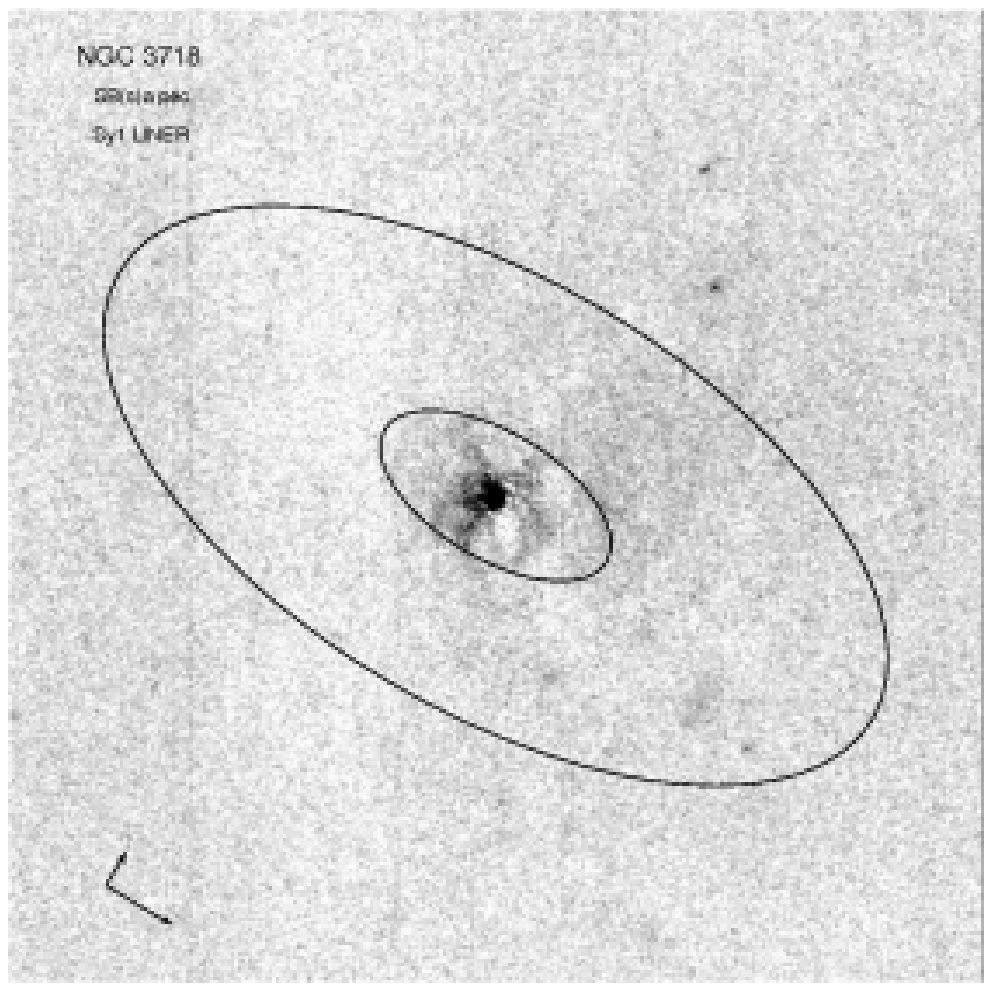}\\
\end{tabular}
\end{center}
\end{figure}

\begin{figure}[h]
\begin{center}
\begin{tabular}{c}
\includegraphics[width=0.4\textwidth]{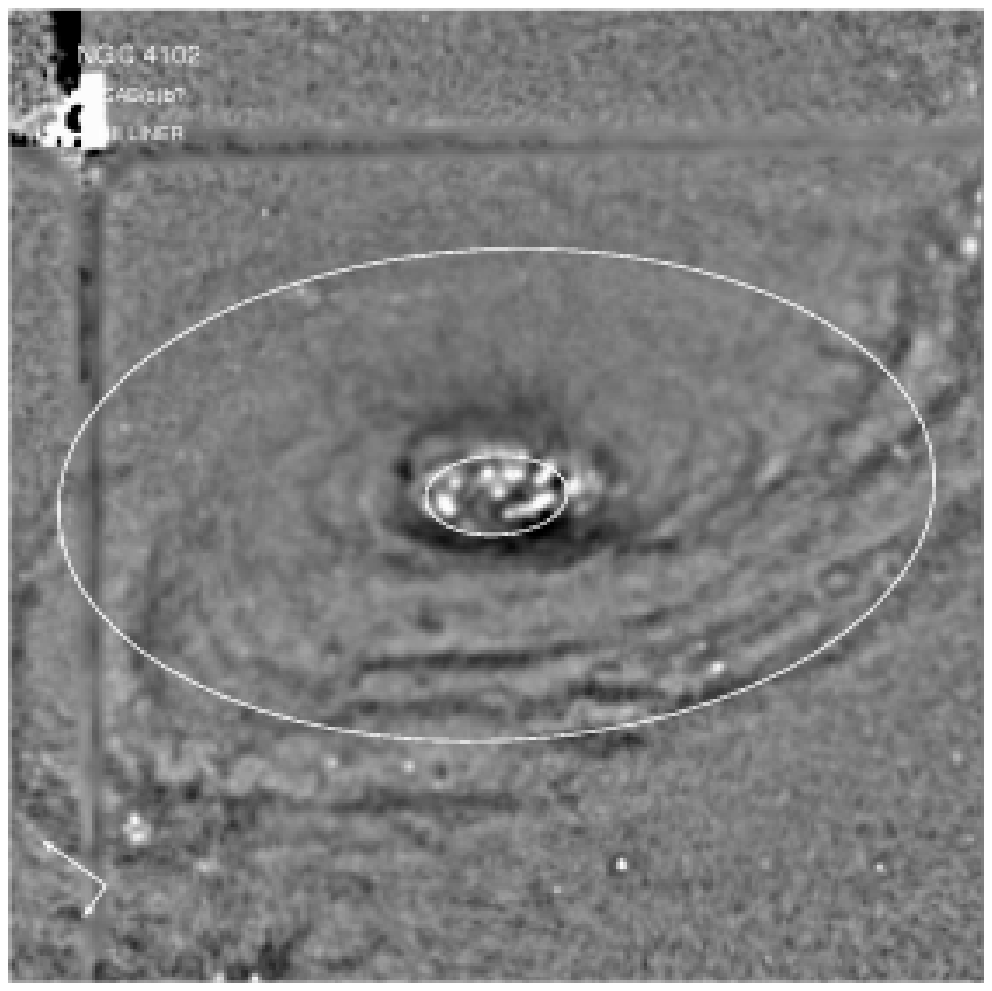}\\
\includegraphics[width=0.4\textwidth]{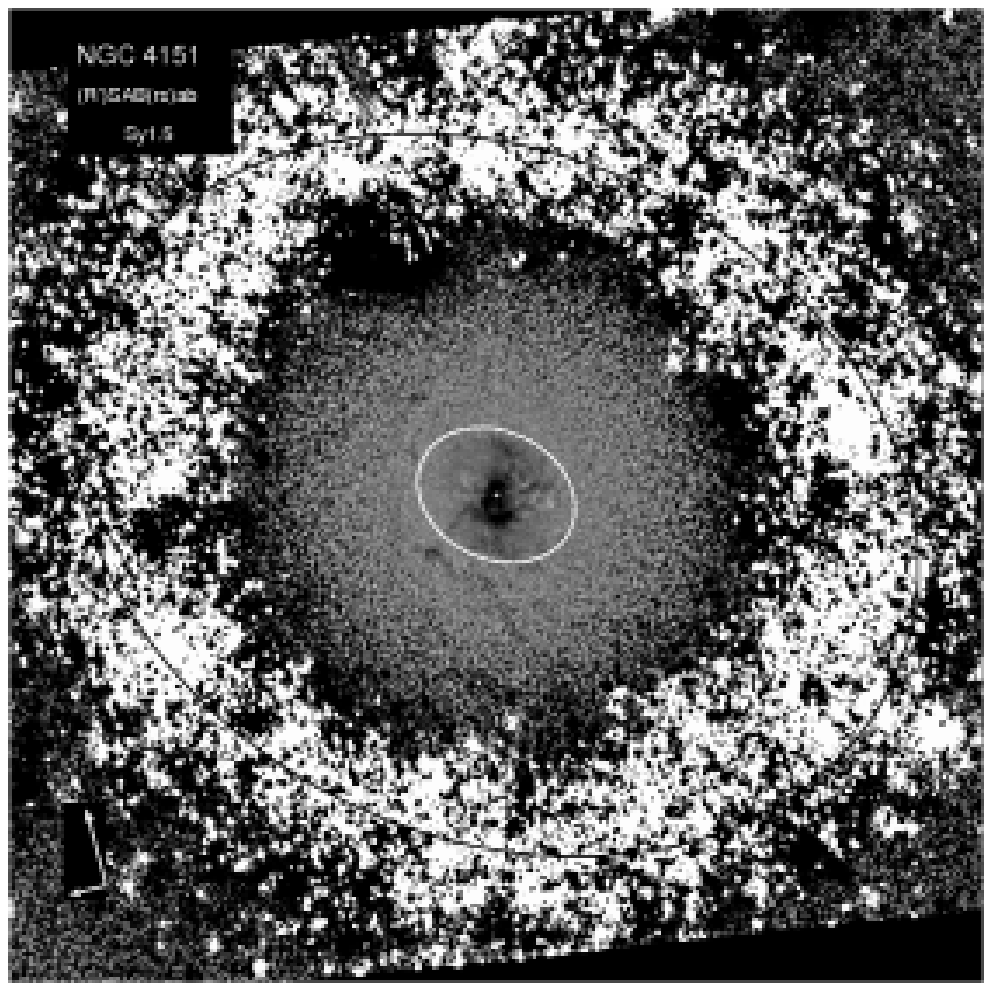}
\includegraphics[width=0.4\textwidth]{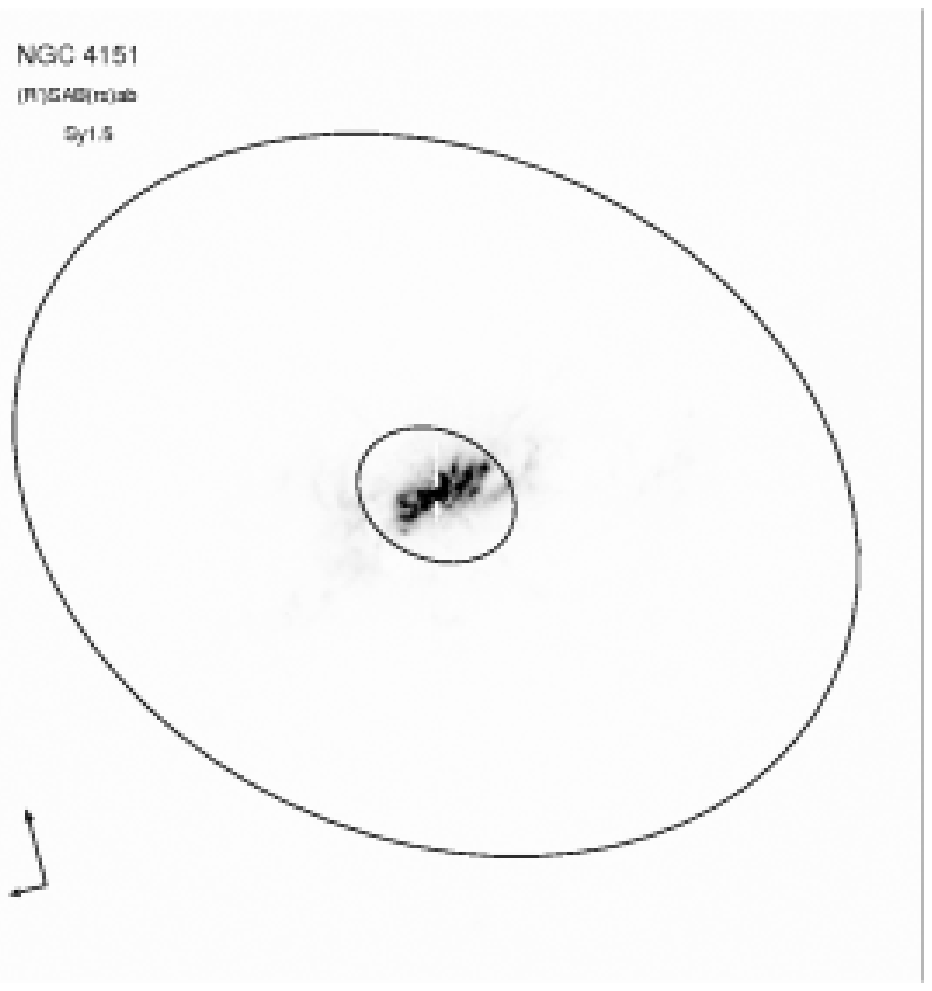}\\
\includegraphics[width=0.4\textwidth]{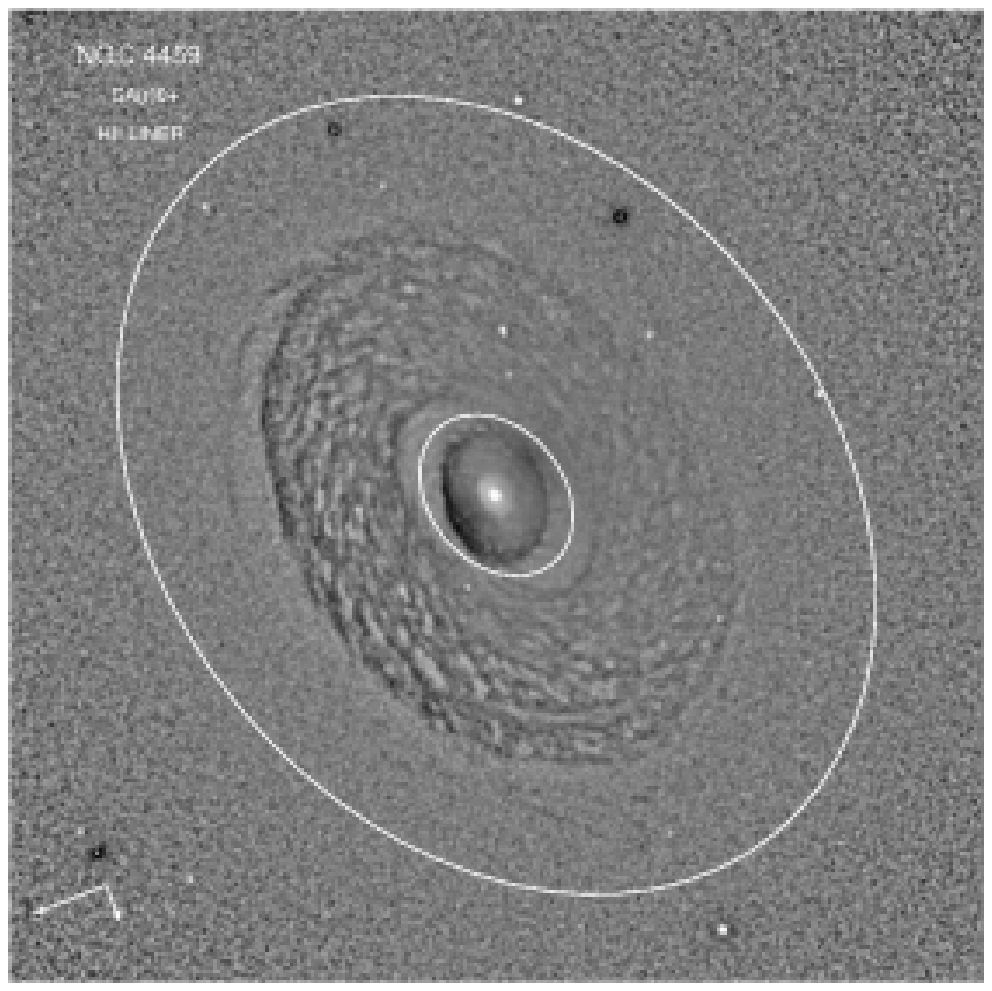}\\
\end{tabular}
\end{center}
\end{figure}

\begin{figure}[h]
\begin{center}
\begin{tabular}{c}
\includegraphics[width=0.4\textwidth]{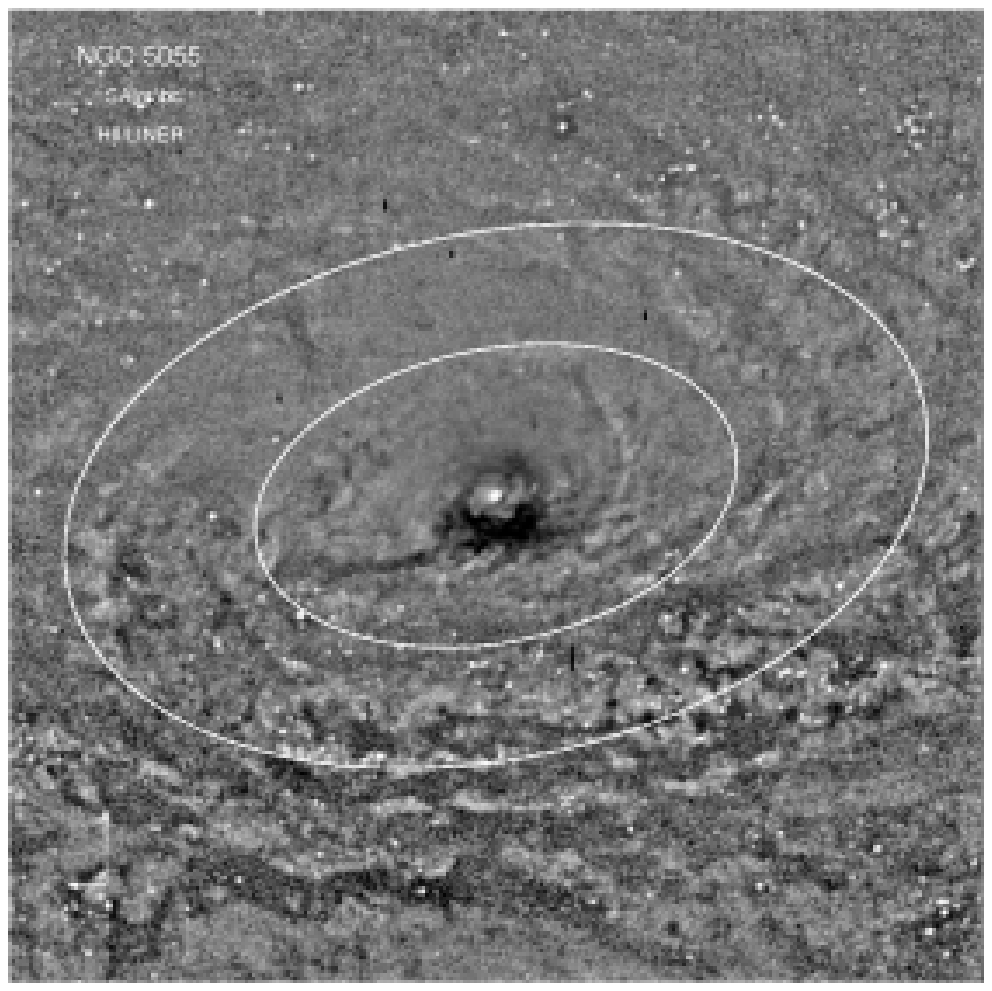}
\includegraphics[width=0.4\textwidth]{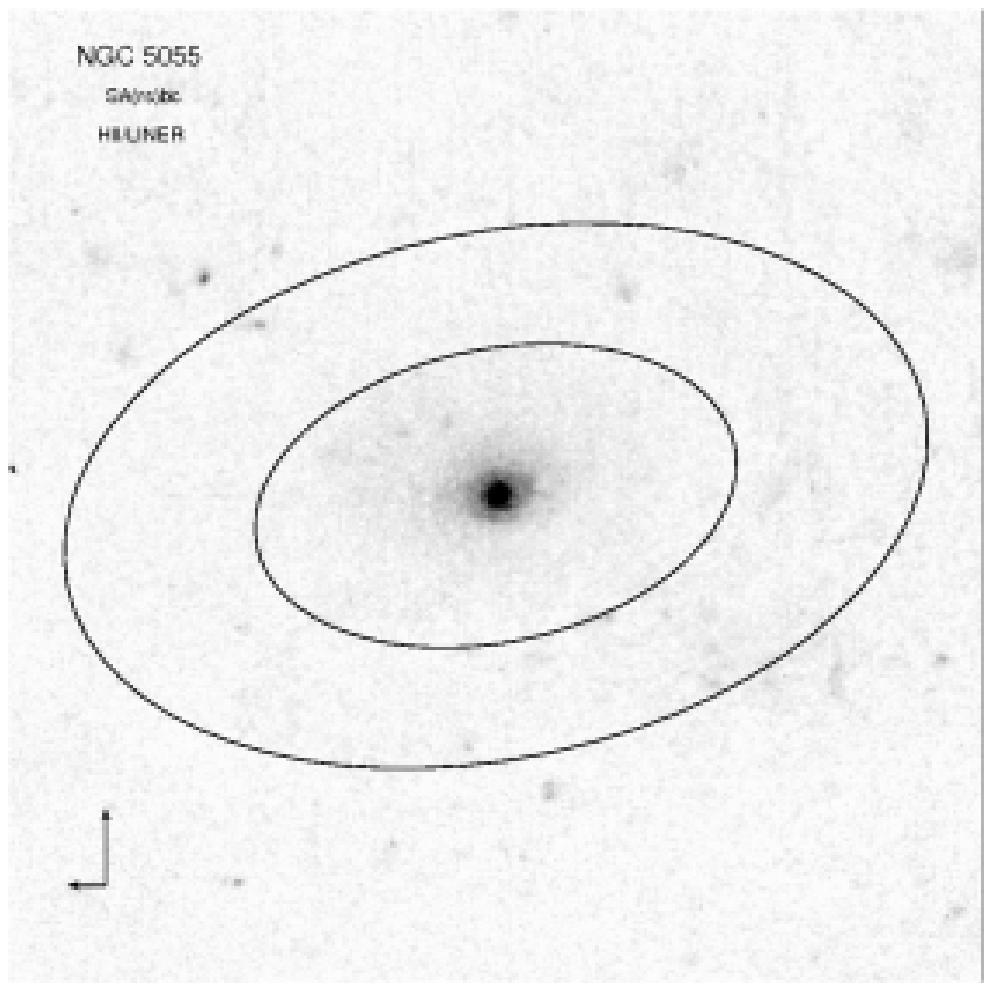}\\
\includegraphics[width=0.4\textwidth]{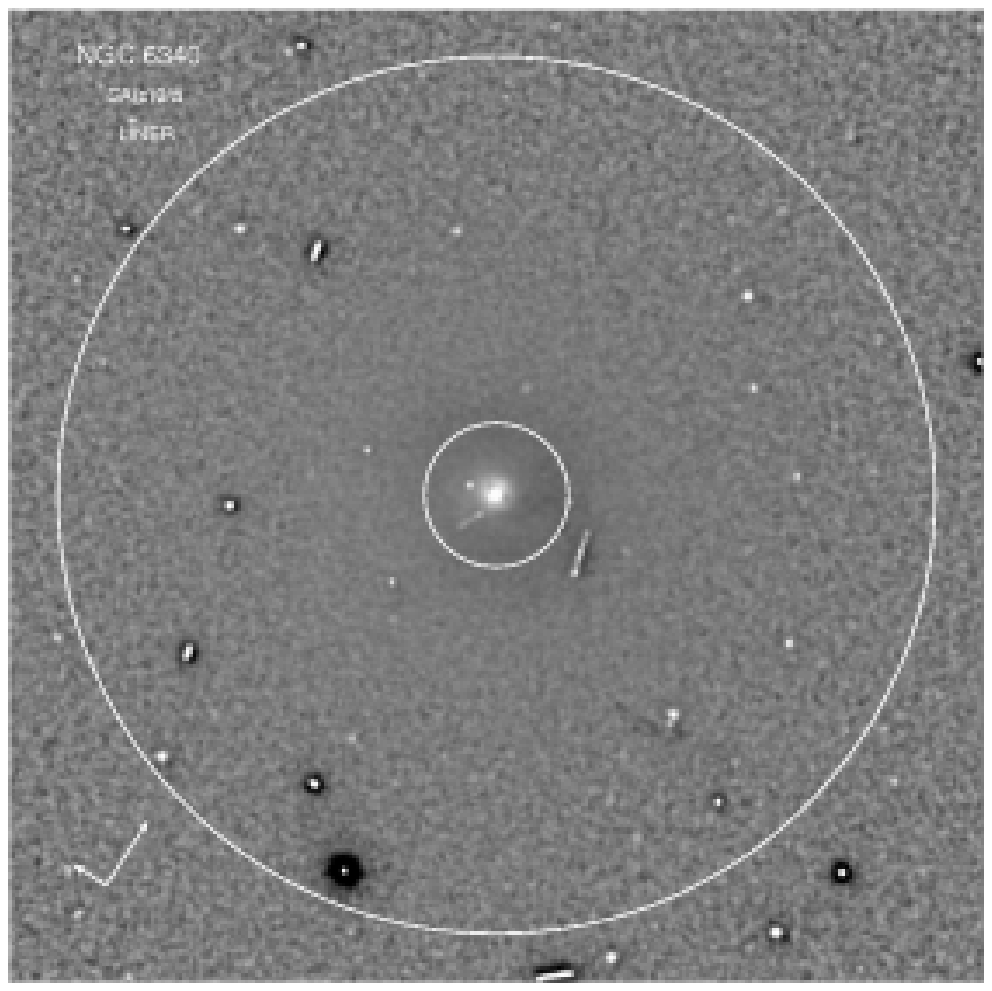}
\includegraphics[width=0.4\textwidth]{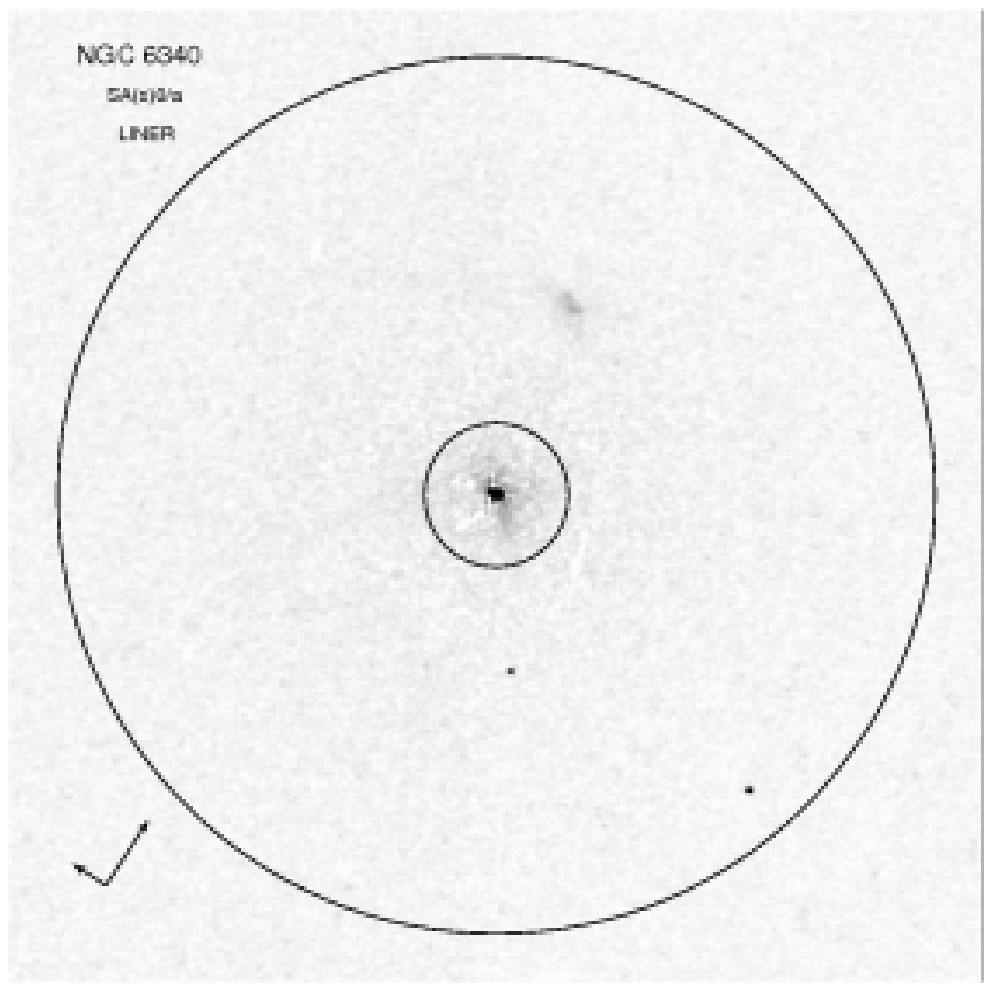}\\
\includegraphics[width=0.4\textwidth]{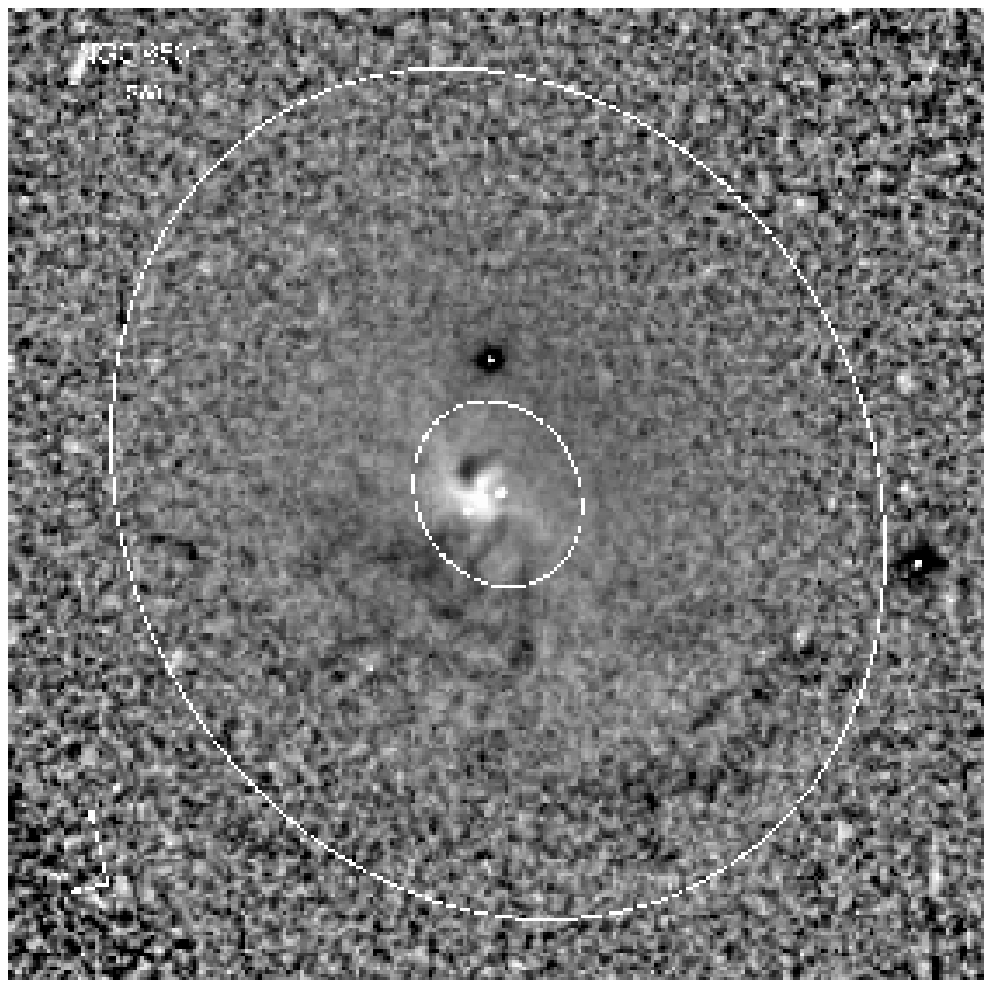}\\
\end{tabular}
\end{center}
\end{figure}

\begin{figure}[h]
\begin{center}
\begin{tabular}{c}
\includegraphics[width=0.4\textwidth]{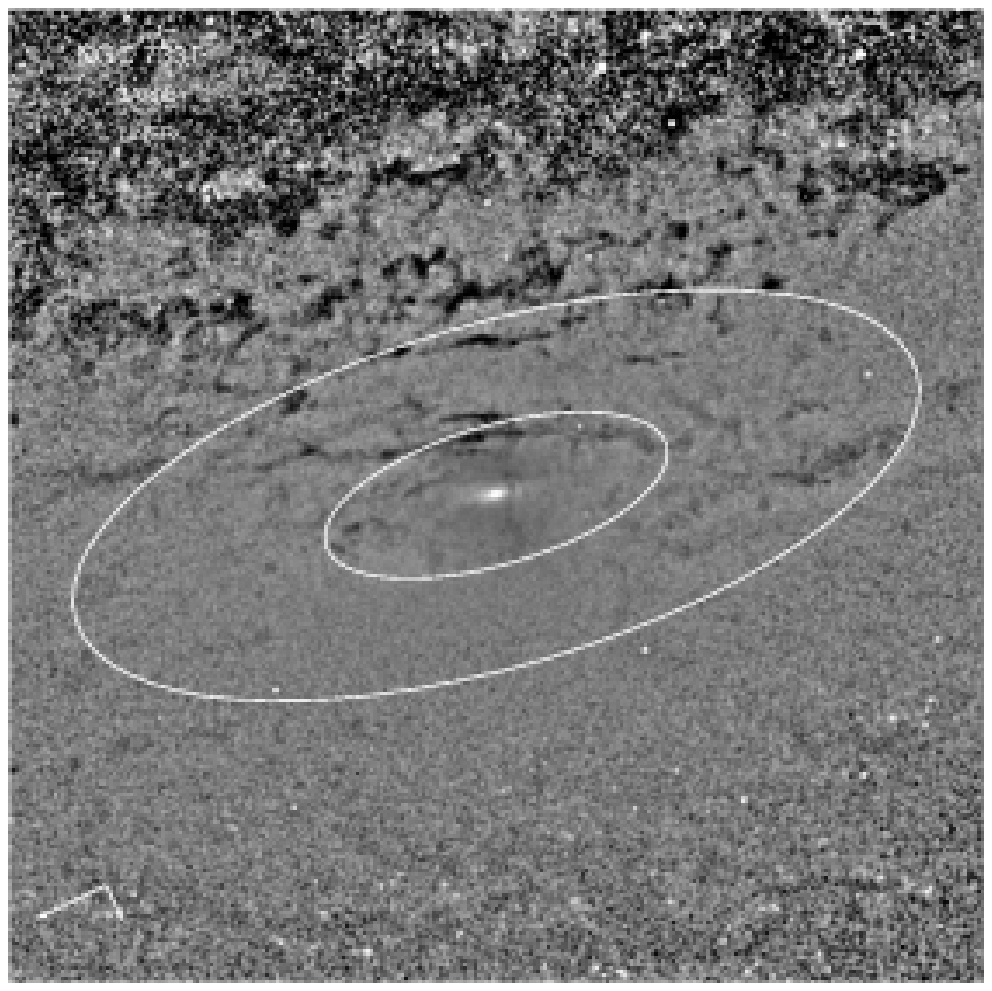}\\
\includegraphics[width=0.4\textwidth]{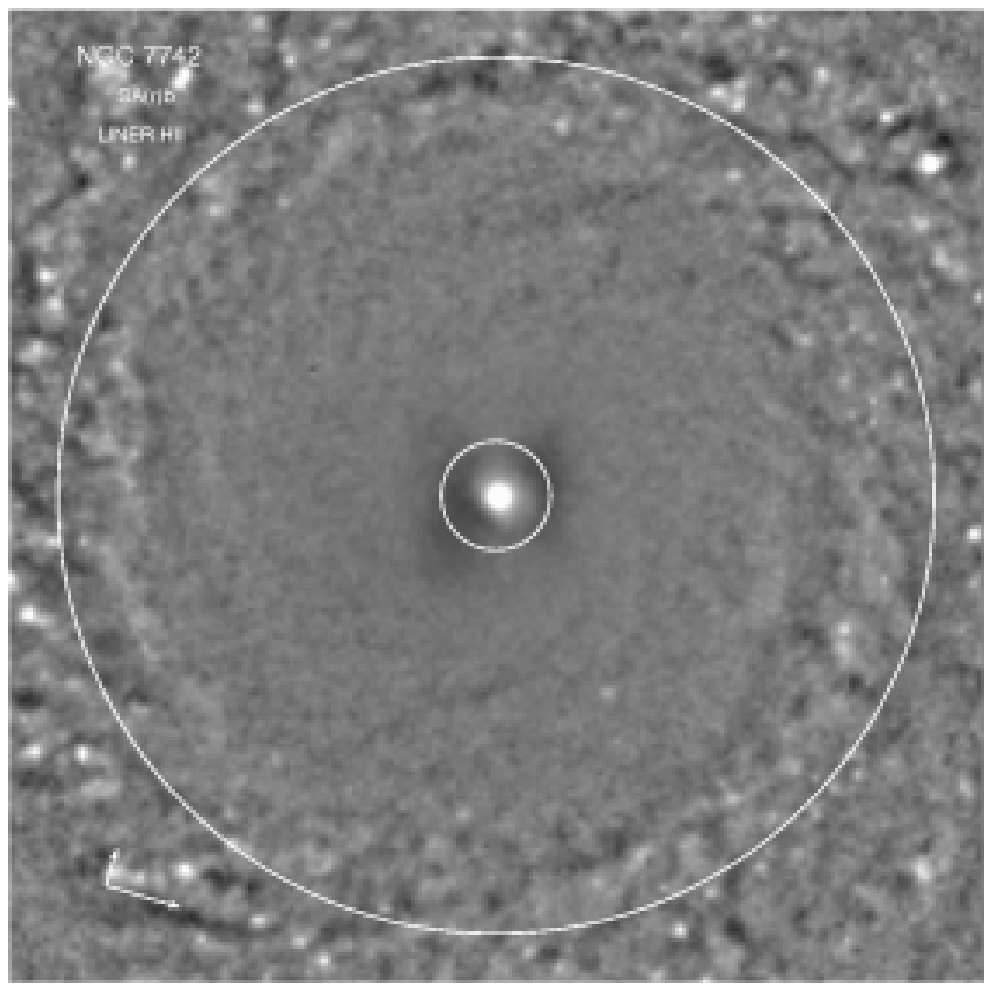}\\
\end{tabular}
\end{center}
\end{figure}

\clearpage

\section{Radial surface brightness profiles of  the $\sigma$-drop
  sample galaxies}
  
Exponential fits are indicated with dashed lines,
while the dotted lines indicate the sky level for each
galaxy. Vertical dashed-dotted lines correspond to the end of the
bar. Where there are two vertical lines, they indicate the lower and
the upper limit to the end of the bar. The squares show the limits of the intervals that we have used to fit the exponential discs.

\begin{figure}[h]
\begin{center}
\begin{tabular}{c}
\includegraphics[width=0.76\textwidth]{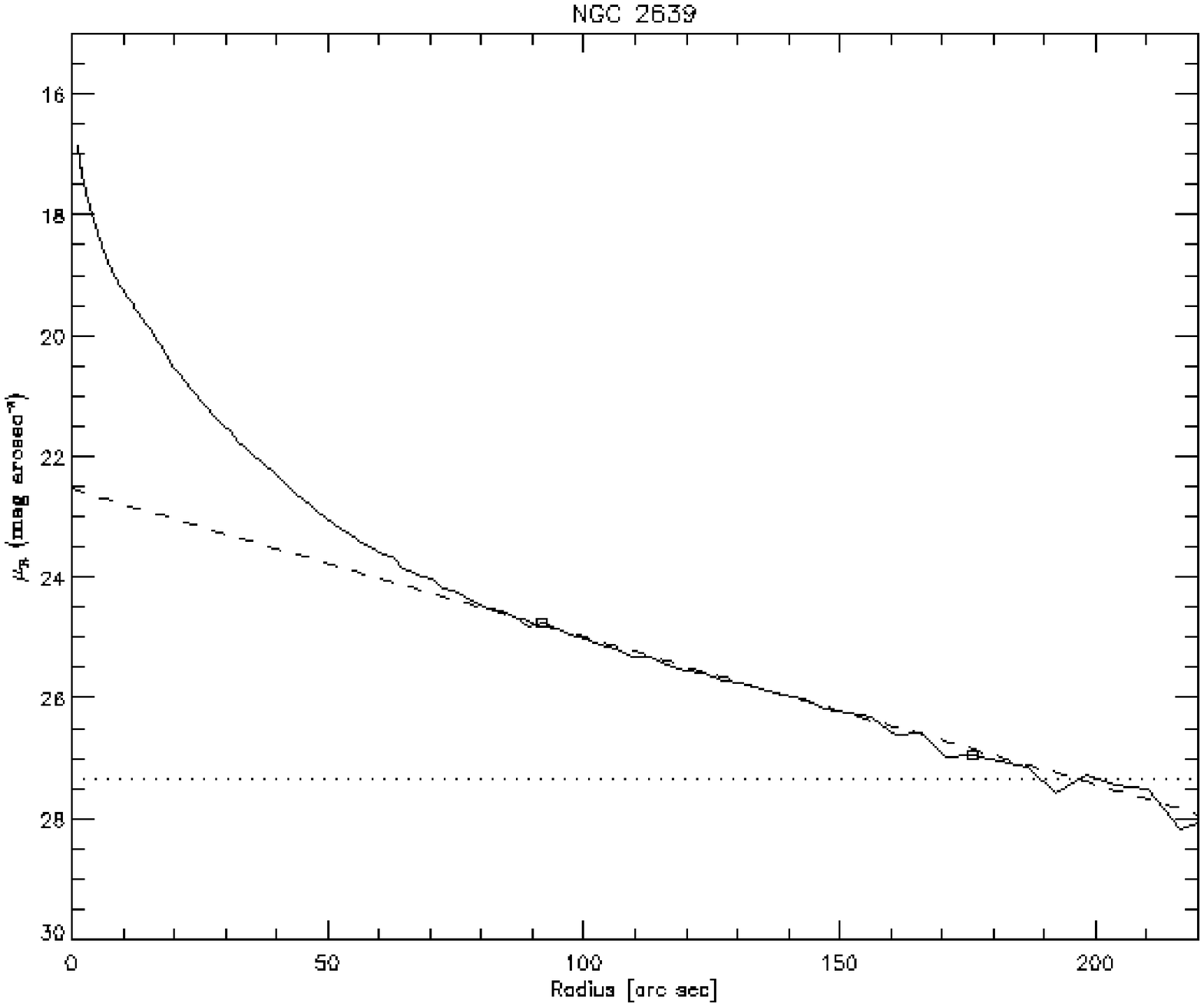}\\
\includegraphics[width=0.76\textwidth]{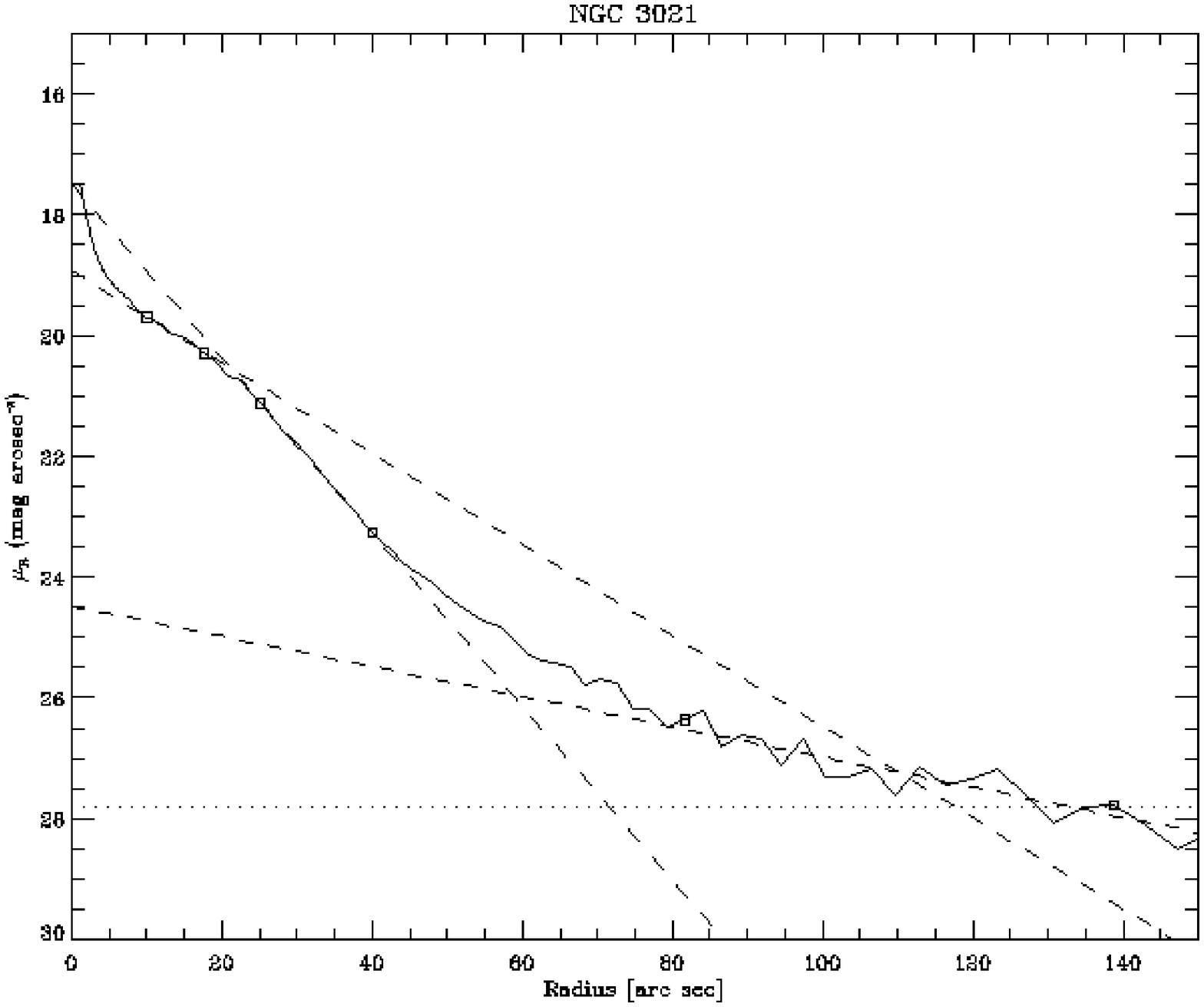}\\
\end{tabular}
\end{center}
\end{figure}

\begin{figure}
\begin{center}
\begin{tabular}{c}
\includegraphics[width=0.76\textwidth]{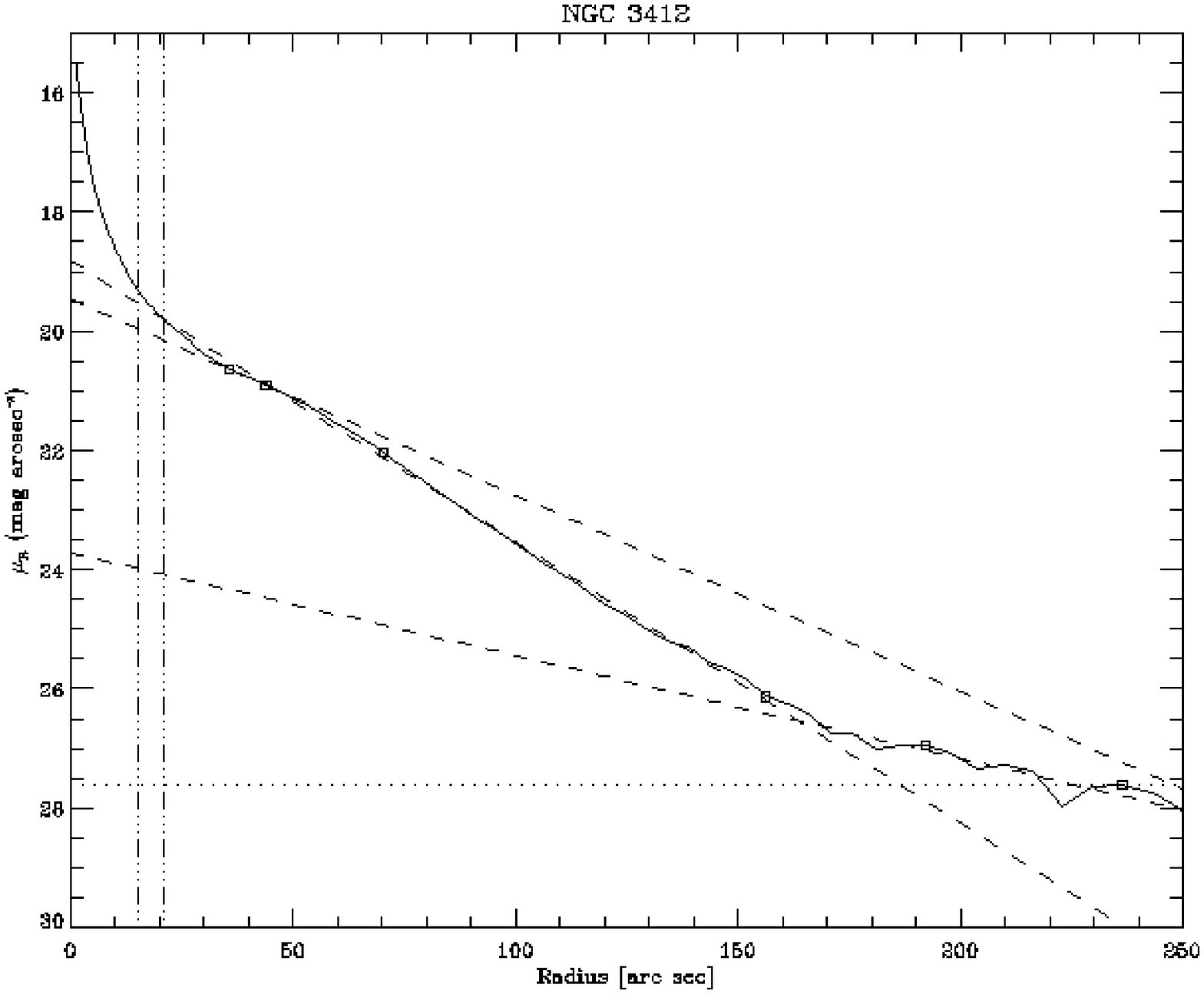}\\
\includegraphics[width=0.76\textwidth]{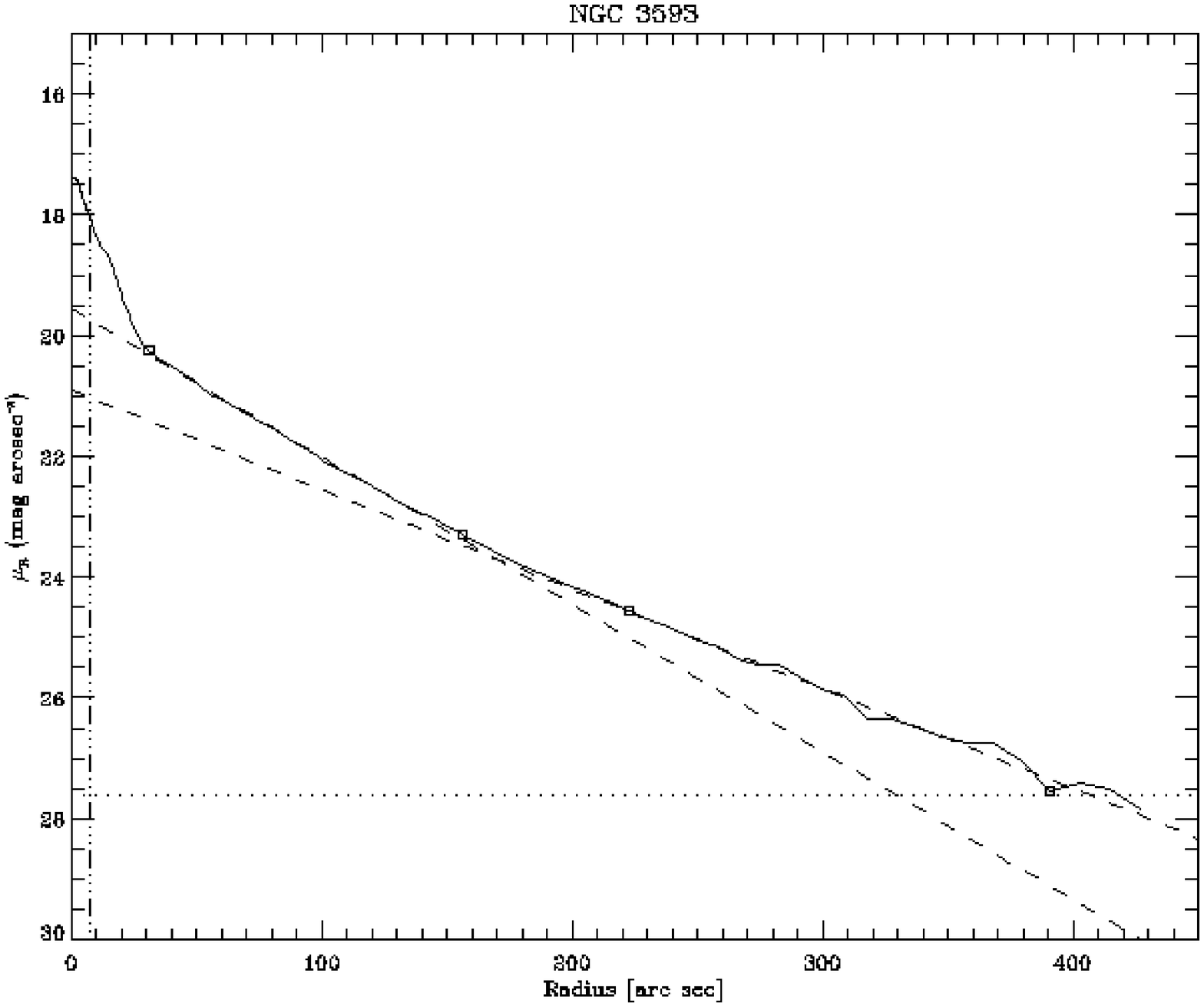}\\
\end{tabular}
\end{center}
\end{figure}

\begin{figure}
\begin{center}
\begin{tabular}{c}
\includegraphics[width=0.76\textwidth]{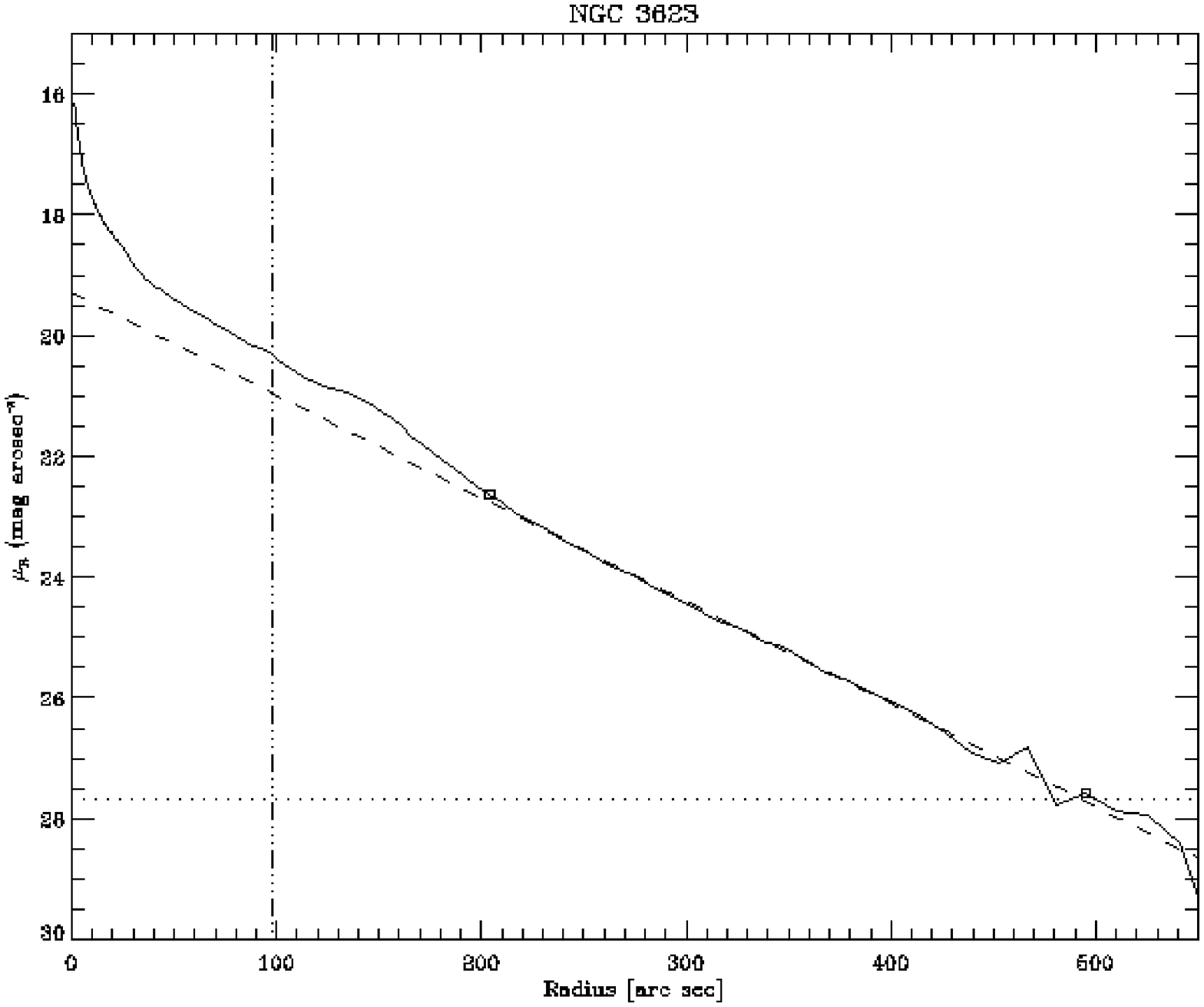}\\
\includegraphics[width=0.76\textwidth]{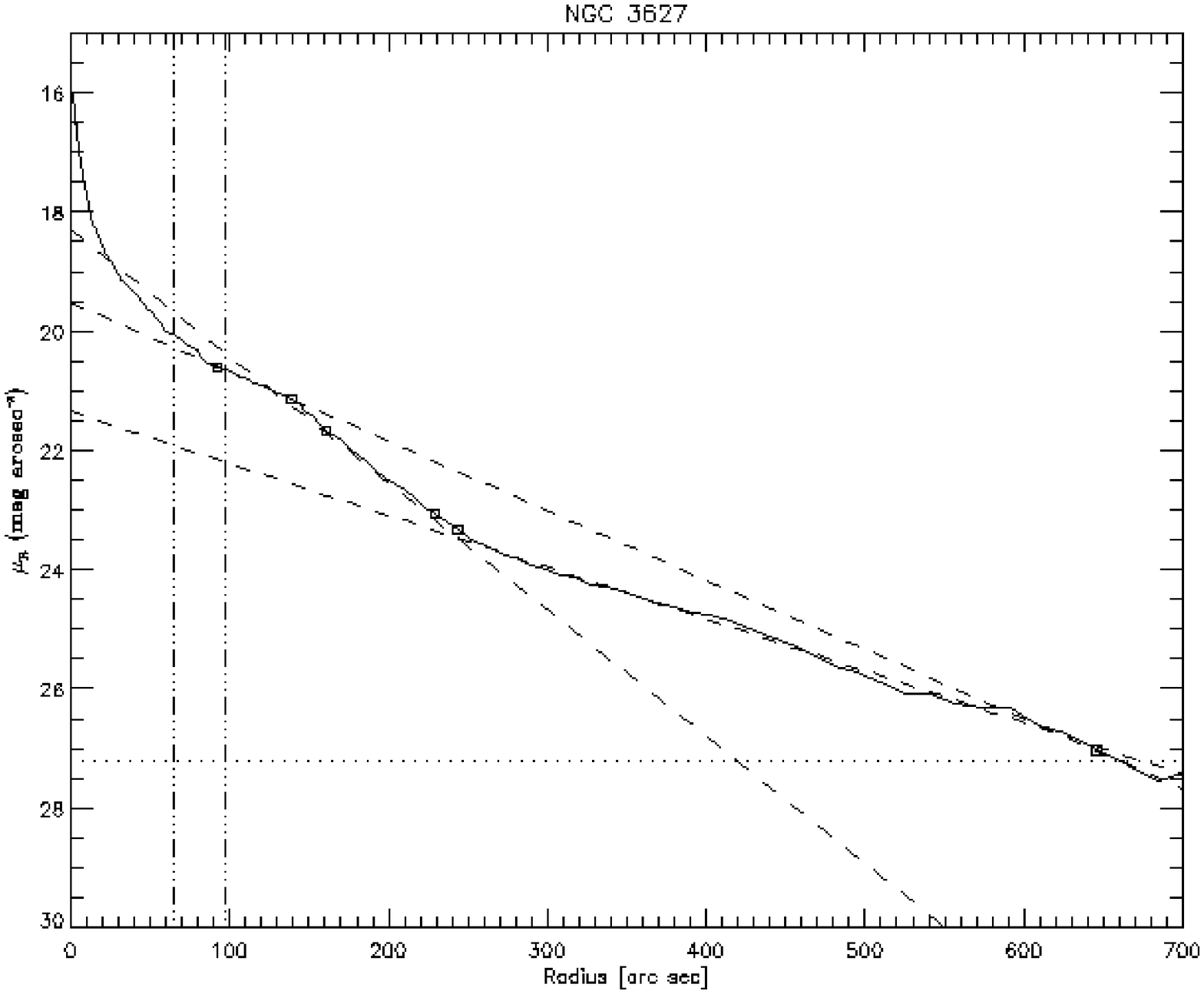}\\
\end{tabular}
\end{center}
\end{figure}

\begin{figure}
\begin{center}
\begin{tabular}{c}
\includegraphics[width=0.76\textwidth]{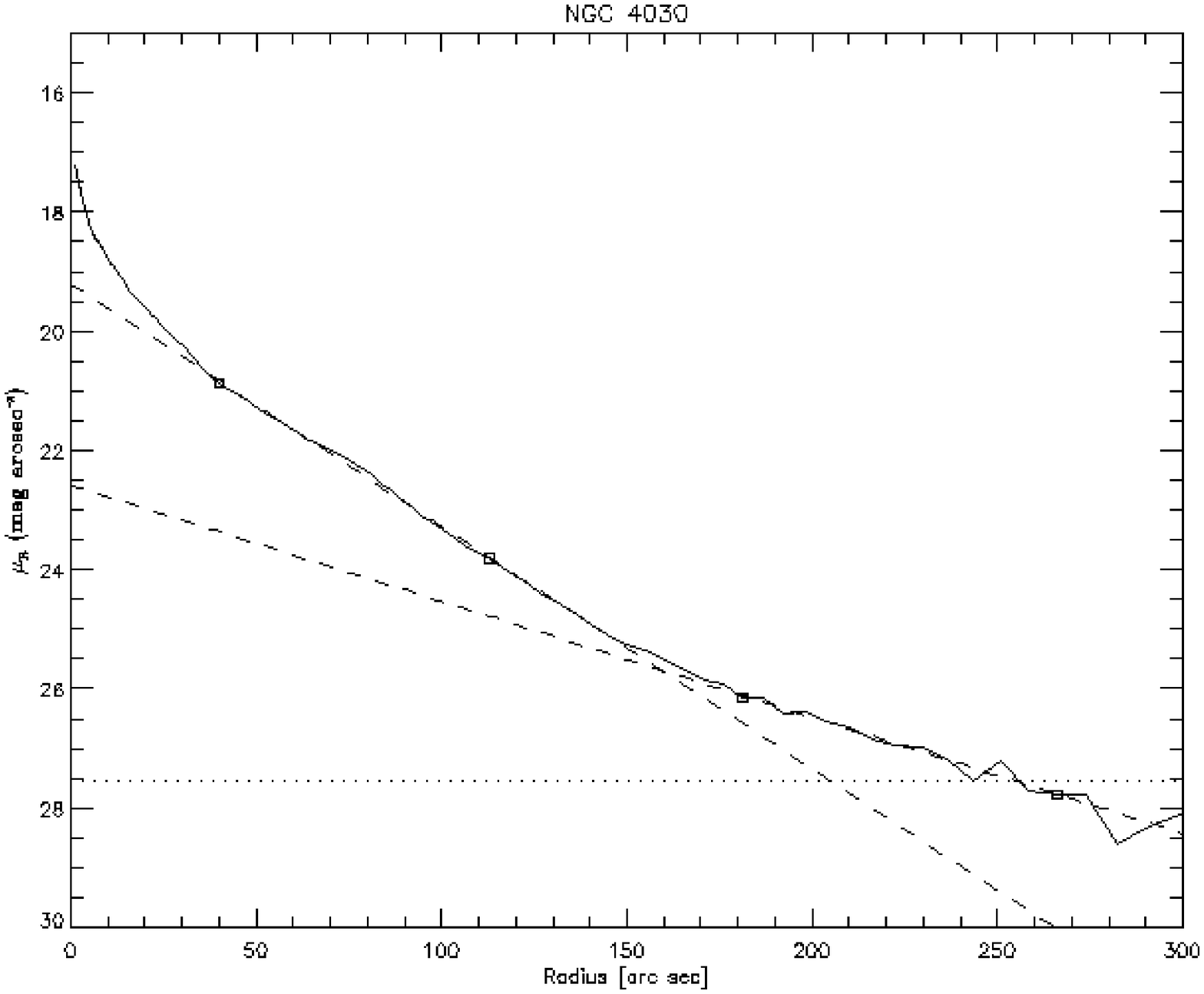}\\
\includegraphics[width=0.76\textwidth]{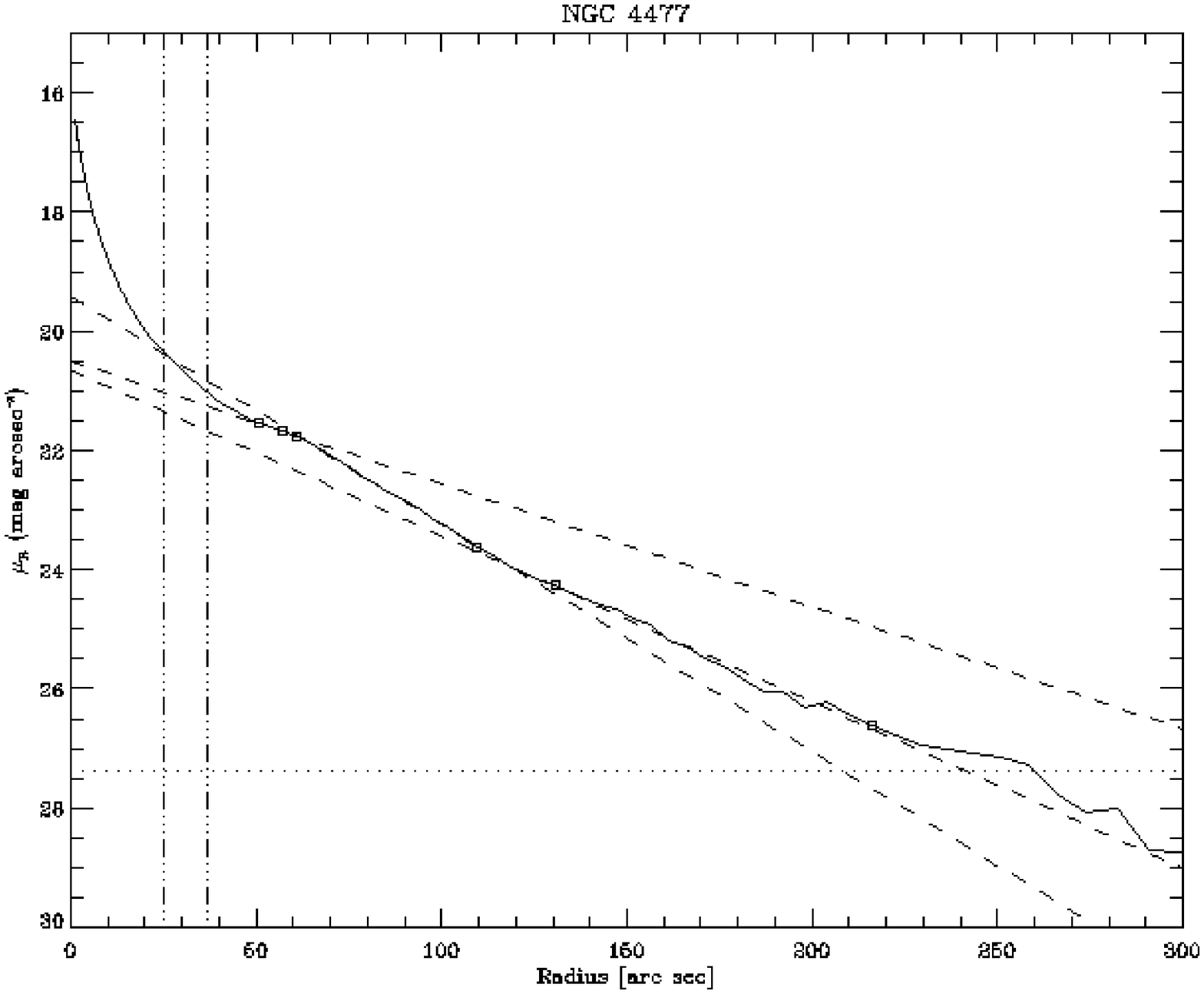}\\
\end{tabular}
\end{center}
\end{figure}

\begin{figure}
\begin{center}
\begin{tabular}{c}
\includegraphics[width=0.76\textwidth]{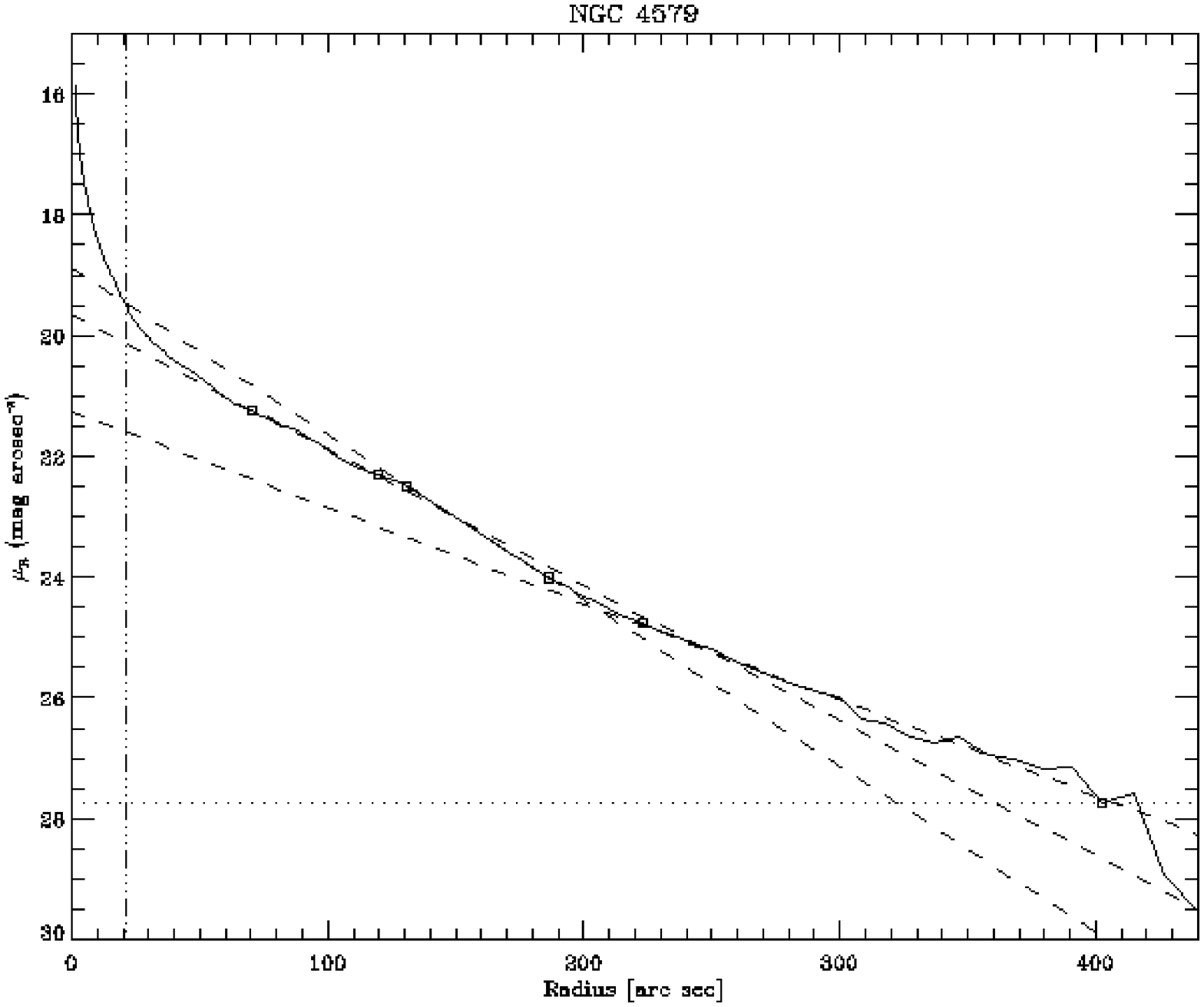}\\
\includegraphics[width=0.76\textwidth]{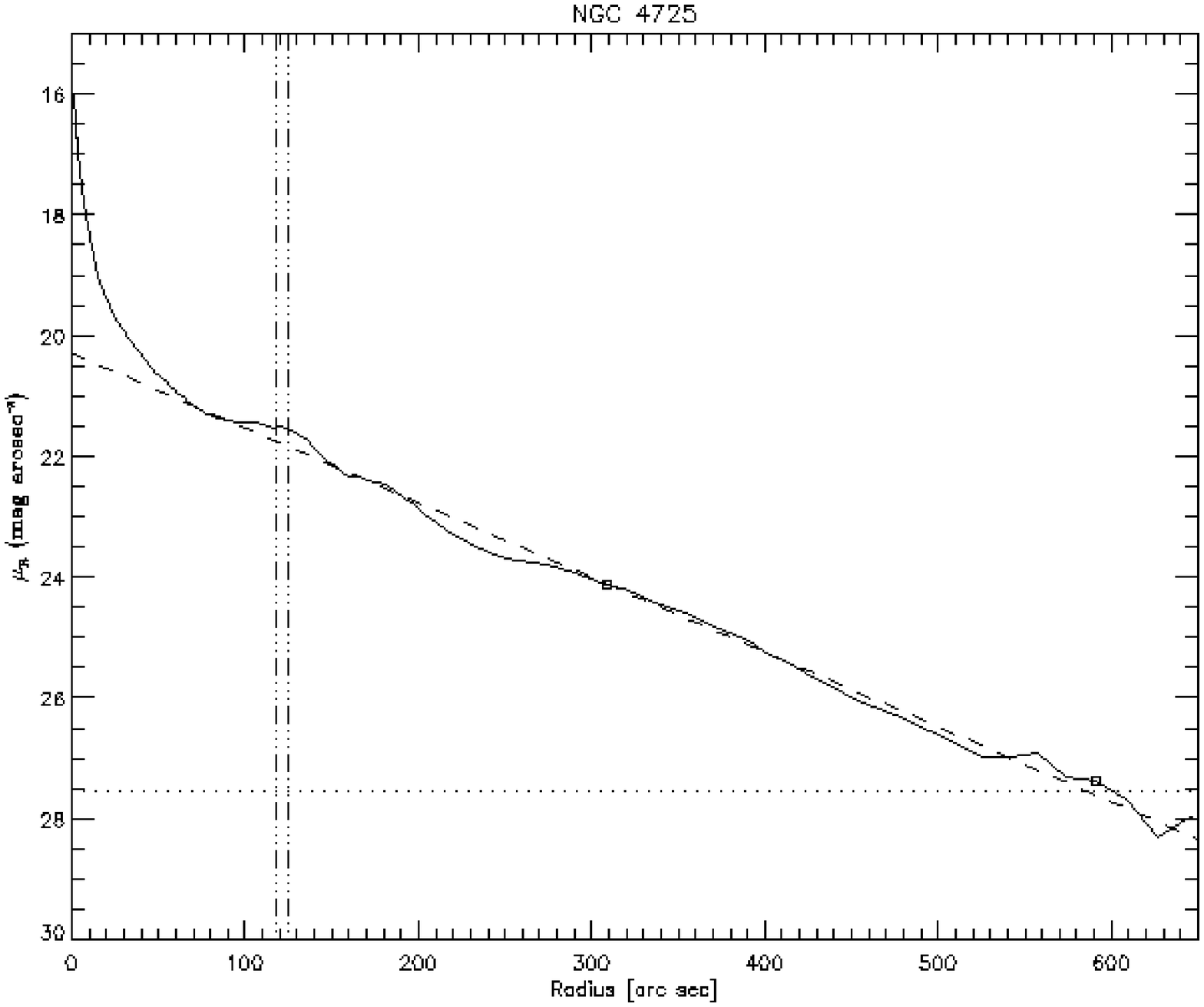}\\
\end{tabular}
\end{center}
\end{figure}

\clearpage

\section{As Appendix~C, now for the  control sample galaxies.}

\begin{figure}[h]
\begin{center}
\begin{tabular}{c}
\includegraphics[width=0.76\textwidth]{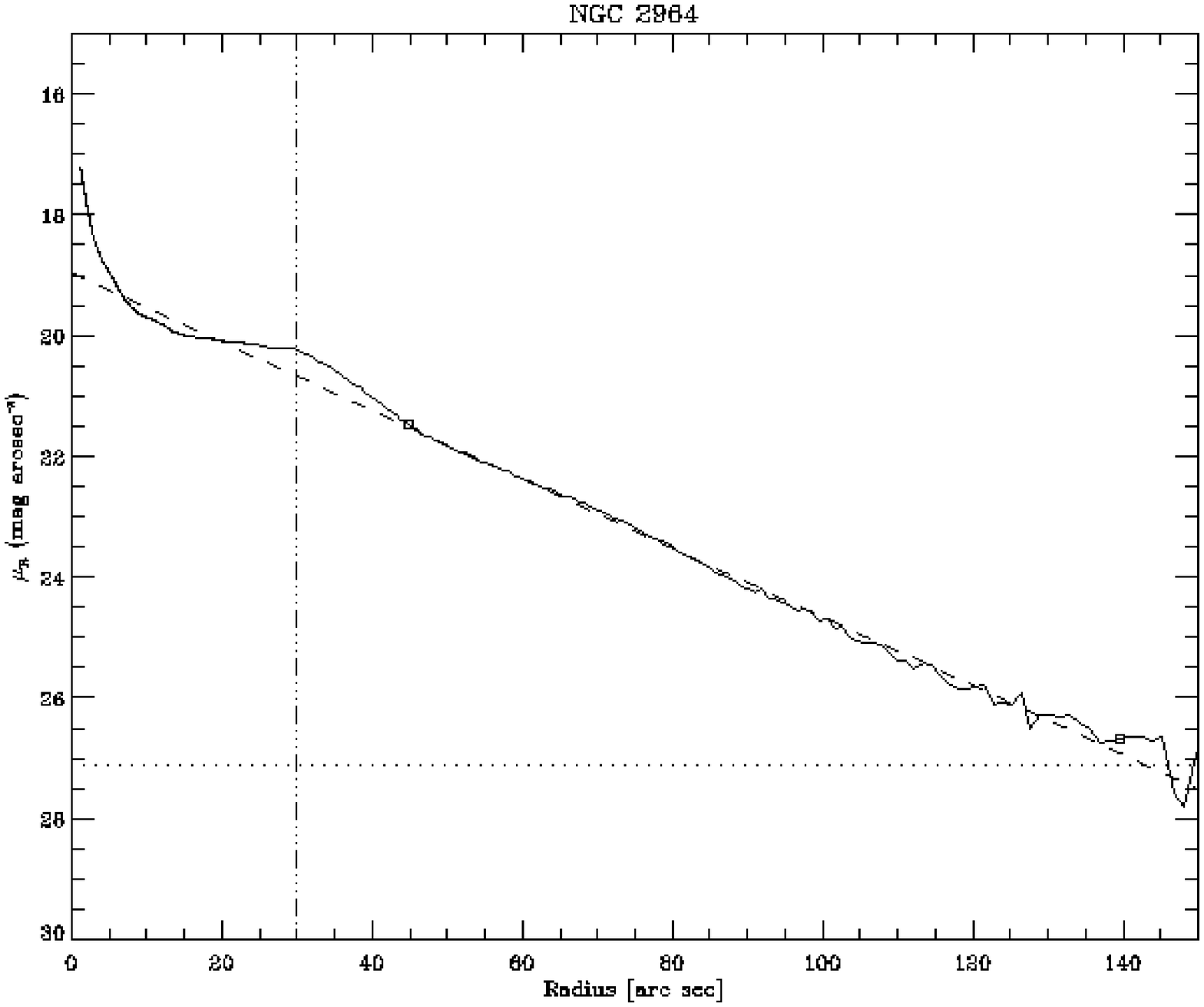}\\
\includegraphics[width=0.76\textwidth]{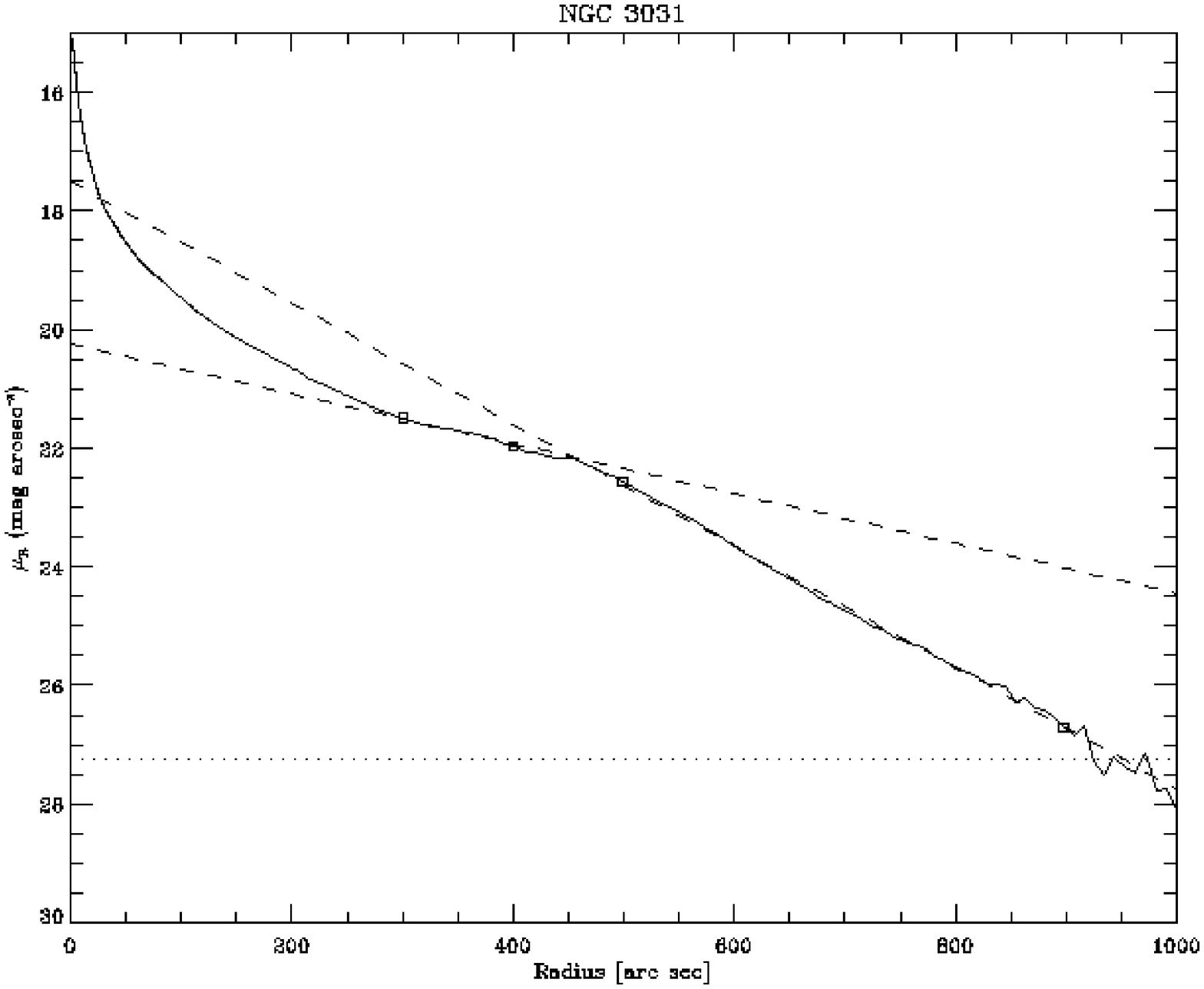}\\
\end{tabular}
\end{center}
\end{figure}

\begin{figure}
\begin{center}
\begin{tabular}{c}
\includegraphics[width=0.76\textwidth]{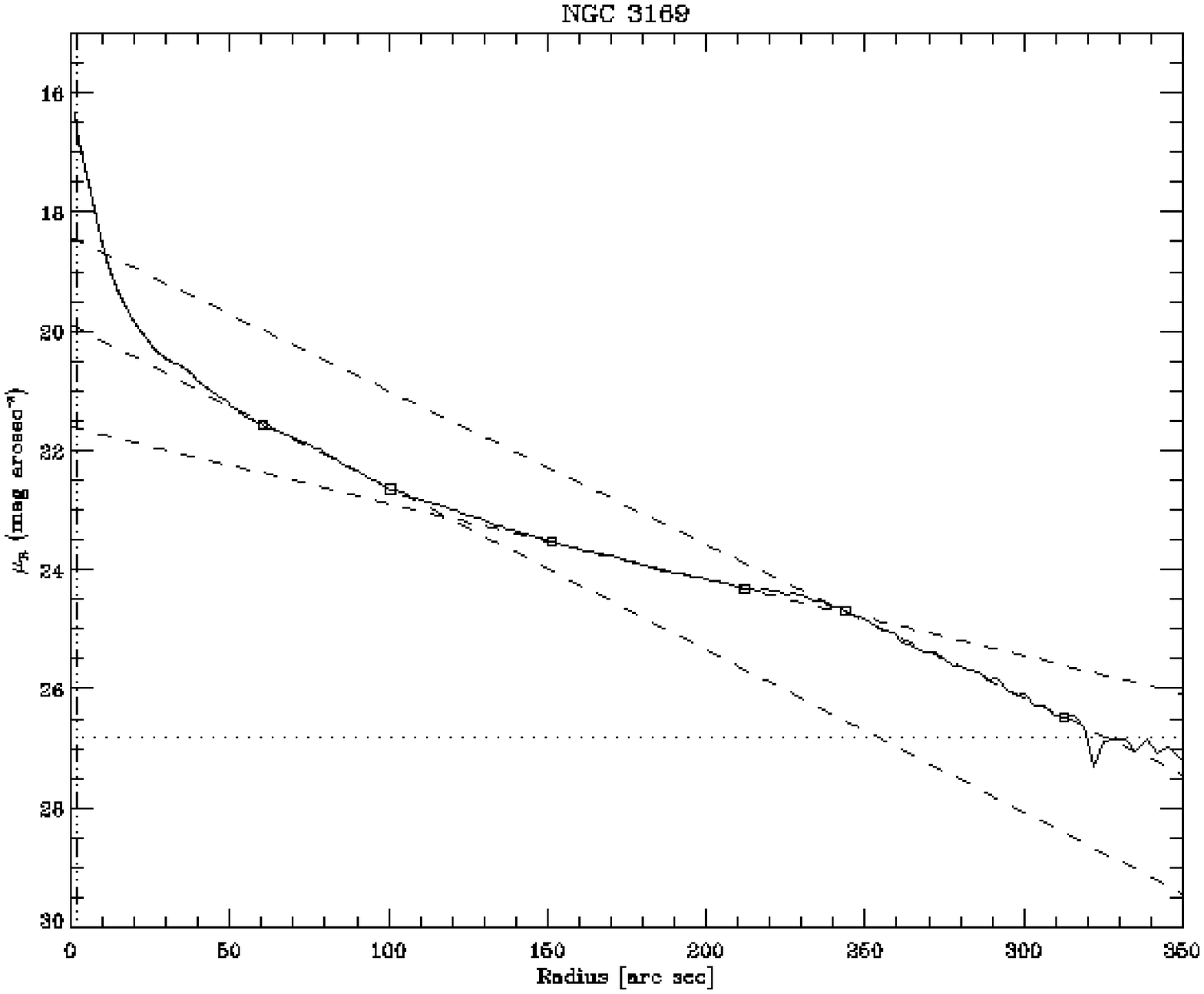}\\
\includegraphics[width=0.76\textwidth]{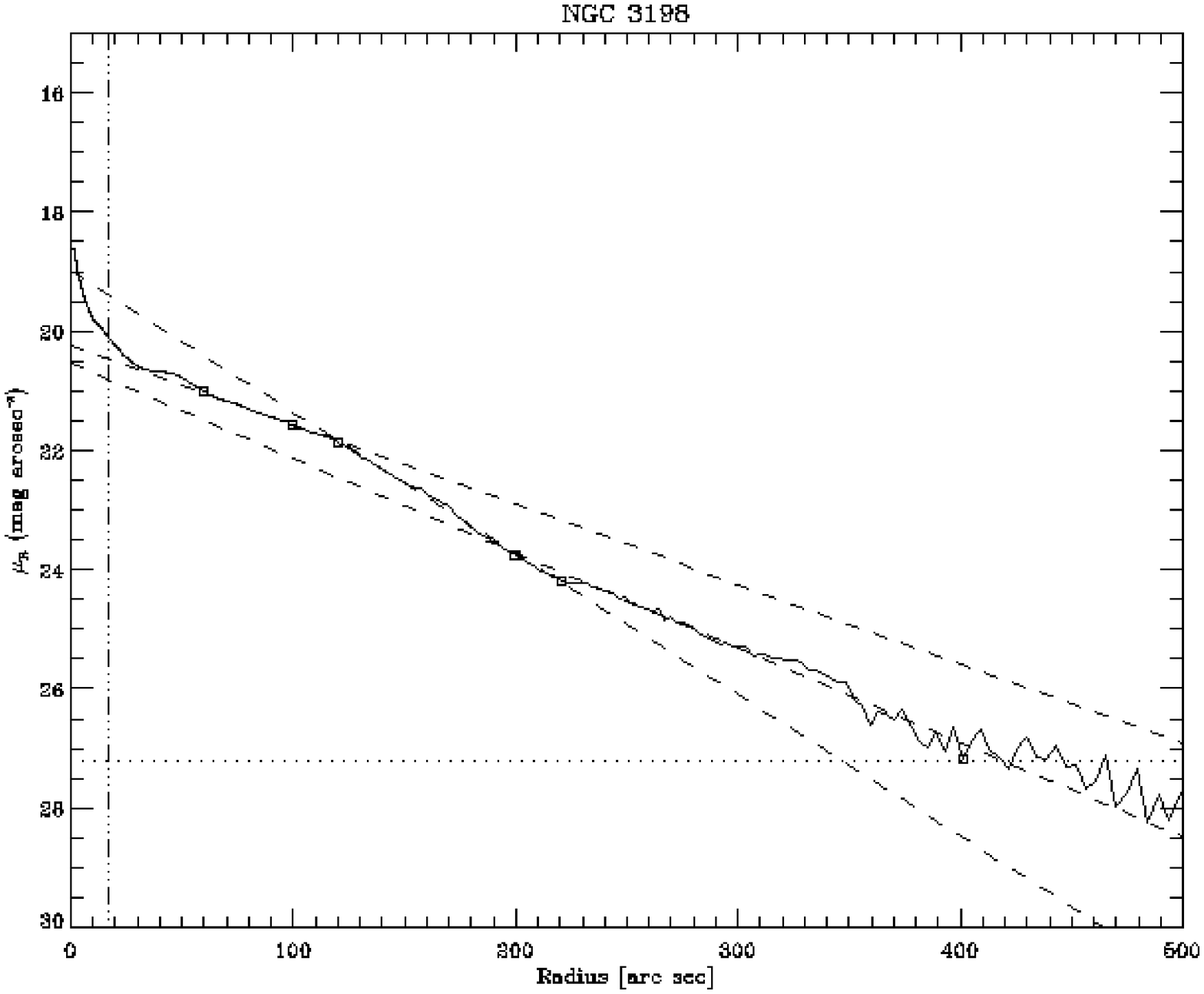}\\
\end{tabular}
\end{center}
\end{figure}

\begin{figure}
\begin{center}
\begin{tabular}{c}
\includegraphics[width=0.76\textwidth]{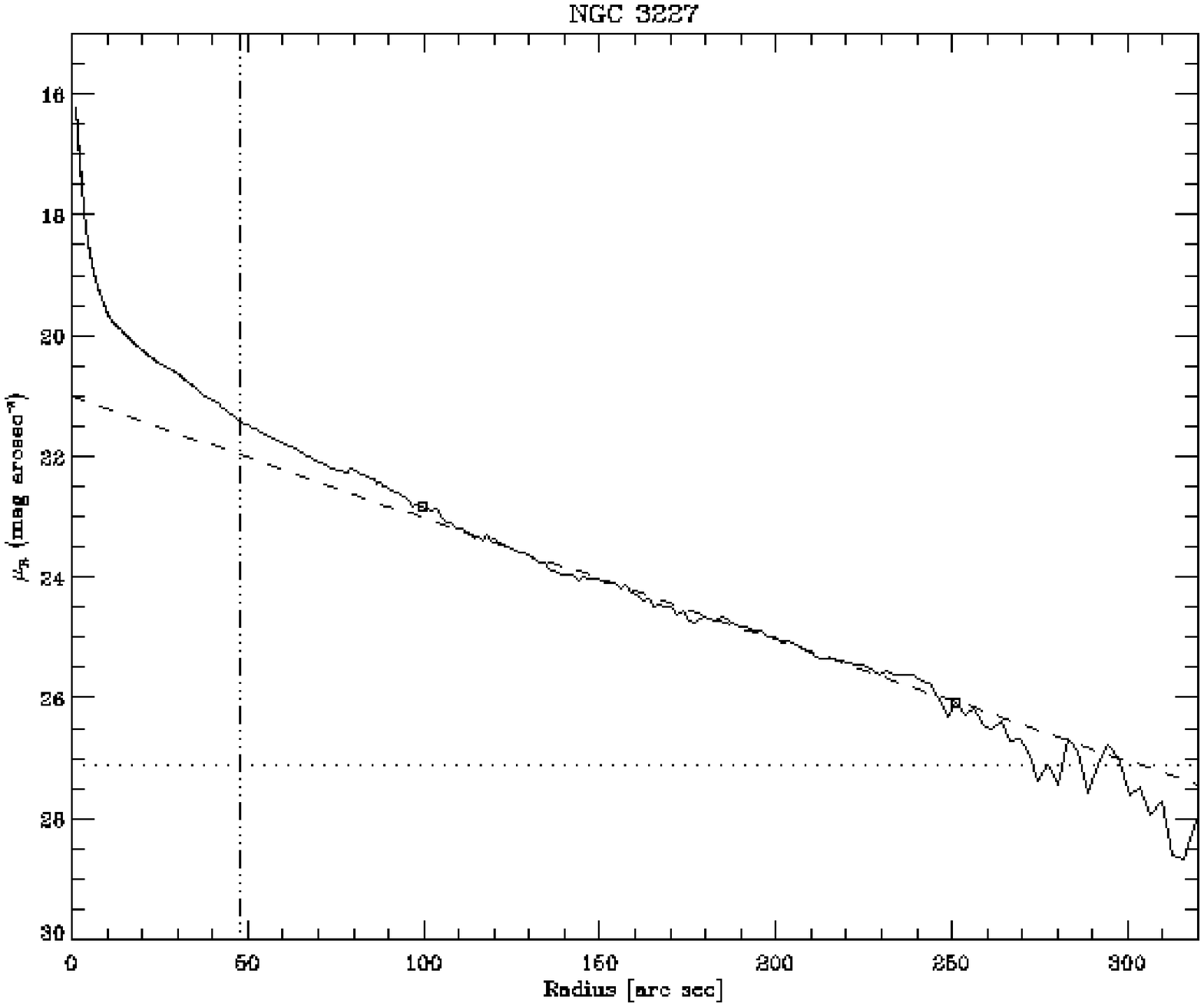}\\
\includegraphics[width=0.76\textwidth]{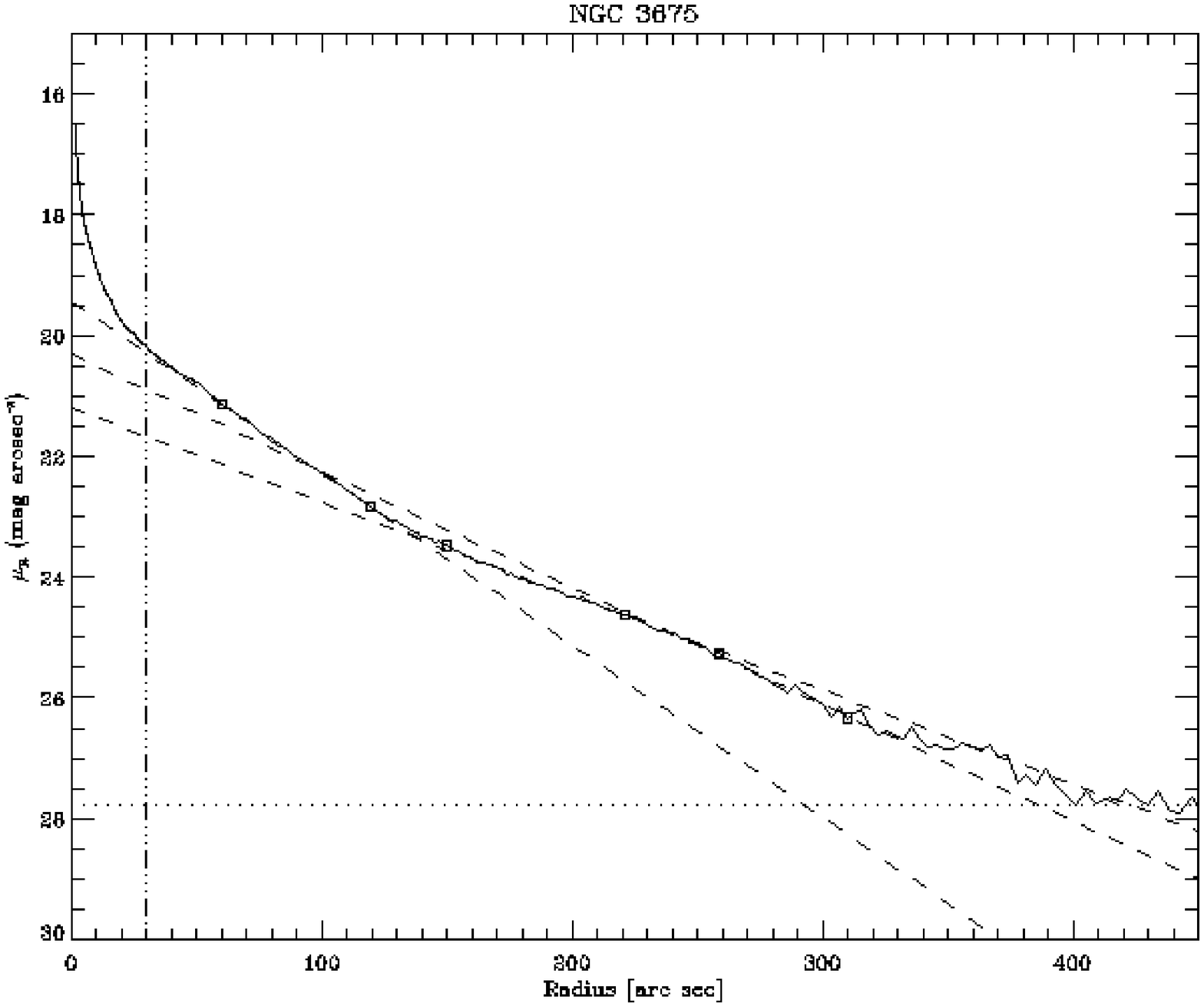}\\
\end{tabular}
\end{center}
\end{figure}

\begin{figure}
\begin{center}
\begin{tabular}{c}
\includegraphics[width=0.76\textwidth]{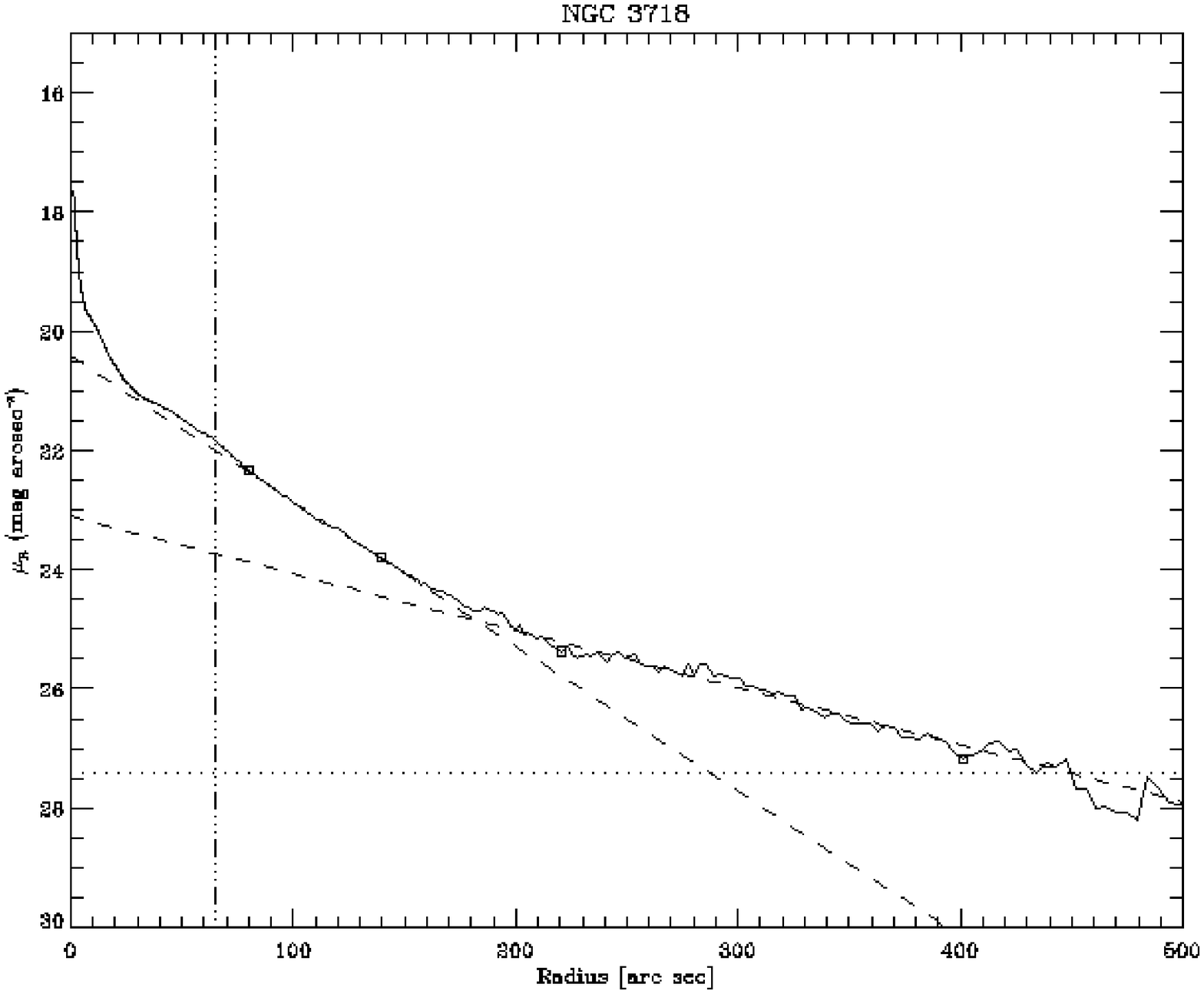}\\
\includegraphics[width=0.76\textwidth]{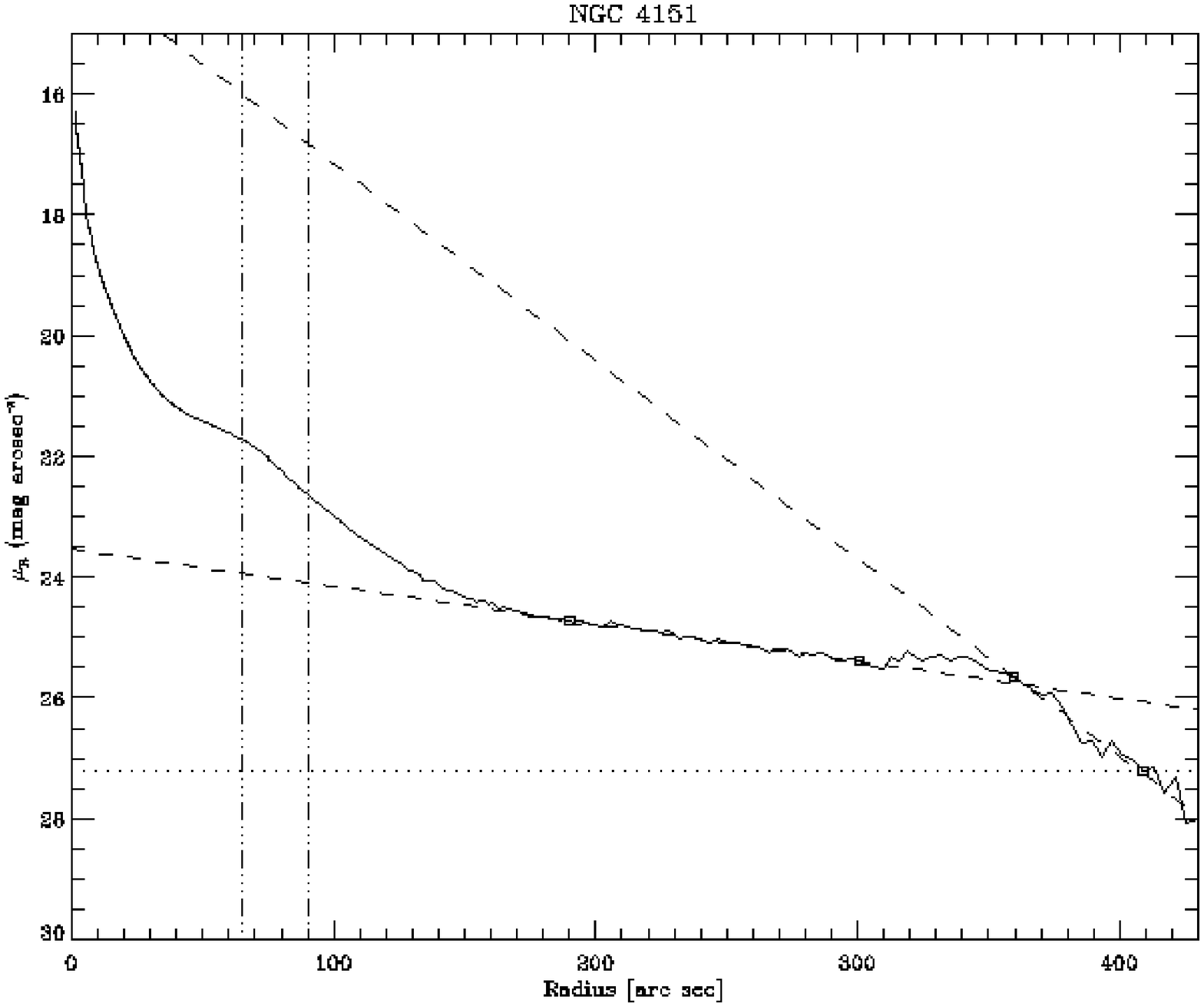}\\
\end{tabular}
\end{center}
\end{figure}

\begin{figure}
\begin{center}
\begin{tabular}{c}
\includegraphics[width=0.76\textwidth]{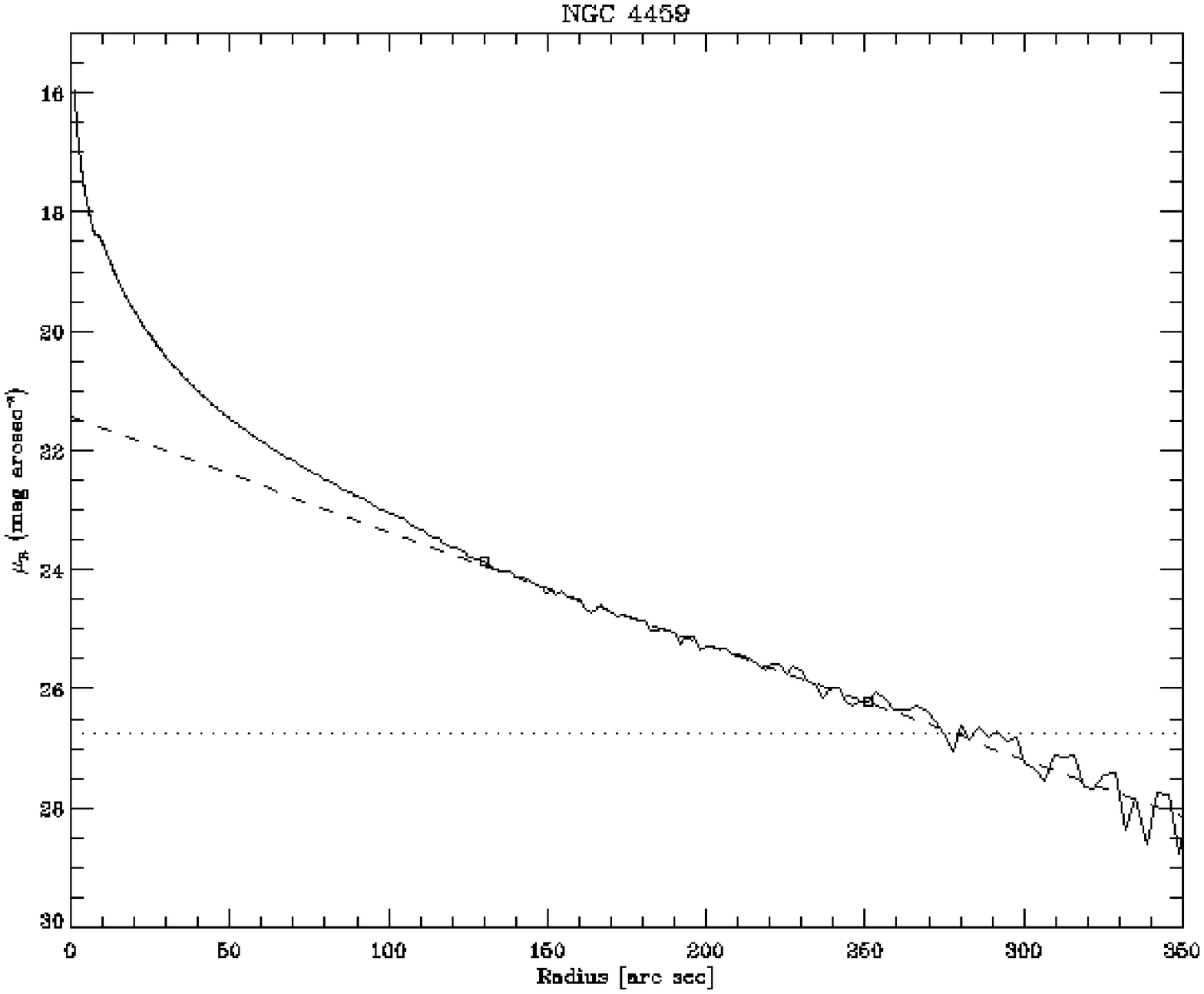}\\
\includegraphics[width=0.76\textwidth]{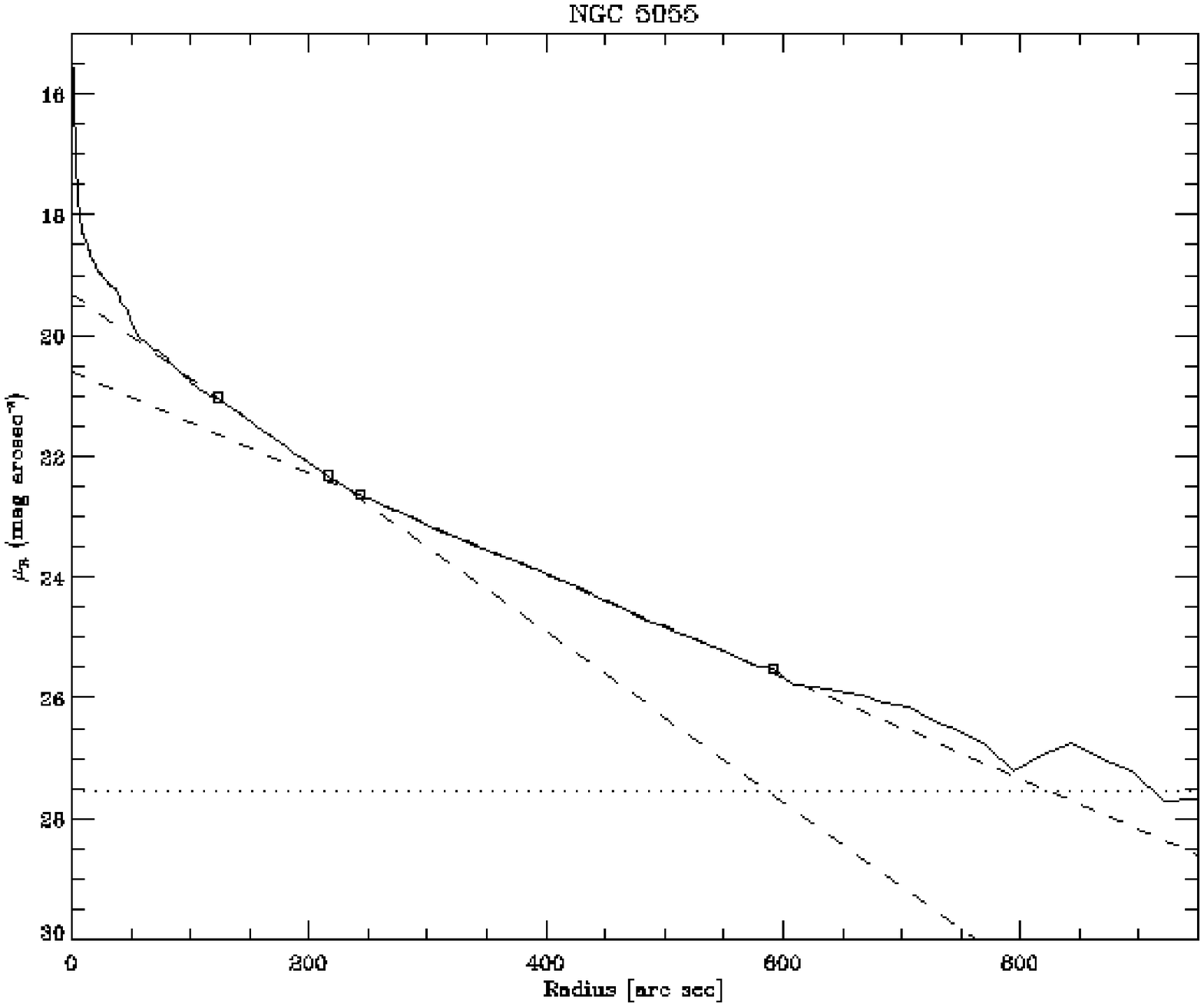}\\
\end{tabular}
\end{center}
\end{figure}

\end{document}